\documentclass[aps,prb,twocolumn,superscriptaddress,showpacs]{revtex4-1}
\usepackage{amsmath,amssymb,amsfonts,amsthm}
\usepackage{graphicx}
\usepackage{bm}
\usepackage{bbm}
\usepackage{dsfont}
\usepackage{xspace}
\usepackage{color}
\usepackage{dcolumn}   
\usepackage{epstopdf}
\usepackage[caption=false]{subfig}
\usepackage{color}
\usepackage[colorlinks=true,linkcolor=blue,urlcolor=blue,citecolor=blue]{hyperref}
\usepackage[normalem]{ulem}

\begin{document}

 \newcommand{\breite}{1.0} 

\newtheorem{prop}{Proposition}
\newtheorem{cor}{Corollary} 

\newcommand{\be}{\begin{equation}}
\newcommand{\ee}{\end{equation}}

\newcommand{\bea}{\begin{eqnarray}}
\newcommand{\eea}{\end{eqnarray}}
\newcommand{\lt}{<}
\newcommand{\gt}{>}

\newcommand{\Reals}{\mathbb{R}}     
\newcommand{\Com}{\mathbb{C}}       
\newcommand{\Nat}{\mathbb{N}}       

\newcommand{\id}{\mathbboldsymbol{1}}    

\newcommand{\Real}{\mathop{\mathrm{Re}}}
\newcommand{\Imag}{\mathop{\mathrm{Im}}}

\def\O{\mbox{$\mathcal{O}$}}   
\def\F{\mathcal{F}}			
\def\sgn{\text{sgn}}

\newcommand{\deo}{\ensuremath{\Delta_0}}
\newcommand{\dea}{\ensuremath{\Delta}}
\newcommand{\ak}{\ensuremath{a_k}}
\newcommand{\ad}{\ensuremath{a^{\dagger}_{-k}}}
\newcommand{\sx}{\ensuremath{\sigma_x}}
\newcommand{\sz}{\ensuremath{\sigma_z}}
\newcommand{\spl}{\ensuremath{\sigma_{+}}}
\newcommand{\smi}{\ensuremath{\sigma_{-}}}
\newcommand{\alk}{\ensuremath{\alpha_{k}}}
\newcommand{\bk}{\ensuremath{\beta_{k}}}
\newcommand{\ok}{\ensuremath{\omega_{k}}}
\newcommand{\vd}{\ensuremath{V^{\dagger}_1}}
\newcommand{\vi}{\ensuremath{V_1}}
\newcommand{\vo}{\ensuremath{V_o}}
\newcommand{\zc}{\ensuremath{\frac{E_z}{E}}}
\newcommand{\xc}{\ensuremath{\frac{\Delta}{E}}}
\newcommand{\xd}{\ensuremath{X^{\dagger}}}
\newcommand{\aok}{\ensuremath{\frac{\alk}{\ok}}}
\newcommand{\tpw}{\ensuremath{e^{i \ok s }}}
\newcommand{\tpe}{\ensuremath{e^{2iE s }}}
\newcommand{\tmw}{\ensuremath{e^{-i \ok s }}}
\newcommand{\tme}{\ensuremath{e^{-2iE s }}}
\newcommand{\epls}{\ensuremath{e^{F(s)}}}
\newcommand{\emis}{\ensuremath{e^{-F(s)}}}
\newcommand{\epl}{\ensuremath{e^{F(0)}}}
\newcommand{\emi}{\ensuremath{e^{F(0)}}}
\newcommand{\g} [1]{{\color{blue}  #1}}
\newcommand{\gr} [1]{{\color{red}  #1}}

\newcommand{\lr}[1]{\left( #1 \right)}
\newcommand{\lrs}[1]{\left( #1 \right)^2}
\newcommand{\lrb}[1]{\left< #1\right>}
\newcommand{\nbt}{\ensuremath{\lr{ \lr{n_k + 1} \tmw + n_k \tpw  }}}

\newcommand{\om}{\ensuremath{\omega}}
\newcommand{\dw}{\ensuremath{\Delta_0}}
\newcommand{\wbp}{\ensuremath{\omega_0}}
\newcommand{\dv}{\ensuremath{\Delta_0}}
\newcommand{\vbp}{\ensuremath{\nu_0}}
\newcommand{\vplus}{\ensuremath{\nu_{+}}}
\newcommand{\vminus}{\ensuremath{\nu_{-}}}
\newcommand{\wplus}{\ensuremath{\omega_{+}}}
\newcommand{\wminus}{\ensuremath{\omega_{-}}}
\newcommand{\uv}[1]{\ensuremath{\mathbf{\hat{#1}}}} 
\newcommand{\abs}[1]{\left| #1 \right|} 
\newcommand{\avg}[1]{\left< #1 \right>} 
\let\underdot=\d 
\renewcommand{\d}[2]{\frac{d #1}{d #2}} 
\newcommand{\dd}[2]{\frac{d^2 #1}{d #2^2}} 
\newcommand{\pd}[2]{\frac{\partial #1}{\partial #2}} 
\newcommand{\pdd}[2]{\frac{\partial^2 #1}{\partial #2^2}} 
\newcommand{\pdc}[3]{\left( \frac{\partial #1}{\partial #2}
 \right)_{#3}} 
\newcommand{\ket}[1]{\left| #1 \right>} 
\newcommand{\bra}[1]{\left< #1 \right|} 
\newcommand{\braket}[2]{\left< #1 \vphantom{#2} \right|
 \left. #2 \vphantom{#1} \right>} 
\newcommand{\matrixel}[3]{\left< #1 \vphantom{#2#3} \right|
 #2 \left| #3 \vphantom{#1#2} \right>} 
\newcommand{\grad}[1]{{\nabla} {#1}} 
\let\divsymb=\div 
\renewcommand{\div}[1]{{\nabla} \cdot \boldsymbol{#1}} 
\newcommand{\curl}[1]{{\nabla} \times \boldsymbol{#1}} 
\newcommand{\laplace}[1]{\nabla^2 \boldsymbol{#1}}
\newcommand{\vs}[1]{\boldsymbol{#1}}
\let\baraccent=\= 

\definecolor{dgreen}{rgb}{0.0, 0.5, 0.0}
\newcommand{\gtext}[1]{\textcolor{dgreen}{{[#1]}}}
\newcommand{\rtext}[1]{\textcolor{red}{{#1}}}
\newcommand{\btext}[1]{\textcolor{blue}{{#1}}}
\newcommand{\vtext}[1]{\textcolor{violet}{{#1}}}
\newcommand{\oldtext}[1]{\textcolor{blue}{\sout{#1}}}


\title{Magnetic noise spectroscopy as a probe of local electronic correlations in two-dimensional systems}

\author{Kartiek Agarwal}
\affiliation{Physics Department, Harvard University, Cambridge, Massachusetts 02138, USA}
\email[]{agarwal@physics.harvard.edu}
\author{Richard Schmidt}
\affiliation{ITAMP, Harvard-Smithsonian Center for Astrophysics, 60 Garden Street, Cambridge, MA 02138, USA}
\author{Bertrand Halperin}
\affiliation{Physics Department, Harvard University, Cambridge, Massachusetts 02138, USA}
\author{Vadim Oganesyan}
\affiliation{Department of Engineering \& Physics, CSI, CUNY, and The Graduate Center of CUNY, New York, NY 10016, USA}
\author{Gergely Zar\'and}
\affiliation{Department of Theoretical Physics and BME-MTA Exotic Quantum Phases Research Group, Budapest University of Technology and Economics, 1521 Budapest, Hungary}
\author{Mikhail D. Lukin}
\affiliation{Physics Department, Harvard University, Cambridge, Massachusetts 02138, USA}
\author{Eugene Demler}
\affiliation{Physics Department, Harvard University, Cambridge, Massachusetts 02138, USA}

\date{\today}
\begin{abstract}
We develop the theoretical framework for calculating magnetic noise from conducting two-dimensional (2D) materials. We describe how local measurements of this noise can directly probe the wave-vector dependent transport properties of the material over a broad range of length scales, thus providing new insight into a range of correlated phenomena in 2D electronic systems. As an example, we demonstrate how transport in the hydrodynamic regime in an electronic system exhibits a unique signature in the magnetic noise profile that distinguishes it from diffusive and ballistic transport and how it can be used to measure the viscosity of the electronic fluid. We employ a Boltzmann approach in a two-time relaxation-time approximation to compute the conductivity of graphene and quantitatively illustrate these transport regimes and the experimental feasibility of observing them. Next, we discuss signatures of isolated impurities lodged inside the conducting 2D material. The noise near an impurity is found to be suppressed compared to the background by an amount that is directly proportional to the cross-section of electrons/holes scattering off of the impurity. We use these results to outline an experimental proposal to measure the temperature dependent level-shift and line-width of the resonance associated with an Anderson impurity.
\end{abstract}
\maketitle
	
\section{Introduction}

Nitrogen Vacancy (NV) centers are atom-like defects in diamond that can be operated as quantum bits~\cite{DuttNV,fuchsNV,MaurerNV} with extremely long coherence times. The transition frequency between NV-spin states is highly sensitive to magnetic fields, which allows accurate measurements of local, static fields. In turn, the relaxation rate of the NV-center is sensitive to magnetic fluctuations at its site and can be used to probe the magnetic noise at its transition frequency. Recent experiments have used these characteristics to demonstrate that NV centers can be used as highly spatially-resolved probes of novel physical phenomena across a range of materials. For instance, NV centers have been employed to image local magnetic textures~\cite{tetienne2015nature} and to probe spin-wave excitations~\cite{francescoNV} at the nanometer scale in ferromagnetic materials, and magnetic noise spectroscopy was used to detect the difference in transport properties between single- and poly-crystalline metallic slabs~\cite{mishaNV}. Magnetic resonance imaging at the single proton level~\cite{ShuskovNV,lovchinsky2016nuclear} has also been performed using these devices. 

In this work, we focus on NV centers as probes of magnetic noise from many-body systems, and we discuss how various non-local transport phenomena can be inferred from such measurements. This analysis, in part, is motivated by the effectiveness of traditional NMR spectroscopy in the study of strongly correlated electronic systems. In NMR studies, electronic spin fluctuations are probed by studying the relaxation of nuclear spins inside the material. These fluctuations are often tied to electronic correlations~\cite{StonerNoise,moriyacorrelationNMR,PinesNMR,Fuldedisorder,shastrydisorder}. For instance, the discovery of the Hebel-Slichter peak---a massive enhancement in relaxation rates of nuclei at temperatures immediately below the superconducting transition temperature~\cite{HebelExp,BCSoriginal,Hebeltheory}---was a defining success of the BCS theory. 
However, bulk NMR measurements probe a certain average (which depends on the crystal structure of the material) of spin fluctuations at all wave-vectors~\cite{SlichterNMR,julien200063}. On the other hand, spatially-resolved NMR measurements~\cite{localNMR} have been limited to micron-scale resolutions~\cite{GloverNMRLimitation}. NV centers, being point defects, can pick up the magnetic noise from spin and current fluctuations in materials in a highly spatially-resolved way. Moreover, unlike traditional linear-response measurements which can often drive a system into the non-linear regime~\cite{vinokur2008superinsulator,altshuleragainstsuperinsulator}, or NMR measurements that require external polarizing fields, NV centers can be used in a minimally-invasive way to measure transport phenomena in materials~\cite{mishaNV}. 

The present theory for magnetic noise near materials is formulated in terms of momentum-dependent reflection/transmission coefficients for s- and p- polarized electromagnetic waves~\cite{Fordelectromagnetism,langsojen,Henkelnoise}. In this work, we first use this formalism to calculate magnetic noise from a conducting two-dimensional (2D) material---which may be a 2D electron gas, graphene, or the gapless surface of a three-dimensional topological insulator~\cite{HasanthreeDtopo}---and we discuss how measurements of this noise directly allow one to probe the wave-vector dependent \textit{transverse} conductivity of the system at all length-scales. To elaborate further, in the conducting materials we consider, the magnetic noise is found to be primarily due to current fluctuations (as opposed to spin fluctuations) inside the material, which in turn are related to the conductivity; it is primarily the transverse current fluctuations (related to the transverse conductivity) that give rise to the noise, because longitudinal fluctuations are damped by efficient screening since they necessarily generate charge fluctuations. Specifically, we find that the noise $N_z$ measured by an NV-center placed at a distance $z_{NV}$ from the material scales as $N_z (z_{NV}) \sim \sigma^T (q = 1/2 z_{NV}) k_BT / z^2_{NV}$, where $\sigma^T(q)$ is the system's wave-vector dependent transverse conductivity, $T$ is the temperature, and $k_B$ the Boltzmann constant. Thus, varying the position of the NV center allows one to measure the complete wave-vector dependent conductivity $\sigma^T (q)$ of the system. These results are discussed in Sec.~\ref{sec:general}. 

(Note that the transverse conductivity should not be confused with the Hall conductivity; it is the response associated with a transverse electric field $\vs{E}^T(\vs{q}) \perp \vs{q}$ that, for a translationally invariant system produces a current $\vs{J}^T(\vs{q}) = \sigma^T(q) \vs{E}^T(\vs{q}) \parallel \vs{E}^T(\vs{q})$. Importantly, $\vs{J}^T (\vs{q}) \cdot \vs{q} = 0$; thus, transverse currents do not create charge imbalances unlike longitudinal currents.)

\begin{figure}
\begin{center}
\includegraphics[width=0.5\textwidth]{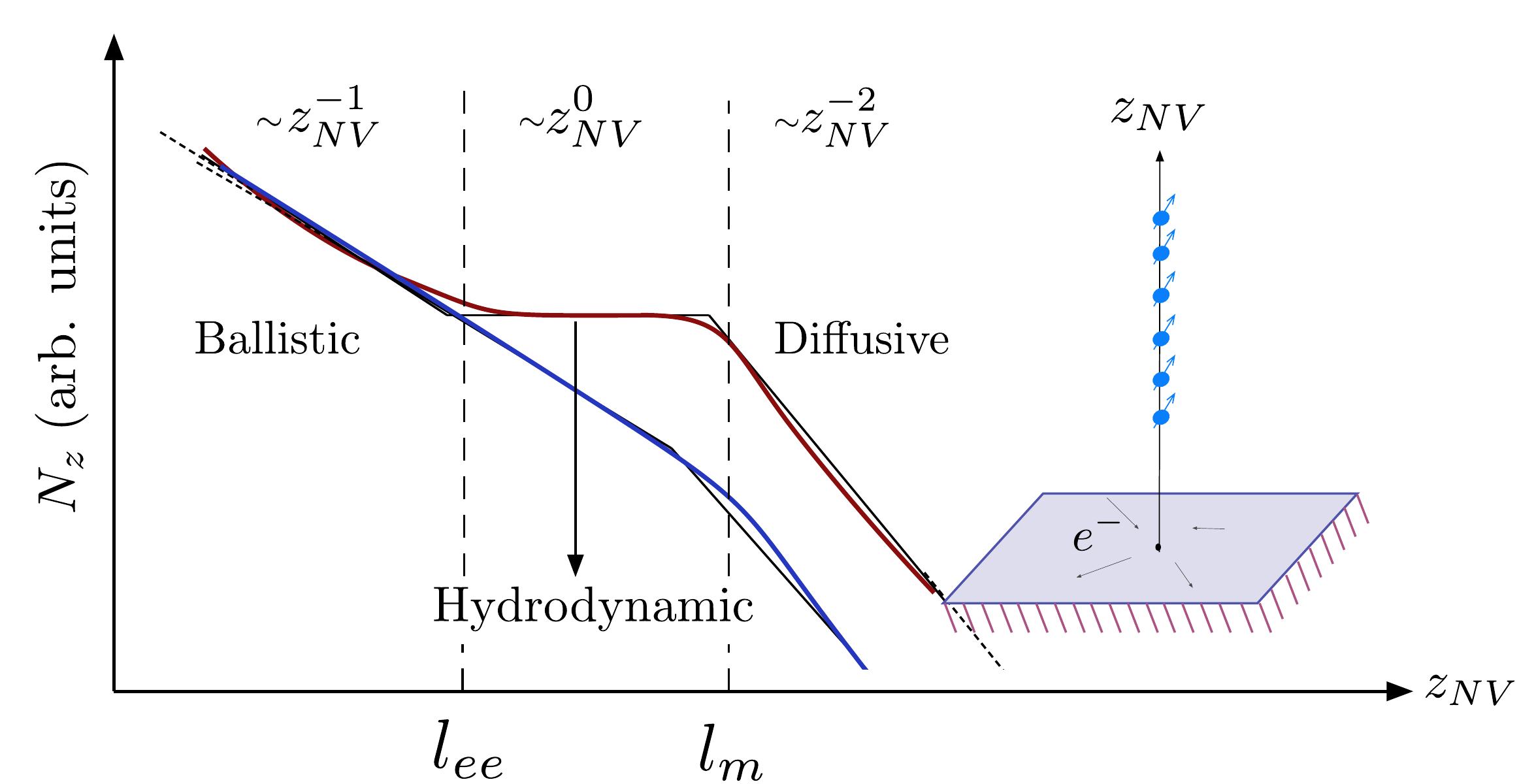}
\caption{Experimental protocol for probing various transport regimes in conducting materials. An array of NV centers (blue dots with arrows) can be placed at varying distance $z_{NV}$ from the 2D material and their relaxation rates can be used to infer the magnetic noise at their location. The curves show, schematically, the magnetic noise as a function of distance from the material in various transport regimes; here $l_m$ and $l_{ee}$ are the mean-free paths due to scattering of electrons due to extrinsic (phonons/impurities) and intrinsic (inter-particle) scattering. The blue curve describes the situation when $l_{ee} > l_m$ and the hydrodynamic regime is absent. For a quantitative discussion of observing this behavior in graphene, see Sec.~\ref{sec:graphenehydrodynamicmain} and Fig.~\ref{fig:graphenenumericalfig}.}
\label{fig:noiseprotocol1}
\end{center}
\end{figure}

As an application of these ideas, we show how various transport regimes in electronic systems, namely ballistic, diffusive and hydrodynamic~\cite{MollenKampHydro,graphenePolini,Bandurinaad0201,CrossnoWFLaw,LucasGrapheneHydro,levitovbackflow,moll2016evidence} regimes have their own unique signatures that can be identified in the scaling of the magnetic noise as a function of the distance from the system [see Fig.~\ref{fig:noiseprotocol1} (a)]. In Sec.~\ref{sec:graphenescalings}, we motivate, using general considerations, why the transverse conductivity, $\sigma^T (q) \sim \text{const.}, 1/q, 1/q^2$ in the diffusive, ballistic and hydrodynamic regimes, respectively. The different $q$-dependent scaling in these regimes gives rise to different scaling of the magnetic noise as a function of the distance from the 2D system. In the hydrodynamic regime, in particular, $\sigma^T(q) \approx \rho^2_0 / (\eta q^2)$ where $\rho_0$ is the charge density of the system and $\eta$ is the viscosity of the (electronic) fluid. Thus, noise measurements can be used to directly infer the viscosity of strongly interacting electronic systems which is of interest due to theoretical predictions of universal bounds~\cite{son2007viscosity,MullerGraphenePerfect}.

 In Sec.~\ref{sec:graphenehydrodynamicmain}, we specialize the discussion to graphene; we present a calculation of the transverse $q$-dependent conductivity of graphene using a Boltzmann kinetic-theory approach incorporating relaxation times describing inter-particle scattering and extrinsic phonon/impurity scattering. This approach yields an analytical result for the DC transverse conductivity of graphene at finite chemical potential and temperature that displays all three transport regimes at various length-scales. We use these results to make quantitative predictions of the magnetic noise from a layer of graphene at varying distances and comment on the feasibility of measuring the viscosity of the electron fluid in graphene via noise spectroscopy. See Figs.~\ref{fig:noiseprotocol1} and~\ref{fig:graphenenumericalfig} for a discussion of some of these results. 
 
\begin{figure}
\begin{center}
\includegraphics[width=0.46\textwidth]{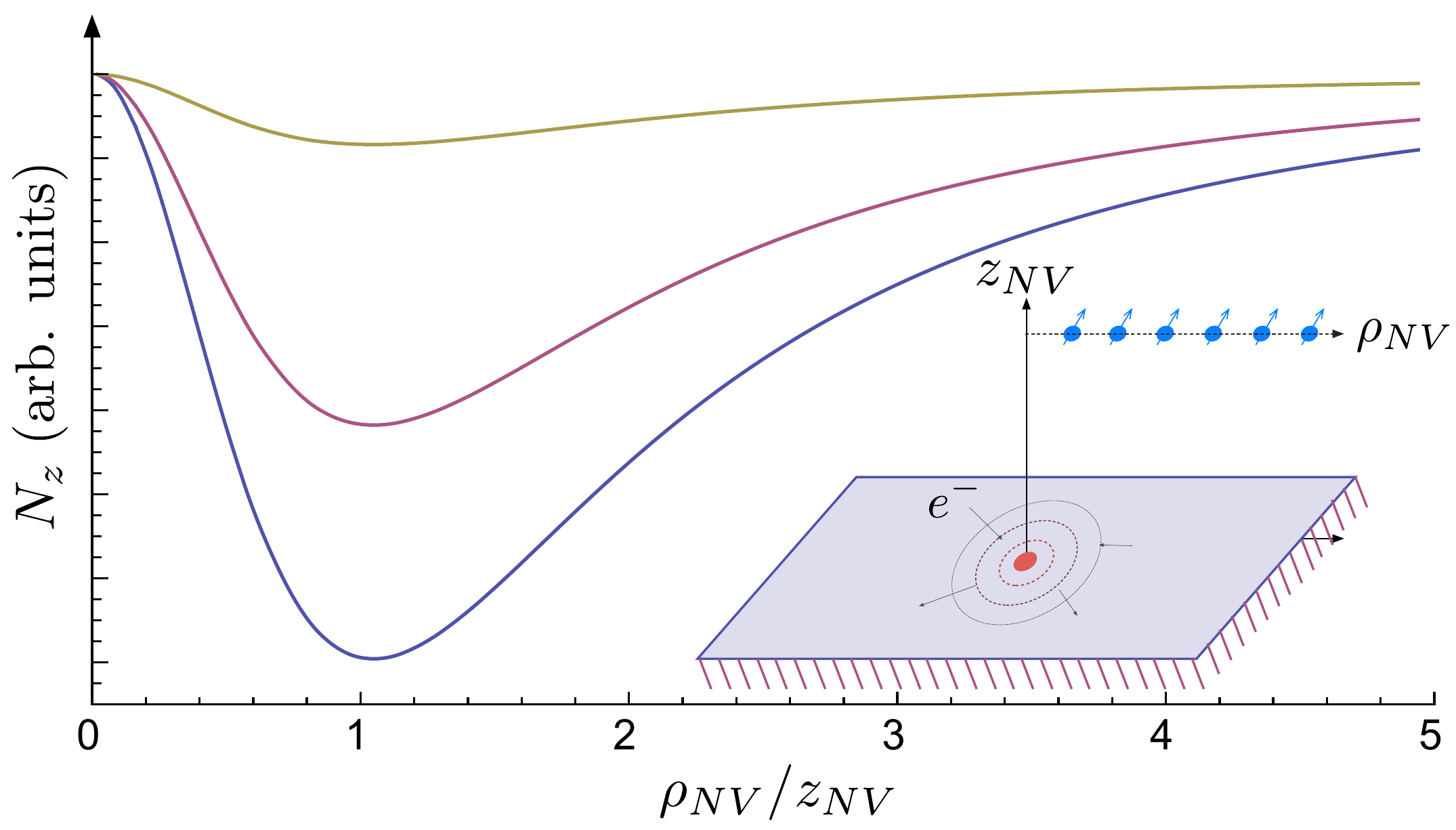}
\caption{Experimental protocol for probing an Anderson impurity (red dot) in a conducting material. An array of NV centers (blue dots with arrows) can be placed at varying lateral separation $\rho_{NV}$ from the impurity, and fixed perpendicular distance $z_{NV}$ to the material. The curves show, schematically, the 
\g{current fluctuation induced} noise as a function of the lateral distance $\rho_{NV}$ (and fixed $z_{NV}$) measured by the NV center. The magnetic noise near the material is suppressed (relative to the background) close to an Anderson impurity below the Kondo temperature $T_K$; the suppression increases as the temperature is lowered (curves from top to bottom), and is 
related to the occupation of the impurity at $T = 0$. The temperature-dependence of the noise can be used infer the level-shift and line-width associated with the resonance of the Anderson impurity within a slave-boson mean-field treatment; see Secs.~\ref{sec:noiseprof} and~\ref{sec:kondonoise} for a detailed discussion and Fig.~\ref{fig:graphenekondofig} for quantitative predictions for this behavior in graphene.}
\label{fig:noiseprotocol2}
\end{center}
\end{figure}

Next, in Sec.~\ref{sec:inhomo}, we extend the existing framework for calculating magnetic noise near materials to allow for calculating the noise near spatial inhomogeneities in a material. Due to loss of translational invariance, the reflection/transmission coefficients become dependent on two momentum variables. We calculate perturbative (linear-response) corrections to the noise due to small spatial inhomogeneities which manifest themselves as corrections to the usual conductivity of the system. As an application, we calculate the two-momentum current-current correlations near an elastic scatterer (lodged inside the two-dimensional conducting material) in Sec.~\ref{sec:twomomenta} and use these correlations to evaluate the noise profile near the impurity in Sec.~\ref{sec:noiseprof}. Naively one expects an interplay between an enhancement in the local density of states which can enhance the current (and hence, the magnetic noise), and the enhanced scattering which reduces the current, near the impurity. Curiously enough, we find that the noise near an isolated impurity is always suppressed compared to the background. Importantly, this suppression provides a direct measure of the scattering properties of the impurity. 

Finally, in Sec.~\ref{sec:kondonoise}, we discuss how magnetic noise measurements near a Kondo impurity, or more generally, a large-U Anderson impurity can be used to infer the energy and line-width of the Kondo resonance within a mean-field slave-boson treatment of the impurity-conduction electron system. In short, the noise suppression measured near the Kondo impurity is found to be proportional to the scattering cross-section of electrons/holes scattering off of the impurity, which in turn yields the spectral weight of the Kondo impurity near the Fermi surface. An illustration of these ideas is shown in Fig.~\ref{fig:noiseprotocol2}; for details, see, in particular, Eq.~(\ref{eq:noiseprofgeneral}), and Figs.~\ref{fig:noiseatsmalldist},~\ref{fig:f2Tplot}; for a quantitative discussion of experimental feasibility of these ideas in metals and graphene, see the discussion in Sec.~\ref{sec:kondoexpfeas} and Fig.~\ref{fig:graphenekondofig}. Recent experiments~\cite{GonzalezGrapheneHAdsorption} on creating isolated, local magnetic moments in graphene using chemical adsorption suggest a possible route to observing the physics we describe. 

We outline a number of promising future directions where we expect these novel probes to have a useful impact, and we conclude in Section~\ref{sec:conclude}. 

\section{Noise from Homogeneous 2D Systems}
\label{sec:general}

In this section, we provide the formalism for computing magnetic fluctuations above a 2D system and show how these are related to its conductivity (and, alternatively, its dielectric properties). The basic reason for the connection between these two quantities can be explained as follows. The magnetic fields generated above the system are related, through a propagation kernel, to the currents in the underlying system. (This kernel is related to the Bio-Savart law, see discussion in Appendix~\ref{sec:biotsavart}.) Therefore, the fluctuations in the magnetic field are tied to the fluctuations in the currents inside the material; the latter are in turn related to the conductivity of the system via fluctuation-dissipation relations. Moreover, the magnetic fluctuations at a distance $z_{NV}$ are most sensitive to currents at wave-vectors of the order of $q \sim 1/z_{NV}$; currents at larger wave-vectors nullify themselves, while those at smaller wave-vectors have smaller phase space. Thus, by tuning the distance of the NV center, the complete finite wave-vector dependence of the conductivity can be found.   

The relaxation rate of the NV center is proportional to the local magnetic noise~\cite{langsojen}, which can be defined~\cite{Agarwal1overf} as a tensor $N_{\alpha \beta} (\omega) = \mathcal{F} \left[ \avg{[B_\alpha(\vs{r}_{NV}, t), B_\beta(\vs{r}_{NV}, t') ]_+} \right]/2$; here $B_\alpha (\vs{r}_{NV})$ is the $\alpha$ component of the magnetic field at the site of the NV center, the operation $\mathcal{F} [.] $ implies a Fourier Transformation, and the notation $[ . , . ]_+$ denotes an anti-commutator of two operators. Note that the NV center's orientation determines the precise contribution of various components of this noise tensor to its relaxation rate~\cite{mishaNV}. In thermal equilibrium, the fluctuation spectrum $N_{AB} (\omega)$, defined by two operators $A$ and $B$ (above, e.g., $B_\alpha$ and $B_\beta$) is related to the corresponding Kubo response-function $\chi_{AB}(\omega)$ via the fluctuation-dissipation relation: $N_{AB} (\omega) = \hbar \coth \left( \hbar \omega / 2 k_B T \right) \text{Im} \left[ \chi_{AB} (\omega) \right]$. Thus, the computation of the fluctuations of the magnetic field can be performed by first calculating the response function~\cite{Agarwal1975} $\chi_{\alpha \beta}(\omega) = \partial B^{tot}_\alpha (\vs{r}_{NV}) / \partial M_\beta (\vs{r}_{NV})$, where $M_\beta (\vs{r}_{NV})$ is an external magnetic dipole set up at the site of the NV-center, pointing in direction $\beta$, and $B^{tot}_\alpha ( \vs{r}_{NV})$ is the total magnetic field (in the direction $\hat{\alpha}$) generated both by the oscillating external magnetic dipole and the reflections from the 2D system. This is the general method that will be used to evaluate the magnetic noise in both the homogeneous and inhomogeneous situations. Below we first consider an application of this approach to the homogeneous case. 

\begin{figure}
 \begin{center}
\includegraphics[width=3in]{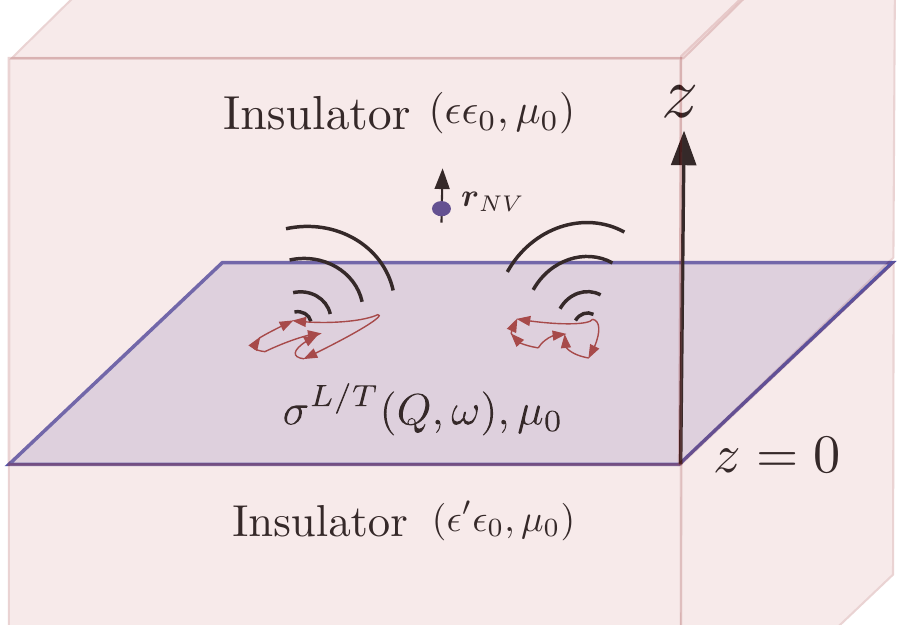}
\caption{Geometry under consideration. The insulator in which the NV center is embedded is characterized by a constant dielectric function $\epsilon$; the layer below the 2D sample is characterized by a constant $\epsilon'$. The 2D sample has some wave-vector and frequency-dependent conductivity $\sigma_{\alpha \beta} (Q, \omega)$ which can be decomposed into a transverse and longitudinal part. Current fluctuations are seen to generate magnetic noise that causes relaxation of the NV center.}
\label{fig:graphenelocal} 
\end{center}
\end{figure}

An external oscillating magnetic dipole moment generates electromagnetic radiation that can be separated into two orthogonal solutions described by an in-plane wave-vector $\vs{Q}$ and a perpendicular-to-plane wave-vector $q^\epsilon_z$ which together satisfy $(q^\epsilon_z)^2 + \vs{Q}^2 = \epsilon \omega^2/c^2$ (assuming the NV-center is buried inside a lossless dielectric with non-dispersive dielectric constants $\epsilon$; see Fig.~\ref{fig:graphenelocal}). The two orthogonal solutions are: s-polarized, if the electric field is parallel to the plane, that is, $\vs{E} \sim \left( \hat{z} \times \hat{Q} \right) e^{i \vs{Q} \cdot \vs{\rho} - i q^\epsilon_z z}$; or p-polarized, if the magnetic field is parallel to the plane, that is, $\vs{B} \sim  \left( \hat{z} \times \hat{Q} \right) e^{i \vs{Q} \cdot \vs{\rho} - i q^\epsilon_z z}$. These waves are partly reflected when they impinge on the 2D material, but keep their form~\cite{LandauLifshitzElectrodynamics} if the system is homogeneous: thus, it is sufficient to describe the reflection, and calculate the total magnetic field by specifying the reflection coefficients $r_s (\vs{Q})$ and $r_p (\vs{Q})$ of the s- and p- polarized waves in this case. 

In terms of these coefficients, the noise (excluding the vacuum, electromagnetic noise) at a distance $z_{NV}$ can be expressed as (see Appendix~\ref{sec:apptrans}) 

\begin{align}
N_z (\omega) &= \hbar \coth \frac{\beta \hbar \omega}{2} \text{Im} \bigg[ \int^\infty_0 \frac{d Q}{2 \pi} Q^3 \frac{i \mu_0}{2 q^\epsilon_z} r_s (Q) e^{2i q^\epsilon_z z_{NV}} \bigg], \nonumber \\
N_x (\omega) &= \hbar \coth \frac{\beta \hbar \omega}{2} \text{Im} \bigg[  \int^\infty_0 \frac{d Q}{2 \pi} Q q^\epsilon_z \frac{ i \mu_0}{4} \nonumber \\
& \left( \frac{\epsilon \omega^2}{(q^\epsilon_z)^2 c^2}  r_p (Q)  - r_s (Q) \right) e^{2 i q^\epsilon_z z_{NV} }
 \bigg]; \nonumber \\
q^\epsilon_z &= i \sqrt{ Q^2 - \epsilon \omega^2/c^2} \; \; \; \text{for} \; Q > \sqrt{\epsilon} \omega/c, \nonumber \\
q^\epsilon_z &= \sqrt{ \epsilon \omega^2/c^2 - Q^2} \; \; \; \text{for} \; Q < \sqrt{\epsilon} \omega/c,  
\label{eq:xznoisersrp}
\end{align}

where $\beta=1/k_BT$, $c$ is the speed of light, $\mu_0$ is the vacuum permeability, $z$ is the perpendicular-to-plane direction and $x$ is an in-plane direction, and we have defined $N_\alpha \equiv N_{\alpha \alpha}$ because the noise tensor is diagonal in this basis for the homogeneous system. 

Let us note that Eqs.~(\ref{eq:xznoisersrp}) are, in principle, generally applicable to calculating noise from any homogeneous material limited to a half-space~\cite{Henkelnoise}. For the three-dimensional case, the precise connection between these reflection coefficients and the transport properties of the material is complicated by the presence of a boundary~\cite{Fordelectromagnetism} and assumptions regarding the properties of the boundary are required. For 2D materials we consider, the reflection coefficients can be computed and directly related to the conductivity of the material. Due to this simplification, and the fact that there is a vast variety of 2D materials with physically interesting phenomena, we refine our study to that of 2D materials sandwiched between two lossless dielectrics, as shown in Fig.~\ref{fig:graphenelocal}.  

We note that, for $c \rightarrow \infty$ and $\omega/(2\pi) = \omega_{NV}/(2\pi) \approx 3 \; \text{GHz}$ being a small frequency corresponding to the NV-center transition, the phase space of magnetic noise due to traveling waves ($q^\epsilon_z > 0$) is limited and most of the noise is due to decaying electromagnetic modes. Consequently, we  can neglect corrections to the noise of the order $\omega/q^\epsilon_z c$. From Eq.~(\ref{eq:xznoisersrp}), it is evident that the noise due to p- polarized modes is negligible due to extra factor of $\omega^2 / (q^\epsilon_z)^2 c^2$. As a consequence, magnetic noise measurements are fairly insensitive to the longitudinal conductivity of the system which only comes into the noise calculation through $r_p$. One can understand why $r_p$ is connected to the longitudinal conductivity via the following argument (see Appendix~\ref{sec:apptrans} for more details): when the magnetic field is parallel to the surface, as in the p-polarized case, the in-plane component of the electric field has to be parallel to the in-plane wave-vector $\vs{Q}$ of the electromagnetic wave; thus, such a field only excites longitudinal currents ($\vs{J}^L (\vs{Q}) \parallel \vs{Q}$). Similarly, if the electric field is s-polarized, it only generates transverse currents (that are perpendicular to $\vs{Q}$); see Fig.~\ref{fig:rsrpfig} for a graphical illustration. These transverse currents are fundamentally different from longitudinal currents because they do not couple to charge fluctuations in the material. 

The reflection coefficients must depend on the transverse/longitudinal conductivity $\sigma^{T/L} (Q, \omega)$ of the material because the currents are set up proportionally to $\sigma^{T/L} (Q, \omega)$, and modify the electromagnetic boundary conditions. In fact, the noise due to p-polarized waves is proportional to $\sigma^L (Q) / \epsilon^{RPA} (Q)$. Thus, the longitudinal currents, which generate charge-density fluctuations, are suppressed due to efficient screening in the conducting material (this suppression is, in fact, related to the suppression $\omega^2 (q^\epsilon_z)^2/c^2$ mentioned above; see Appendix~\ref{sec:apptrans}). The noise due to s-polarized waves is approximately proportional to $\sigma^T(Q)$, and is not suppressed. Detailed calculations are provided in Appendix \ref{sec:apptrans}; here we present only results essential for the further analysis:

\begin{align}
&r_s (Q,\omega, \epsilon = \epsilon') = \frac{-1}{1 + \frac{2 q^\epsilon_z}{\mu_0 \omega \sigma^T}}  \approx - \frac{\omega \sigma^T (Q, \omega) \mu_0}{2 q^\epsilon_z}, \nonumber \\ 
&N_z (\omega) \approx \frac{k_B T \mu^2_0}{16 \pi z^2_{NV}} \int^\infty_0 dx \; x e^{-x} \text{Re} \left[ \sigma^T \left(\frac{x}{2 z_{NV}}, \omega \right) \right], \nonumber \\
&N_x (\omega)= N_y (\omega) = \frac{N_z(\omega)}{2} + \mathcal{O} \left[ \frac{\omega z_{NV}}{c} \right], 
\label{eq:noisetrans}
\end{align}
where, as before, $N_z$ is the magnetic noise in a direction perpendicular to the plane, and $N_x = N_y$ is the in-plane noise. The noise results are correct to order $\mathcal{O} [ \omega z_{NV} / c ]$, and have been presented for $\epsilon = \epsilon'$. Note that the measure $x e^{-x}$ picks out the wave-vector $q \sim 1/2z_{NV}$ as previously advertised. In principle, since the above can be interpreted as a Laplace transformation, a complete set of measurements obtained by varying $z_{NV}$ can be used to deduce the conductivity of the material at any wave-vector. We also note that, since the NV-center frequency $\omega_{NV}$ is small, the measurement essentially picks out the DC conductivity. 

These results can be understood as follows. The noise is a single volume integral (assuming correlations at one length scale are dominant) over two kernels relating the currents in the plane to the magnetic fields at the site of the NV center. Thus, $N (\omega) \sim z^2_{NV} \left( 1/z^2_{NV} \right)^2 \left|J \right|^2$, where $|J|^2$ is the amplitude of current fluctuations at the scale $q \sim 1/z_{NV}$ and frequency $\omega = \omega_{NV}$. Noting that $\left|J \right|^2  = \omega \sigma \coth{\beta \omega/2}$, we directly arrive at the result (besides constant factors). It is also important to note that, in the above analysis, we excluded magnetic noise from spin fluctuations. One can show that this assumption holds as long as the system is not extremely localized, ($k_F l_m \lesssim 1$, where $l_m$ is the mean-free path of the electrons in the system and the NV-center is not brought closer than the inter-particle distance $\sim 1/k_F$ to the material. A detailed justification is presented in Appendix~\ref{sec:magfromspins}. 

\section{Noise scaling as a function of distance in the ballistic, hydrodynamic, and diffusive regimes. }
\label{sec:graphenescalings}

The transverse conductivity $\sigma^T (Q, \omega)$ of metallic systems is typically momentum-independent, as denoted by $\sigma_0$, at long length scales, that is, in the diffusive regime for which $Q \ll 1/l_m$, where $l_m$ is the mean-free path for electrons determined by scattering off of impurities or phonons. By contrast, in the ballistic regime, $Q \gg 1/l_m$, the transverse conductivity scales as $\sim \sigma_0 / (Q l_m)$. A simple way to understand this scaling is that the conductivity (and current fluctuations) is proportional to the scattering time of electrons. In the ballistic limit, $Q \gg 1/l_m$, the current-current correlations do not have access to the time-scale associated with scattering of electrons off of impurities. Instead, `scattering' is determined by the time electrons take to whiz past regions of size $1/Q$ over which the applied electric field oscillates. Thus, one can replace the scattering time by $1/Qv_F$, yielding the required scaling. 

A novel scaling regime of the conductivity appears when the electron-electron scattering length $l_{ee}$ is shorter than $l_m$, the scattering length due to impurities, phonons and other extrinsic sources of relaxation. In this situation, a hydrodynamic description of the electron fluid holds at length scales $l \gtrsim l_{ee}$. Such a regime is hard to achieve in normal metals where impurity and phonon scattering is practically always dominant, but can be achieved in graphene when operated near the charge neutrality point~\cite{Bandurinaad0201,CrossnoWFLaw,LucasGrapheneHydro} where it has been observed by measurements of non-local resistances~\cite{levitovbackflow,Bandurinaad0201}, and a breakdown of the Wiedemann Franz Law~\cite{HartnollHydro,CrossnoWFLaw}. One motivation for observing such hydrodynamic flow is the measurement of viscosity $\eta$, among other novel properties of quantum critical fluids, which are predicted to have universal limits~\cite{son2007viscosity,MullerGraphenePerfect}. 

In the hydrodynamic regime, one can describe the momentum relaxation (of the system) by the Navier-Stokes equation for incompressible ($\boldsymbol{\nabla}.\vs{u}$ = 0; as in transverse flow) fluid motion:
\be
\rho_m \left(\partial_t + \frac{1}{\tau} \right) \vs{u} - \eta \boldsymbol{\nabla}^2 \vs{u} = -e \rho_0 \vs{E}, 
\ee  
where $\vs{u} (x), \rho_m, \rho_0, \tau, \vs{E}$ are the local fluid velocity, mass density, net charge density, momentum relaxation time (due to impurities, phonons, etc.), and the externally applied electric field, respectively. The equation can be straightforwardly solved for a transverse solution $\vs{u} \cdot \vs{q} = \vs{E} \cdot \vs{q} = 0$ at wave-vector $\vs{q}$. Then, computing the charge current $\vs J = \rho_0 \vs{u} = \sigma^T (\vs{q}) \vs{E}$ leads to the result for the conductivity $\sigma^T (q) \approx \rho^2_0 / (\eta q^2)$. Thus, the wave-vector dependence of the transverse conductivity in the hydrodynamic regime is different from both the usual ballistic and diffusive limits. This results in an unusual distance-independent scaling of the noise originating from the electronic system in this regime. Moreover, we note that  the noise is only dependent on viscosity and the charge density $\rho_0$, and the latter can be measured independently. This allows a direct measurement of the viscosity of the system without requiring additional fitting parameters for thermodynamic entities. 

The consequence of these scaling limits is that the distance dependence of the noise measured by the NV-center exhibits three different scaling regimes: $N_z (z_{NV}) \sim 1/z^2_{NV}$ for $z_{NV} \gg l_m$, $N_z (z_{NV}) \sim \text{const.}$ for $l_{ee} \ll z_{NV} \ll l_m$ and $N_z (z_{NV}) \sim 1/z_{NV}$ for $z_{NV} \ll l_m, l_{ee}$. Note that, if $l_m$ is less than $l_{ee}$ then the intermediate, hydrodynamic regime, is not observed. See Fig.~\ref{fig:noiseprotocol1} for a qualitative illustration of these ideas, and Fig.~\ref{fig:graphenenumericalfig} for a quantitative illustration of these regimes in graphene.  

\section{Transport regimes in graphene}
\label{sec:graphenehydrodynamicmain}

We now specialize the discussion of the above ideas to graphene. In this section, we derive the conductivity of graphene in a phenomenological Boltzmann approach that incorporates two relaxation times: $\tau_{ee}$, which sets the time-scale for inter-particle collisions, and $\tau$, which sets the time-scale for momentum relaxation due to an external bath (which may due to phonons or impurities). As we will see, these two time-scales set the length scales $l_{ee} = v_F \tau_{ee}$ and $l_m = v_F \tau$ that determine the various transport regimes, namely, diffusive, hydrodynamic, and ballistic regimes in graphene. We leave the discussion of determining these time-scales and other numerical estimates to Appendix~\ref{sec:graphenenumerical}. 

Our analysis is largely motivated by the Boltzmann approach of Refs.~\cite{FritzsigmaQ,MullerGraphenePerfect}. We will be interested in the regime where chemical potential and temperature are large and of similar magnitude; a large temperature allows one to decrease the inter-particle scattering time, thus sending the system into the hydrodynamic regime, while a large chemical potential helps increase the charge density and makes current fluctuations easier to measure experimentally. In this case, we can assume a local distribution $f_{\vs{k}, \lambda} (\vs{r}) = \avg{ \xi^\dagger_{\vs{k}, \lambda} \xi_{\vs{k}, \lambda}} (\vs{r})$ of Dirac electrons of each of the $N = 4$ species (consisting of two valley and two spin states) in graphene in the bands $\lambda = +1$ (electron-like) and $\lambda = -1$ (hole-like) with dispersion $\epsilon_{\vs{k}, \lambda} = \lambda v_F |\vs{k}|$. (Here we have denoted the creation operator of the Dirac electron at momentum $\vs{k}$ and band $\lambda$ by $\xi^\dagger_{\vs{k}, \lambda} $.) Further, we can neglect off-diagonal single particle expectation values of the form $\avg{\xi^\dagger_{\vs{k}, \mp} \xi_{\vs{k}, \pm}}$ since these terms are only important in the high-frequency (see, for example~\cite{FritzsigmaQ}) regime $\omega \sim 2 v_F k_F, k_B T$ which will not be of interest to us here. 

Then, the semi-classical Boltzmann equation reads 
\begin{align}
\big( \partial_t + \vs{v}_{\vs{k}, \lambda}& \cdot \boldsymbol{\nabla}_x - e \vs{E} \cdot \boldsymbol{\nabla}_k \big) f_{\vs{k},\lambda} \nonumber\\&= - I_{\text{coll}, \vs{k}, \lambda} [f] - I_{\text{imp}, \vs{k}, \lambda} [f],
\label{eq:boltzeq}
\end{align}
where $\vs{E}$ is the applied electric field, $e$ is the electron charge, $\boldsymbol{\nabla_{x}}$ and $\boldsymbol{\nabla_{k}}$ are gradients in the real and wave-vector space, $\vs{v}_{\vs{k}, \lambda} = \lambda \hat{\vs{k}}$ is the group velocity of the Dirac fermions at momenta $\vs{k}$ and band-index $\lambda$, and $I_{\text{coll}, \vs{k}, \lambda} [f]$ is the collision integral associated with the scattering of the electrons and holes off of one-another, while $I_{\text{imp}, \vs{k}, \lambda} [f]$ describes scattering of electrons off of an external bath. 

To solve the Boltzmann equation, we consider a solution of the following form:

\begin{align}
f_{\vs{k}, \lambda} (\vs{r}) &= f^0_{\vs{k}, \lambda} (\vs{u}, \mu, \vs{r}) + f^1_{\vs{k}, \lambda} (\vs{r}) ; \nonumber \\
f^0_{\vs{k}, \lambda} (\vs{u}, \mu, \vs{r}) &= \frac{1}{e^{\beta ( \epsilon_{\vs{k}, \lambda} - \mu - \vs{u} \cdot \vs{k} )} + 1}, \nonumber \\ 
\end{align}

where, $\vs{u}$ and $\mu$ are the position-dependent local velocity and chemical potential of the system. Further, we demand $\frac{1}{\mathcal{V}} \sum_{\lambda, \vs{k}}  f^1_{\vs{k}, \lambda} = 0$ and $\frac{1}{\mathcal{V}} \sum_{\lambda, \vs{k}} \vs{k}  f^1_{\vs{k}, \lambda} = 0$, with $\mathcal{V}$ being the system's area. This implies that the deviation of the distribution from $f^0$ is not associated with excess charge or momentum. This imposition does not constrain the solution since an appropriate choice of local velocity $\vs{u} (\vs{x}) $ and chemical potential $\mu (\vs{x})$ can describe the total momentum and charge of the system. 

To make further progress, we assume the following form for the collision integrals $I_{\text{coll}, \vs{k}, \lambda} [f]$ and $I_{\text{imp}, \vs{k}, \lambda} [f]$:

\begin{align}
I_{\text{coll}, \vs{k}, \lambda} [f]  &= - \frac{f^1_{\vs{k}, \lambda}}{\tau_{ee}}, \nonumber \\
I_{\text{imp}, \vs{k}, \lambda} [f]  &= - \frac{f_{\vs{k}, \lambda} - f^0_{\vs{k}, \lambda} (\vs{u} = 0)}{\tau} .
\label{eq:collapprox}
\end{align}

The above choices are motivated as follows. Inter-particle collisions conserve total momentum and total charge; thus, we reqiure $\sum_{\lambda, \vs{k}} \vs{k} I_{\text{coll},\vs{k},\lambda} [f] = 0$ and $\sum_{\lambda, \vs{k}} I_{\text{coll},\vs{k},\lambda} [f] = 0$. These conditions are clearly satisfied by the collision integral under the assumptions on $f^1_{\vs{k},\lambda}$. Moreover, we expect these inter-particle collisions to relax the system to a state of `fluid motion', as captured by $f^0_{\vs{k}, \lambda}$; thus, the relaxation rate should be proportional to the deviation $f^1_{\vs{k}, \lambda}$ as we have considered. Next, unlike inter-particle collisions, interactions with an external bath of phonons and impurities alter the total momentum of the system, while (typically) conserving total charge. 
These interactions attempt to relax the system's state of motion to that of stationarity captured by the distribution $f^0_{\vs{k}, \lambda} (\vs{u} = 0)$. Our choice of $I_{\text{imp}, \vs{k}, \lambda} [f]$ captures both these facets; namely, the relaxation rate is proportional to the deviation of the distribution of electrons from the stationary distribution, and it conserves the total charge (to order $\mathcal{O} [(u/v_F)^2]$) while allowing for relaxation of the total momentum of the system. In what follows, we will show that the above Boltzmann equation yields a conductivity that is of the Drude form~\footnote{A note on terminology. Beyond the inter-particle scattering length, the particles are already `diffusing', that is, local density relaxes as a power-law in time; this is because inter-particle collisions also conserve charge. One can observe this by considering the situation where no Electric field is applied, thus $\vs{u} = \vs{E} = 0$; then substituting the current from Eq.~(\ref{eq:currenteq}) in the continuity equation, Eq.~(\ref{eq:conteq}), readily yields Ficks' law with a diffusion constant $D = v^2_F \tau_{ee} / 2$. In the case that $\tau_{ee} \gg \tau$, and consequently, inter-particle collisions are irrelevant, the above calculation can be repeated to yield a Ficks' law with the diffusion constant $D = v^2_F \tau /2$. The hydrodynamic regime is thus delineated from the diffusive regime by the presence of a current relaxation rate that is wave-vector dependent (which arises due to viscous damping). We deem the transport as diffusive when this viscous damping is unimportant and the conductivity is given by the usual $q$-independent Drude form, as is traditional.} (and independent of $q$) for $q v_F \tau \ll 1$, ballistic form (scaling as $1/q$) for $q v_F \tau_{ee} \gg 1$, and of the hydrodynamic form (scaling as $1/q^2$) at intermediate length scales.

\begin{figure}
\begin{center}
\includegraphics[width = 3.5 in]{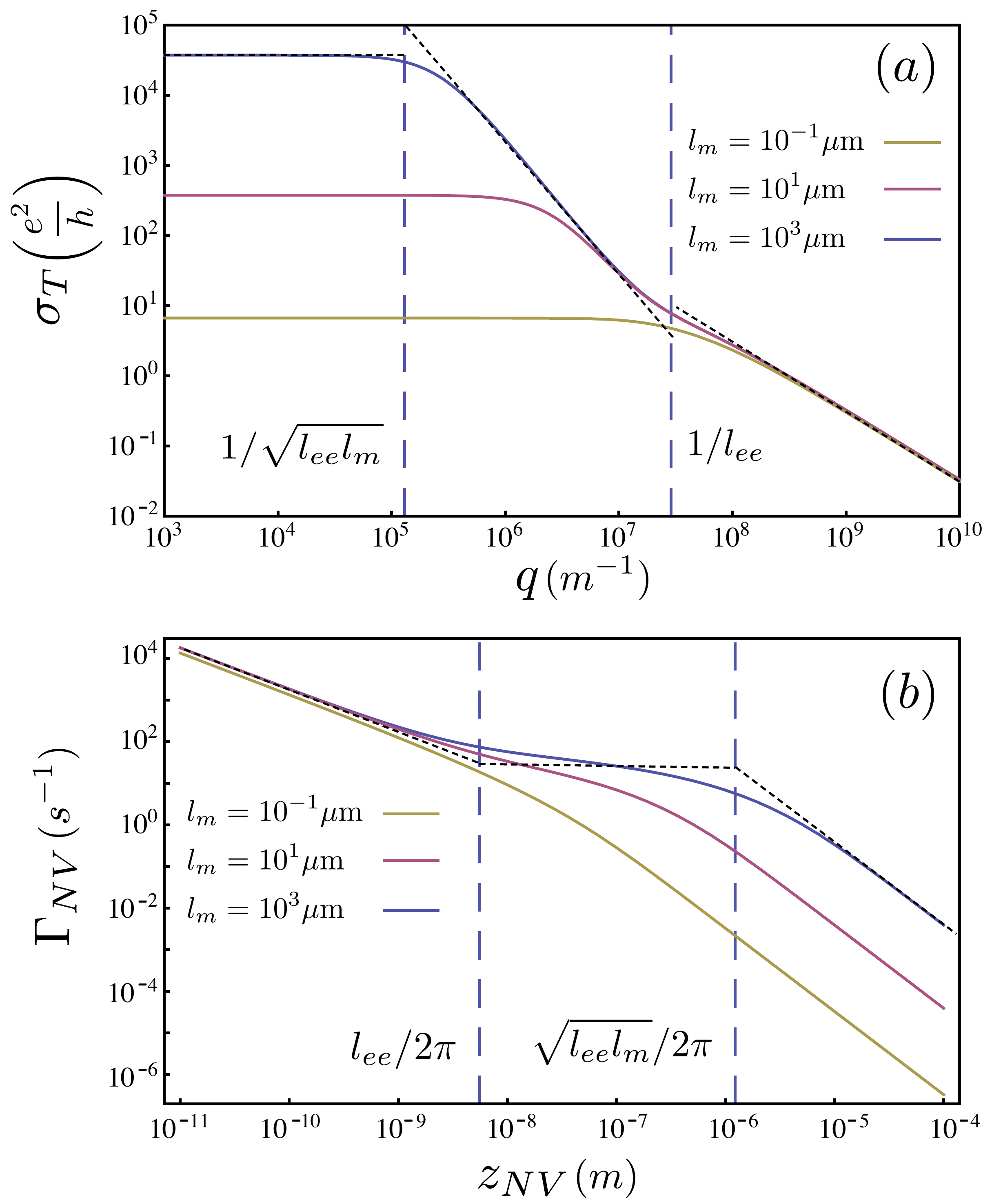}
\caption{(a) The wave-vector dependent conductivity of graphene at a chemical potential $\mu$ equal to the electronic temperature $k_B T = 1000$K, for samples with different mean free paths $l_m$ (set by phonons, or disorder). The long-dashed vertical lines correspond to $l_m = 1000 \mu$m and serve as a guide between the diffusive, hydrodynamic and ballistic regimes (left to right) for this case. (b) The corresponding relaxation rates of an NV center susceptible to noise only in the $z$-direction, as a function of the distance from the graphene layer. The short-dashed lines are different power-law fits in $q$ and $z_{NV}$ in the different transport regimes, as per the discussion in the main text.}
\label{fig:graphenenumericalfig}
\end{center}
\end{figure}

One can now obtain an equation on the conservation of charge density and momentum flux density by integrating the Boltzmann equation with the measure $\frac{N}{\mathcal{V}} \sum_{\vs{k}, \lambda}$ and $\frac{N}{\mathcal{V}} \sum_{\vs{k}, \lambda} \vs{k}$ on both sides. ($N = 4$ captures the two spin and two valley species of Dirac fermions in graphene.) This yields (to first order in $\vs{u}$) for charge conservation, 

\begin{align}
& \partial_t \rho_0 + \boldsymbol{\nabla} \cdot \vs{J} = 0, \; \text{where} \nonumber \\
& \rho_0 = \frac{N}{\mathcal{V}} \sum_{\vs{k}, \lambda} f^0_{\vs{k}, \lambda} (\vs{u} = 0) -  \frac{N}{\mathcal{V}} \sum_{\vs{k}} 1 \nonumber \\
&= \frac{N (k_BT)^2}{2 \pi \hbar^2 v_F^2} \left( \text{Li}_2 (- e^{-\beta \mu}) -  \text{Li}_2 (- e^{\beta \mu} ))\right), \nonumber \\
& \vs{J} = \rho_0 \vs{u} + \frac{N}{\mathcal{V}} \sum_{\vs{k}, \lambda} \vs{v}_{\vs{k}, \lambda} f^1_{\vs{k}, \lambda},  \nonumber \\
\label{eq:conteq}
\end{align}

and for momentum conservation, 

\begin{align}
&(\epsilon + P) (\partial_t + 1/\tau) \vs{u} + \vs{u} \partial_t (\epsilon + P) + \boldsymbol{\nabla} P + \vs{F}_\eta = -e \vs{E} \rho_0;  \nonumber \\
&\epsilon = \frac{N}{\mathcal{V}} \sum_{\vs{k}, \lambda} \epsilon_{\vs{k}, \lambda} f^0_{\vs{k}, \lambda} (\vs{u} = 0) -  \frac{N}{\mathcal{V}} \sum_{\vs{k}} \epsilon_{\vs{k}, -1},  \nonumber \\
&P = \frac{N}{2\mathcal{ V}} \sum_{\vs{k}, \lambda} \vs{k}. \vs{v}_{\vs{k}, \lambda} f^0_{\vs{k}, \lambda} (\vs{u} = 0) -  \frac{N}{2\mathcal{V}} \sum_{\vs{k}} \epsilon_{\vs{k}, -1} = \frac{\epsilon}{2}, \nonumber \\
& \epsilon + P = - \frac{6 N (k_B T)^3}{\pi \hbar^2 v^2_F} \left(  \text{Li}_3 (- e^{-\beta \mu}) +  \text{Li}_3 (- e^{\beta \mu} ) \right), \nonumber \\
&\vs{F}_\eta = \frac{N}{\mathcal{V}} \sum_{\vs{k}, \lambda} \vs{k} \vs{v}_{\vs{k}, \lambda} \cdot \boldsymbol{\nabla}_x f^1_{\vs{k}, \lambda}.  \nonumber \\
\label{eq:momentumeq}
\end{align}

In the above equations, the quantities $\rho_0, \vs{J}, \epsilon, P$ can be interpreted as the charge density, charge-current density, energy density and pressure respectively. $\text{Li}_\text{n}$ refers to the $\text{n}^{\text{th}}$ polylogarithm function. Note that we have defined $\rho_0$ and $J$ without the factor of electron charge, $-e$. 

 We see that the current $\vs{J}$ is composed of a term corresponding to the usual flow of a net charge $\rho_0$ with velocity $\vs{u}$ and a term directly proportional to $\vs{E}$ which can be non-zero even if the total charge is zero: this corresponds to the finite current carried by thermally excited electrons and holes as identified in Ref.~\cite{HartnollHydro}. This current carries information about the relaxation time $\tau_{ee}$ since it is proportional to the non-equilibrium distribution $f^1_{\vs{k},\lambda}$. The term $\vs{F}_\eta$ in the momentum-conservation equation describes the viscous forces acting on the system; this interpretation will become clearer below. Note that the term reflecting momentum convection $\propto (\epsilon + P) \vs{u} \cdot \boldsymbol{\nabla} \vs{u}$ appears at a higher order in $\vs{u}$ and has been neglected. 

We now evaluate the extra contributions to the current $\vs{J}$ and the viscous force $\vs{F}_\eta$ using the relaxation time approximation as described in Eq.~(\ref{eq:collapprox}). We work under the following condition: $\omega$ and $\vs{E}$ are to be understood as parametrically small quantities so that we may neglect any terms of the kind $ u E \sim E^2 $ or $u \times \omega \sim \omega E$ or $\omega^2$ or $E^2$. Then, we may solve for $f^1$ using the re-arrangement of Eq.~(\ref{eq:boltzeq}):

\begin{align}
f^1_{\vs{k},\lambda} &= - \frac{\tau'_{ee}}{1 + \tau'_{ee} \vs{v}_{\vs{k}, \lambda} \cdot \boldsymbol{\nabla}_x }\nonumber\\ &\cdot[ \left( \partial_t + 1/\tau + \vs{v}_{\vs{k}, \lambda} \cdot \boldsymbol{\nabla}_x - e \vs{E} \cdot \boldsymbol{\nabla}_k \right) f^0_{\vs{k},\lambda} ], \nonumber \\
\label{eq:f1boltz}
\end{align}
where $1/\tau'_{ee} \equiv 1/\tau_{ee} + 1/\tau$. 

Before we proceed further and derive the full wave-vector dependence of the conductivity, we derive the equations of motion of the Dirac fluid in graphene in the hydrodynamic regime: $q v_F \tau_{ee} \ll 1$ and $ q v_F \tau \gg 1$. In this regime, $\tau'_{ee} \approx \tau_{ee}$ and $\tau'_{ee} \vs{v}_{\vs{k},\lambda} \cdot \boldsymbol{\nabla}_x$ in the denominator of Eq.~(\ref{eq:f1boltz}) can be neglected. We then find 
\begin{align}
-e \vs{J} &= -e \rho_0 (1 + \omega \tau_{ee} ) \vs{u} + \sigma_0 ( \vs{E} - \boldsymbol{\nabla}_x \mu/e), \nonumber \\
\vs{F}_\eta &= - \left( \eta \boldsymbol{\nabla}^2 \vs{u} + \zeta \boldsymbol{\nabla} \left(\boldsymbol{\nabla} \cdot \vs{u} \right) \right), 
\label{eq:currenteq}
\end{align}
where 
\begin{align}
\sigma_0 &= -\frac{e^2}{\hbar} \frac{N \tau_{ee}}{\hbar} \frac{(\hbar v_F)^2}{2 \mathcal{V}} \sum_{\vs{k}, \lambda} \partial_{\epsilon_{\vs{k},\lambda}} f^0_{\vs{k},\lambda} (\vs{u} = 0) \nonumber \\
 &= \frac{e^2}{h} \frac{N \tau_{ee}}{\hbar} \left( k_BT \log ( 1 + e^{\beta \mu} ) - \mu/2 \right), \nonumber \\ 
\eta &= \frac{(\epsilon + P) \tau_{ee}}{4} \nonumber \\
&= \frac{-3 N \tau_{ee} (k_B T)^3}{2 \pi \hbar^2 v_F^2} \left( \text{Li}_3 ( - e^{-\beta \mu} ) + \text{Li}_3 (- e^{\beta \mu} ) \right), \nonumber \\
\zeta &= 2 \eta.
\label{eq:visc}
\end{align}

We can readily interpret $\eta$ and $\zeta$ as the shear and bulk viscosity, respectively, owing to the form of the hydrodynamic equations. $\sigma_0$ is an extra, intrinsic conductance of graphene in the hydrodynamic regime that exists even when the chemical potential is tuned to zero. One can Fourier-transform the momentum conservation equation and find a transverse solution at wave-vector $\vs{q}$ satisfying $\vs{u} \cdot \vs{q} = \vs{E} \cdot \vs{q} = 0$ to arrive at the wave-vector dependent transverse conductivity of the system: 

\be
\sigma^T (q) = \sigma_0 + \frac{e^2 \rho^2_0 / (\epsilon + P)}{ - i \omega + \frac{1}{\tau} + \frac{\eta}{\epsilon + P} (qv_F)^2 } \approx \sigma_0 + \frac{\rho^2_0}{\eta q^2}.
\label{eq:hydroqtranscond}
\ee

Here we note that the hydrodynamic regime persists at length scales $q^{-1} = l$ in the range $(l_m l_{ee})^{1/2} \lesssim l \lesssim l_{ee}$. For $l \gtrsim \sqrt{l_m l_{ee}}$, we note that the momentum relaxation rate $1/ \tau$ dominates over the viscous relaxation rate in the pole of Eq.~(\ref{eq:hydroqtranscond}). For $q^{-1} = l \lesssim l_{ee}$, we expect the viscosity to depend strongly on the momentum $q$ [due to our omission of the term $q v_F \tau_{ee}$ in the denominator of Eq.~(\ref{eq:f1boltz})]. 

We now discuss the solution of the full wave-vector dependent transverse conductivity by solving for the viscous force $\vs{F}_\eta$ and the current $\vs{J}$ with the complete result for $f^1_{\vs{k},\lambda}$ in Eq.~(\ref{eq:f1boltz}). We find 

\begin{align}
\sigma^T_q &= Y(q) \sigma_0+\frac{e^2 \rho^2_0 v^2_F  \tau'_{ee}}{\epsilon + P} \nonumber\\
&\quad\times\frac{(1 - X(q))(1 - X(q) + (i \omega - 1/\tau)\tau'_{ee} Y(q))}{X(q) + (1- X(q)) (- i \omega + 1/\tau) \tau'_{ee}} \nonumber \\
\vs{u} &= - e \vs{E} \tau'_{ee} v^2_F \nonumber\\
&\quad\times\frac{\rho_0}{\epsilon + P} \frac{1 - X(q)}{(1+ X(q))(-i \omega + 1/\tau) \tau'_{ee} + X(q) }, \nonumber \\
X(q) &=  \frac{(q v_F \tau'_{ee})^2}{ (q v_F \tau'_{ee})^2 + 2 + 2 \sqrt{1 + (q v_F \tau'_{ee})^2}}, \nonumber \\
Y(q) &= \frac{2}{1 + \sqrt{1 + (q v_F \tau'_{ee})^2}}.
\label{eq:final}
\end{align}

The various transport regimes can be quickly surmised from the behavior of $\sigma^T_q$ in Eq.(\ref{eq:final}) in the limiting cases. In the diffusive regime, $q v_F \ll 1/\tau, 1/\tau_{ee}$ and $\omega \rightarrow 0$, we can set $X(q) = 0$. Then, $\vs{u} (q) = -e \vs{E} \tau v^2_F \frac{\rho_0}{\epsilon + P}$ and conductivity $\sigma^T_q \approx e^2 \rho^2_0 \tau v^2_F/ (\epsilon + P)$ is independent of $q$. (There is a small additional part to the conductivity proportional to $\tau'_{ee}$, and also independent of $q$ that we neglect.) In the hydrodynamic regime, $ 1/\tau \ll q v_F \ll 1/\tau_{ee}$, we have $X(q) = (qv_F \tau'_{ee})^2/4$, $Y(q) \approx 1$, which yields $\sigma^T_q = \frac{e^2 \rho^2_0}{\eta q^2} + \sigma_0$ as in Eq.~(\ref{eq:hydroqtranscond}). In the ballistic regime, $q v_F \gg 1/\tau_{ee}$ and we have $1 - X(q) \approx \frac{2}{\tau'_{ee} q v_F}$; we find that the velocity $\vs{u}$ vanishes as $1/qv_F$ and the conductivity becomes $\sigma^T_q = \sigma_0 / (q v_F \tau'_{ee})$. Importantly, the time-scale $\tau_{ee}$ vanishes from the results and the conductivity scales as $1/(qv_F)$. These results confirm our expectations for the $q$-dependence of the conductivity in various transport regimes. 

In Fig.~(\ref{fig:graphenenumericalfig}) we use the results in Eq.~(\ref{eq:final}) along with values of $\tau_{ee}$ as discussed in Appendix~\ref{sec:graphenenumerical} to compute the complete $q$-dependent conductivity of graphene and numerically validate our expectations. We also calculate the relaxation rate of the NV-center measuring only noise $N_z$ in the $z$-direction from Eqs.~(\ref{eq:noisetrans}). We expect that a relatively clean sample of graphene in which the electrons are heated to large temperatures~\cite{hotcarrierreview} of the order of $\sim 500$K may be used to experimentally investigate these effects. At these temperature, we note that $l_{ee} \sim 100$ nm, while $l_m \sim 10 \mu$m (if limited by acoustic phonons, as is likely in graphene on hBN substrates). This provides an order of magnitude in length (NV-center distance) $q^{-1} \in [\sqrt{l_{ee} l_m}, l_m]$ to observe hydrodynamic behavior.  

\section{Noise from Inhomogeneous Systems}
\label{sec:inhomo}

In the previous sections, we considered all forms of irregularities in the material to be `disorder-averaged'. This assumption becomes less acceptable in the near-field regime where the distance of the NV-center is comparable to the mean-free path (or smaller) \emph{and} an individual impurity may significantly alter the noise profile from the background. In particular, if the impurity has interesting physical properties (such as a temperature-dependent level shift, line-width or other scattering properties), as in the case of Kondo impurities, this extra contribution can be used to study the physics of an isolated impurity in detail. To address such situations, Eqs.~(\ref{eq:xznoisersrp}) and (\ref{eq:noisetrans}) need to be modified to allow for calculation of noise when the conductivity has two-momentum corrections (due to the breaking of the translational symmetry). 

We address this question in a perturbative framework or linear-response framework in which we treat the system as having small two-momentum corrections to the conductivity on top of the background single-momentum-dependent (translationally-invariant) conductivity. We follow the previous sections and perform first the linear-response calculation in which we determine the total magnetic field at the site of the NV-center in the presence of a magnetic dipole at the same site. 

This magnetic dipole generates an electromagnetic field, say $\vs{E}_0 (\vs{Q}, q_z)$, with an in-plane momentum $\vs{Q}$ and perpendicular-to-plane momentum $q_z$; for the ease of presentation we consider a single wave-vector at a time, although the dipole generates fields at all wave-vectors which impinges on the 2D material. Now, besides generating a current $\vs{J} (\vs{Q})$ at wave-vector $\vs{Q}$, the electric field generates weaker source currents with different in-plane wave-vectors $\vs{Q}'$, that is, $\vs{J}_s (\vs{Q'}) =  \sum_{\alpha \beta} \hat{\alpha} \sigma_{\alpha \beta} ( \vs{Q}', \vs{Q}) E_{0,\beta} (\vs{Q})$. These source currents will then generate additional outgoing electromagnetic waves $\vs{E}_1 (\vs{Q}', q'_z)$ whose amplitude must be determined self-consistently to first order in perturbation theory (in particular, in the presence of the additional induced current $\vs{J}_1 (\vs{Q}') = \sigma_0 (\vs{Q}') \vs{E}_1 (\vs{Q}')$ which is of the same order as $\vs{J}_s (\vs{Q'})$). These reflected fields will modify the noise at the NV center and their amplitude must be evaluated. 

In order to perform these calculations, we first solve the problem of \emph{outgoing} electromagnetic radiation (since there are no incoming waves associated with the two-momentum corrections) due to a source current $\vs{J}_s( \vs{Q})$, in the 2D material, which may be transverse or longitudinal. The two cases, as before, generate different polarizations of outgoing radiation: the transverse (longitudinal) source current produces only s- (p- ) polarized fields. The detailed solution of this boundary-value problem are presented in Appendix~\ref{sec:appnontrans}. 

We again note that, since the (decaying solutions of the) electromagnetic field emanating from the magnetic dipole are primarily s- polarized, the generating electric field and current inside the 2D material will predominantly be transverse polarized. However, the currents that are generated at different momenta $\vs{Q}'$ will have both transverse and longitudinal components. We again assume that the longitudinal part of these currents is suppressed due to screening and will generate even smaller noise corrections than the transverse part; thus, we neglect these contributions. (Note that this last assumption is not required to evaluate the noise in the $perpendicular$-direction.) Thus, the two-momentum corrections to the conductivity of interest are effectively correlations of transverse currents at different momenta $\vs{q}_1$ and $\vs{q}_2$. In particular, we require $\text{Re} [ \sigma_{T,T} (\vs{q}_1, \vs{q}_2) ]$ where 

\begin{align}
\text{Re} [ \sigma_{T/L,T/L} (\vs{q}_1, \vs{q}_2) ] &= \frac{\text{Im} [ \Pi_{T/L,T/L} (\vs{q}_1, -\vs{q}_2, \omega + i0^+) ]}{ \omega}, \nonumber \\
\Pi_{T/L,T/L} (\vs{q}_1 , -\vs{q}_2,\tau - \tau') &= \avg{ T_\tau [J_{T/L} (\vs{q}_1, \tau) J_{T/L} (- \vs{q}_2, \tau')]}, \nonumber \\
J_T (\vs{q}_1) &= \sum_\alpha (\hat{z} \times \hat{q}_1 )_\alpha J_\alpha, \nonumber \\
J_L (\vs{q}_1) &= \sum_\alpha (\hat{q}_1)_\alpha J_\alpha.
\end{align}

(Thus, $\Pi_{T/L,T/L}$ are the retarded two-momentum transverse and longitudinal current-current response functions, defined with an additional negative sign; see Appendix~\ref{sec:appnontrans}.) The details of the calculation of the correction to the reflection coefficients and the complete calculation of the noise due to these corrections is carried out in detail in Appendix~\ref{sec:appnontrans}. Here we simply quote the important final results of these calculations: 

\begin{align}
&N_{z} (\vs{r}_{NV}) = \frac{\mu^2_0 k_B T}{2} \int_0^\infty \frac{q_1 d q_1 d \theta_1}{(2\pi)^2}  \int_0^\infty \frac{q_2 d q_2 d \theta_2}{(2\pi)^2}  \nonumber \\
&\quad\quad \times e^{i \rho_{NV} (q_1 \cos \theta_1 - q_2 \cos \theta_2 ) - (q_1 + q_2 ) z_{NV} }\nonumber\\
&\quad\quad \times \text{Re} \left[ \sigma_{T,T} (\vs{q}_1,\vs{q}_2) \right], \nonumber \\
&N_{\hat{n}_1\hat{n}_2} (\vs{r}_{NV}) = \frac{\mu^2_0 k_B T}{2} \int_0^\infty \frac{q_1 d q_1 d \theta_1}{(2\pi)^2}  \int_0^\infty \frac{q_2 d q_2 d \theta_2}{(2\pi)^2}  \nonumber \\
&\quad\quad \times e^{i \rho_{NV} (q_1 \cos \theta_1 - q_2 \cos \theta_2 ) - (q_1 + q_2 ) z_{NV}} (\hat{q}_1.\hat{n}_1) (\hat{q}_2.\hat{n}_2) \nonumber \\
 &\quad\quad \times \text{Re} \left[ \sigma_{T,T} (\vs{q}_1,\vs{q}_2) \right],
\label{eq:nontransnoise}
\end{align}
where $\hat{n}_1$ and $\hat{n}_2$ are in-plane directions; we have assumed that the NV-center sits at a distance $z_{NV}$ away from the 2D surface, and a distance $\rho_{NV}$ in the radial direction away from the the `origin' on the 2D surface. (In the next section, an impurity is assumed to reside at this origin.) The azimuthal angles $\theta_1$ and $\theta_2$ of $\vs{q}_1$ and $\vs{q}_2$ are measured with respect to a fixed (but arbitrarily chosen) axis on the 2D surface. Off-diagonal correlations of the form $N_{z \hat{n}_1}$ are negligible, and we again adopt the notation $N_z \equiv N_{zz}$. Note that the above results are correct to order $\mathcal{O} [ \omega z_{NV} / c ]$, and under the assumption that $|\epsilon_{RPA} (q \ll k_F)| \gg 1$. Thus, we assume that the NV-center is at a distance much greater than $1/k_F$ from the 2D material which allows us to neglect complications coming from charge fluctuations that become visible at distances smaller than the screening length in the material. Under these approximations, the results of Eqs.~(\ref{eq:nontransnoise}) are fairly straightforward extensions of the translationally-invariant result: we see that inserting $\text{Re} [ \sigma_{T,T} (\vs{q}_1, \vs{q}_2) ] = \text{Re} [ \sigma_{T,T} (\vs{q}) ] \delta (\vs{q}_1 - \vs{q}_2)$ reproduces the results of the homogeneous system. 

\section{Current-Current correlations near an isolated impurity}
\label{sec:twomomenta}

We now calculate the two-momentum corrections to the conductivity due to an isolated impurity; for a sense of the experimental setup, see Fig.~\ref{fig:noiseprotocol2}. We will focus on situations where we can treat the impurity in a non-interacting framework (and can hence, neglect vertex corrections) so that it only serves as a elastic point scatterer with momentum-independent scattering matrix elements (which may depend on energy). These matrix elements may themselves be computed from a more complicated interacting model such as the Anderson model (for a discussion on computing these properties of the Kondo model at temperatures below the Kondo temperature $T_K$, see~\cite{hewson1997kondo}). We further assume that the impurity is located in a background that can be accounted for by assuming a line-width $\frac{1}{2 \tau}$ (for instance, due to phonons) to the electron Green's functions; for simplicity, we assume this line-width is due to isotropic scattering in the background so that (ladder-type) vertex corrections associated with it are absent. Note that $\tau$ can be experimentally determined by bulk conductivity measurements. Alternatively, experiments can be performed in the near-field regime wherein the NV center is placed much closer to the impurity as compared to the mean-free path $l_m = v_F \tau$. In this limit, the `excess' noise near the impurity does not depend on $\tau$ and our approximations pertaining to the background should not matter significantly. 

Then, in the presence of the extra impurity at position $\vs{r}_f$, and frequency-dependent scattering matrix element $T_f (i \omega_n)$, the electron Green's functions can be written as

\begin{align}
G_0 (\vs{q} , i \omega_n) &= \frac{1}{i \omega_n - \epsilon_{\vs{q}} + \frac{i}{2\tau} \text{sign} ( \omega_n )}, \nonumber \\
G (\vs{q}_1, \vs{q}_2, i \omega_n ) &= G_0  (\vs{q}_1, \vs{q}_2, i \omega_n )  + G_1  (\vs{q}_1, \vs{q}_2, i \omega_n ), \nonumber \\
G_1 (\vs{q}_1, \vs{q}_2, i \omega_n ) &= e^{i ( \vs{q}_1 - \vs{q}_2 ). \vs{r}_f} \times \nonumber \\
&G_0 (\vs{q}_1, i \omega_n ) T_f (i \omega_n) G_0 (\vs{q}_2, i \omega_n ).
\label{eq:gcgreen}
\end{align}

\begin{figure}
\begin{center}
\includegraphics[width=3.5in]{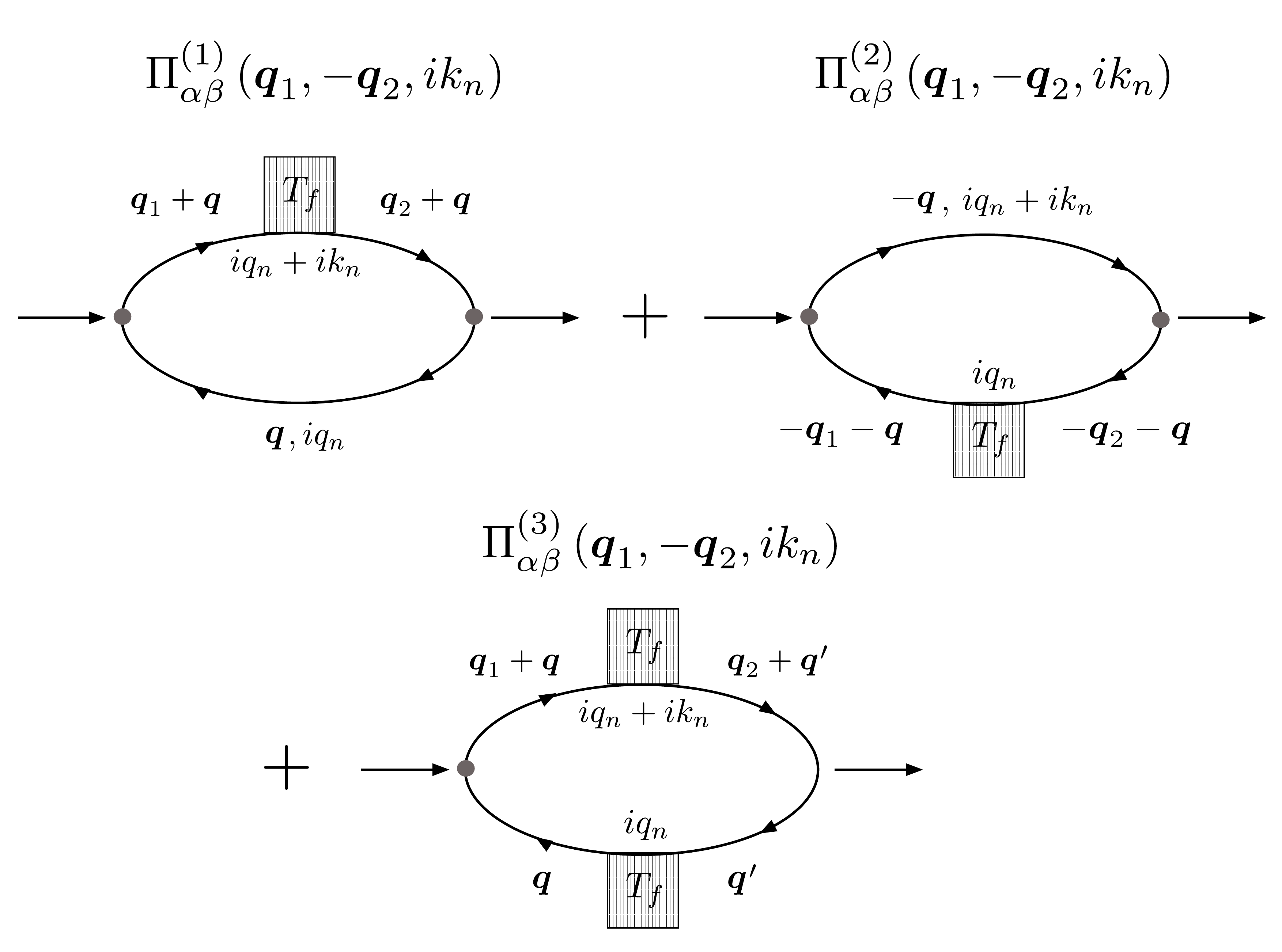}
\caption{The corrections to the current-current correlation function. Note that the diagram with impurity-scattering for both the particle and the hole evaluates to zero for transverse current correlations; see Appendix~\ref{sec:appkondo}.}
\label{fig:currentcorrdiags}
\end{center}
\end{figure}

From here on, we set $\vs{r}_f = 0$, without loss of generality. The two-momentum corrections to the conductivity are described by the diagrams in Fig~\ref{fig:currentcorrdiags}. (Note that the diagram involving impurity-scattering of both the particle and the hole vanishes for transverse correlations.) These represent instances where either the particle or hole in the particle-hole pair carrying the current scatters off of the impurity and yields novel two-momentum correlations. 

The imaginary part of the transverse current-current correlations, $\text{Im} \left[ \Pi_T (\vs{q}_1,\vs{q}_2,\omega + i 0^+) \right]$ and consequently the real part of the two-momentum corrections to the conductivity can be evaluated (from the diagrams in Fig.~\ref{fig:currentcorrdiags}) to yield

\begin{align}
\text{Re} \left[ \sigma_{T,T} (\vs{q}_1,\vs{q}_2, \omega_{NV}) \right] &\approx 4 \sigma_0 \tau F_1 [ x_1,x_2,\theta_1 - \theta_2] F_2 [T], \nonumber \\
F_2[T] &= \frac{-1}{\pi} \int d \omega [-n_F'(\omega)] \text{Im} [ T_f (\omega) ], 
\label{eq:kondocondsimp}
\end{align}

\begin{figure}
\begin{center}
\includegraphics[width = 3in]{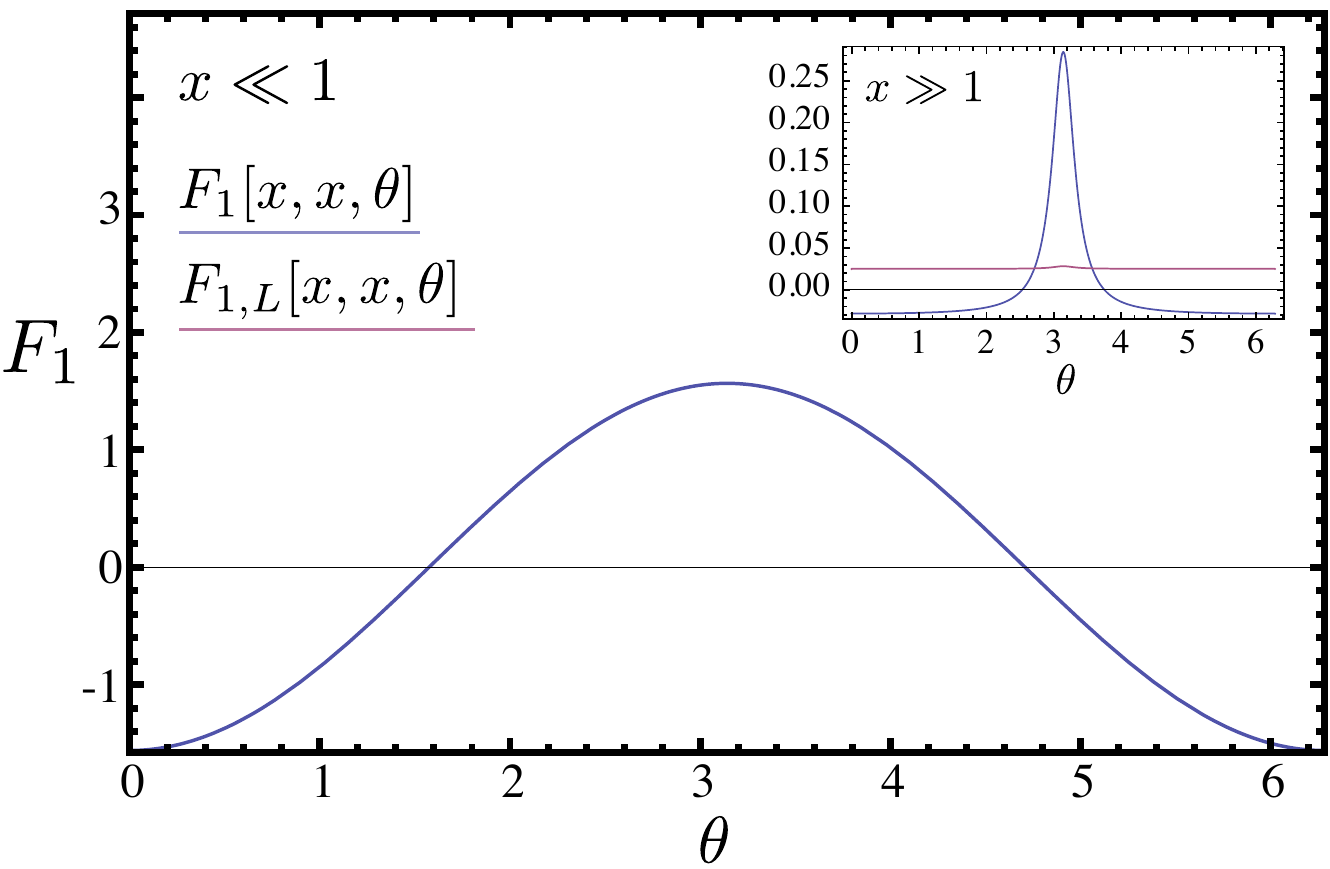}
\caption{$F_1$ ($F_{1,L}$) which characterize the angular dependence of transverse (longitudinal) two-momentum current-current correlations are plotted as a function of the relative angle $\theta$ between the two momenta of equal magnitude. The main figure shows the angular dependence for small momenta $q$, characterized by $x = v_F q \tau = 0.1$, while the inset is for large momenta, with $x = 10$. At small momenta, the transverse and longitudinal correlations coincide; they are both suppressed for forward scattering and enhanced for back scattering. At large momenta, $x \gg 1$, as seen in the inset, transverse and longitudinal current correlations differ significantly: longitudinal correlations are uniformly enhanced while transverse correlations appear to be uniformly suppressed except for back scattering where they are enhanced.}
\label{fig:kondocurrentangle}
\end{center}
\end{figure}

where $\sigma_0 = e^2 v^2_F \nu(0) \tau / 2$ is equal to the DC, uniform-field conductivity of the 2D system, $n'_F$ is the derivative of the Fermi function, $x_{1,2} = v_F q_{1,2} \tau$, $\theta_1$ and $\theta_2$ are azimuthal angles of the in-plane momenta $\vs{q}_1$ and $\vs{q}_2$, $F_1$ is an analytically determined dimensionless function (explicit form in Appendix~\ref{sec:appkondo}) which contains geometric information (besides an amplitude) about the current-current correlations near the impurity, and $F_2[T]$ is proportional to the scattering cross-section of the process in which a particle (or hole) scatters off of the isolated impurity (since it depends on the imaginary part of the T-matrix $T_f$). The above results hold under the assumption that $\omega = \omega_{NV} \approx 3$ GHz is the smallest scale in the problem; in this case, the two diagrams yield the same result leading to an extra factor of $2$ in Eq.~(\ref{eq:kondocondsimp}). Details of this calculation are provided in Appendix~\ref{sec:appkondo}.

Now we physically motivate this result. First, we note that the diagrams that we have considered calculate the generation of two-momentum current-current correlations due to the process in which a constituent particle or hole of the particle-hole pair (carrying the current) is scattered off of the isolated Kondo impurity. Thus, we can estimate the current-current fluctuations by counting the number of scattering events $N_s$, in time $\tau$ (the characteristic scale at which these fluctuations will decay), and the amplitude $A_s$ of current-current fluctuations that these contribute to. Then, $\text{Im} \left[ \Pi_T (\vs{q}_1, -\vs{q}_2, \omega) \right] = \omega \text{Re} [\sigma_{T,T} (\vs{q}_1, \vs{q}_2, \omega) ] \sim N_s A_s \tau$. 

First, we compute the number $N_s$ of scattering events that generate these current-current correlations. Keeping in mind the notation of Eqs.~(\ref{eq:kondocond}), we note that particle-hole pairs are created with a certain density of states $\nu(0)$ and weight $n_F (\omega') - n_F (\omega + \omega') \approx -\omega n_F'(\omega')$. Thus, $\nu(0) \omega$ is the effective particle-hole density at energy difference $\omega$, and approximate energy $\omega'$. A `tube' of these particle hole pairs of length $v_F \tau$ and thickness of the scattering cross-section scatter off of the Kondo impurity in time $\tau$. The scattering cross-section (of particles/hole at frequency $\sim \omega'$) is given by (using the optical theorem in two-dimensions~\cite{sakuraiQM}) $\sim \text{Im} T_f(\omega')/v_F$. With these details, the total number of scattering events, in the time $\tau$ can be estimated to be $N_s \sim \omega \nu(0) v_F \tau\frac{F_2[T]}{v_F}$.  

Next, we determine $A_s$. Heuristically computing the full angular and momentum dependence of $A_s$ is difficult. We note that for Eq.~(\ref{eq:kondocondsimp}) to hold, we require $A_s =  v^2_F F_1[\theta]$. Here we rationalize this result in some limits. Since particle-hole pairs are created with momenta $k\sim k_F$, they have velocities $\sim v_F$. This explains the factors of $v^2_F$. $F_1[\theta]$ can then be interpreted as a complicated interference amplitude between particle/hole plane-wave states with momenta $\vs{q} + \vs{q}_1$, $\vs{q} + \vs{q}_2$, and $\vs{q}$ ($q \sim k_F$ but otherwise arbitrary) and decaying due to the line-width $1/\tau$. The angular dependence of this function is complicated by the fact that we calculate transverse current correlations and will be discussed later, but the amplitude can be estimated. At small momenta, $v_F \vs{q}_1 \tau, v_F \vs{q}_2 \tau \ll 1$, the decay of the wave-function dominates the interference amplitude; consequently, $F_1[\theta] \sim 1$. At large momenta, the oscillations cause a decay in the amplitude, of the order of $1/x^2$, where $x$ is some combination of $v_F q_{1,2} \tau$. Reassuringly, in this limit, the current-current correlations do not depend on the background scattering-time $\tau$. We note that our qualitative expectations for $F_1[x_1,x_2,\theta]$ agree with the behavior of this function in the small and large momenta limits, respectively [see Eq.~(\ref{eq:f1forms})]. 

In Fig.~\ref{fig:kondocurrentangle}, we produce the geometric form of the transverse current correlations (by plotting $F_1[x_1 = x_2, \theta]$ for small and large momenta). At small momenta, $x_1, x_2 \ll 1$, the correlations are proportional to $- \cos (\theta_1 - \theta_2)$. This can be understood by the fact that, at small momenta, electron Green's functions do not retain geometric information since this is suppressed by the large, isotropic relaxation rate $1/\tau$. The geometrical dependence of correlations comes from the fact that the transverse part of the current carried by the electron-hole pair comes with amplitudes $\sin (\theta_q - \theta_1) \sin (\theta_q - \theta_2)$ which, averaged over $\theta_q$, yields the desired result. Note that the negative result at $\theta_1 = \theta_2$ is due to the fact that the impurity effectively reduces current-current correlations in the forward direction because of it's role in scattering the particles in all directions.  At large momenta, the angular dependence is mostly negated because the electron Green's functions are sharp, and the integral over $\theta_q$ picks out special angles for the average momentum $\vs{q}$ of the particle-hole pair that makes the particle-hole pair as on-shell as possible. This breaks down for back-scattering, that is, when $\theta_1 = \theta_2 \pm \pi$, since, in order to enforce the on-shell condition, $\vs{q} \perp \vs{q}_1, \vs{q}_2$ and the electron's Green's functions are again dominated by their imaginary part. 

Next, we focus on the function $F_2[T]$ which is solely related to the impurity's properties and which, in particular, will show interesting temperature dependence for the case of a Kondo impurity. 

\section{Noise profile near an isolated impurity}
\label{sec:noiseprof}
\begin{figure}
\begin{center}
\includegraphics[width = 3.5in]{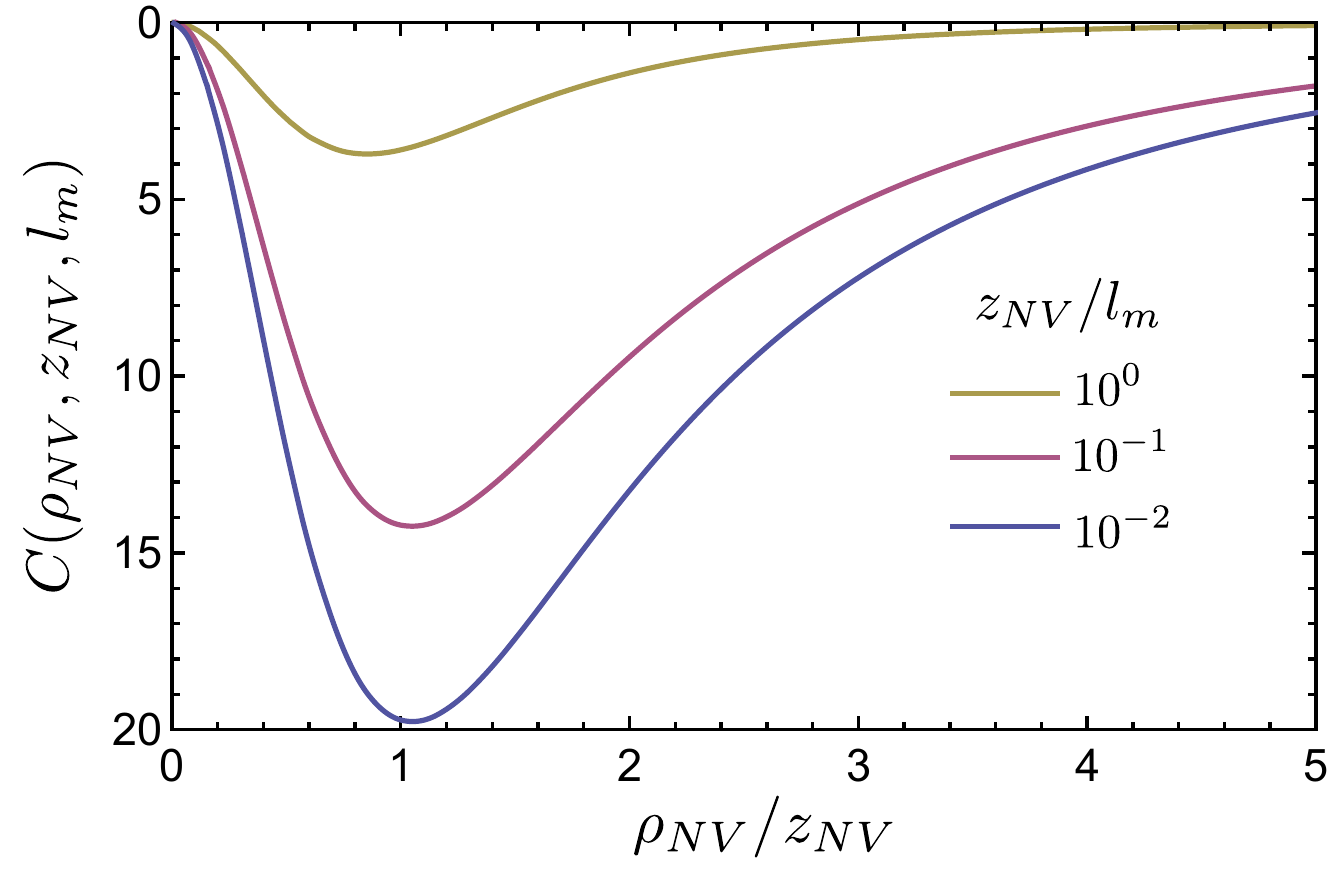}
\caption{The dimensionless curve $C (\rho_{NV}, z_{NV}, l_m)$ is plotted as function of $\rho_{NV}$ (in units of $z_{NV}$) for $z_{NV} =$ $l_m$, $0.1 l_m$, and, lastly, $0.01 l_m$, for which the curve attains its universal form associated with the limit $z_{NV} \ll l_m$.}
\label{fig:noiseatsmalldist}
\end{center}
\end{figure}

We now calculate the noise (due to current fluctuations) near a single impurity using Eqs.~\ref{eq:nontransnoise} and Eqs.~\ref{eq:kondocondsimp}. (See also Fig.~\ref{fig:noiseprotocol2} for an illustration of the experimental setup.) In the far-field limit where $z_{NV} \gg l_m$, the form of $F_1 [x_1 \ll 1, x_2 \ll 1, \theta] \approx - (\pi/2) \cos(\theta)$ allows us to analytically calculate the noise in all directions (see Appendix~\ref{sec:appkondo} for the results). In the near-field limit, $z_{NV} \ll l_m$, which is of greatest experimental interest, we rely on a numerical computation. The 
noise in the perpendicular-to-plane direction is given by: 

\begin{align}
N_z &= N_{z,\text{back.}} - \frac{\mu^2_0 k_B T e^2 \nu(0)}{16\,\hbar\,  \pi^4 z^2_{NV}} F_2[T] \; C (\rho_{NV}, z_{NV}, l_m), \nonumber \\
\label{eq:noiseprofgeneral}
\end{align}

where $N_{z, \text{back.}}$ is the background noise contribution due to the conductivity of the system in the absence of the impurity, and $C$ is a dimensionless function whose form is known analytically for $z_{NV} \gg l_m$, and it attains a universal ($l_m$ independent) form in the limit  $z_{NV} \ll l_m$ which can be determined numerically (see Fig.~\ref{fig:noiseatsmalldist}). Thus, as expected, the noise due to the impurity is independent of the mean-free path at distances $z_{NV} \ll l_m$. 

We note (from the results of Fig.~\ref{fig:noiseatsmalldist} and analytical results in the far-field regime) that the noise is unchanged (from the background) precisely above the impurity; this is because we only consider s-wave scattering (by considering a momentum-direction independent T-matrix) and by symmetry, these waves do not generate magnetic noise on top of the impurity. 

More surprisingly, the noise near the impurity is suppressed compared to the background. A simple-minded explanation is that in the far-field regime, where noise depends only on low-momentum conductivity, we can expect the Kondo impurity to act as any other impurity in that it simply reduces the effective scattering time, which reduces the conductivity, and consequently the magnetic noise. 
However, it is not immediately obvious that the `excess' noise must be negative in the near-field regime. In particular, in Fig.~\ref{fig:kondocurrentangle}, we see that unlike transverse current fluctuations, longitudinal fluctuations are enhanced at large momenta (or near-field) due to the presence of the Kondo impurity. These fluctuations, however, do not contribute to the noise because they are quickly screened at length scales of the 
inverse Fermi wave-vector. (For a discussion on the calculation of the longitudinal current-current fluctuations, see Appendix~\ref{sec:appkondo}.)

We can re-cast the noise calculation in a way which separates positive and negative contributions to the noise: 
\begin{widetext}
\begin{align}
N_z (\omega \rightarrow 0) \sim \int \frac{ d^2 \vs{q}}{(2\pi)^2} \int d \omega  (-n'_F (\omega)) A_f (\omega) A_0(\vs{q},\omega) & \bigg\{ \left| \int \frac{d^2 \vs{q}_1}{(2 \pi)^2} e^{- |\vs{q}_1| z_{NV} + i \vs{q}_1 \cdot \vs{\rho}_{NV} } \text{Re} \left[ G_0 (\vs{q} + \vs{q}_1) \right] \left(\vs{q} - \frac{\vs{q} \cdot \vs{q}_1}{q^2_1} \vs{q}_1\right) \right|^2 \nonumber \\
 &-  \left| \int \frac{ d^2 \vs{q}_1}{(2 \pi)^2} e^{- |\vs{q}_1| z_{NV} + i \vs{q}_1 \cdot \vs{\rho}_{NV} }\text{Im} \left[ G_0 (\vs{q} + \vs{q}_1) \right] \left(\vs{q} - \frac{\vs{q} \cdot \vs{q}_1}{q^2_1} \vs{q}_1\right) \right|^2 \bigg\}. \nonumber \\
\end{align}
\end{widetext}

This representation makes it obvious that in the far-field regime, where the important momenta satisfy $v_F q_1 \tau \ll 1$, the `excess' noise is negative because the imaginary part of the Green's function has a larger amplitude. In the opposite limit, the fact that the net result is still negative relies intricately on the factors $|\vs{q} - \vs{q} \cdot\hat{\vs{q}_1} \vs{q}_1|^2$; evidence for this comes from the fact that, unlike transverse fluctuations, longitudinal fluctuations are enhanced in the large momentum limit.  

\section{Noise from a Kondo or large-U Anderson impurity}
\label{sec:kondonoise}

In this section, we specialize the previous discussion to a Kondo or more generally, a large-U Anderson impurity. Such an impurity can exhibit strong scattering at low temperatures $T \lesssim T_K$ which can separate it from other weak or small-angle scatterers in the near-field regime. Moreover, the scattering properties are temperature dependent (unlike a simple potential scatterer), and as we discuss below, can be studied in the experiment we propose. 

Before we discuss in detail the noise profile due to current fluctuations near a Kondo impurity, we remark that at short distances, the noise from spin flips at the site of the impurity can become significant. This noise scales as $\sim 1/z^6_{NV}$ (due to the fact that the magnetic field from the spins themselves decay as $1/z^3_{NV}$) as opposed to the $1/z^2_{NV}$ [see Eq.~(\ref{eq:noiseprofgeneral})] scaling of noise from modified current fluctuations near the impurity. At most experimentally accessible distances, and for impurities with a Kondo temperature in the few Kelvin range, this noise turns out to be numerically smaller than the current noise. An estimate for the crossover scale $z_c$ below which this noise becomes dominant is provided in Appendix~\ref{sec:magfromspins}. Thus, we will focus on current noise in what follows. 

The results of Secs.~\ref{sec:twomomenta} and~\ref{sec:noiseprof} can generally be used to calculate the two-momentum current-current correlations when impurity scattering is primarily elastic; these can therefore be applied to an Anderson impurity in the regime $T \lesssim T_K$~\cite{zarandinelastic}. In what follows, we calculate the quantity $F_2[T]$ in Eq.~(\ref{eq:kondocondsimp}) for a large-U Anderson impurity in a mean-field slave boson approach. Together with the results of Sec.~\ref{sec:noiseprof}, this determines the complete noise profile near an Anderson impurity. We also comment on the experimental feasibility of our results. 

At low temperatures, $T \ll T_K$, a mean-field slave-boson (equivalently, a large-N expansion) approach [as discussed in section (7.5) in Ref.~\cite{hewson1997kondo}] can be used to arrive at a single pole approximation for the impurity Green's function. In this approximation, the impurity orbitals are replaced by independent fermionic operators along with the introduction of a slave boson operator 
mediating the exchange interaction.
In the mean-field approximation, the slave boson condenses below the Kondo transition temperature, and an effective quadratic model is found describing the scattering of conduction electrons off of the local impurity states. The scattering matrix element is found to be proportional to the condensed fraction of the slave-boson, and thus, it also controls the line-width of the impurity orbitals $\Delta$. The T-matrix for electron-impurity scattering is given by $T(\omega) = \frac{\Delta}{\pi \nu(0)} G_f (\omega)$ where $G_f (\omega) = 1/(\omega - \epsilon_f + i \Delta)$ is the retarded impurity Green's function. Assuming a flat conduction band of width $2D$, the renormalized hybridization parameter $\Delta(T)$ and resonance energy $\epsilon_f(T)$ satisfy the self-consistent equations (found by minimizing the free energy of the system)
\begin{align}
&n_f (T) = \int^D_{-D} d \omega n_F (\omega) \frac{-1}{\pi} \text{Im} \left[ G_f (\omega) \right],   \label{eq:slaveboson}
\\
&\int^{D}_{-D} d \omega n_F (\omega) \text{Re} \left[ G_f (\omega) \right] \approx - \text{log} \left( \frac{\sqrt{\epsilon_f (0)^2 + \Delta (0)^2}}{D} \right),
\nonumber
\end{align}
where $n_f [T]$ is the occupation of the impurity orbital at temperature $T$ (not to be confused with $n_F(\omega)$, the Fermi function). These equations can be solved given two input parameters, which we can choose to be, $n_f (T = 0) = \frac{1}{2} - \frac{1}{\pi} \tan^{-1}\left( \frac{\epsilon_f (0) }{\Delta(0)} \right)$, and the Kondo temperature $T_K$, which we define as $k_B T_K = \sqrt{\epsilon^2_f(0) + \Delta^2(0)}$. 

The function $F_2[T]$ is determined in terms of the temperature-dependent line-width and resonance energy as

\begin{figure}
\begin{center}
\includegraphics[width = 3.5in]{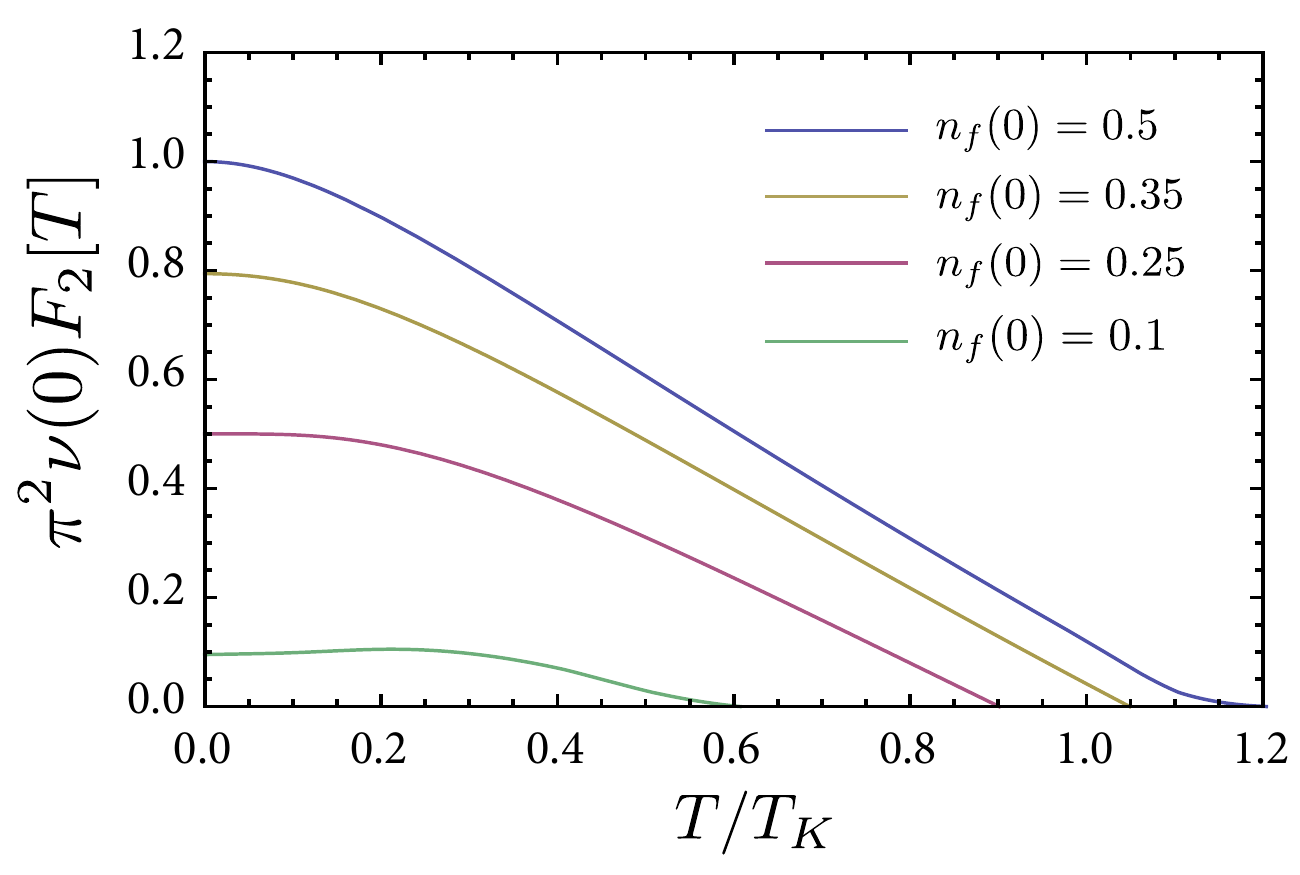}
\caption{The function $F_2[T]$ (in units of $1/\pi^2 \nu(0)$), calculated in the slave-boson mean-field approximation using Eqs.~\ref{eq:slaveboson}, is plotted as a function of the temperature (in units of the Kondo temperature $T_K$) for various values of the zero-temperature occupation of the Anderson impurity, $n_f (T  = 0)$, as discussed in the main text. At $T = 0$, the occupation of the impurity $n_f (0)$ determines the phase shift ($2 \pi n_f (0)$) acquired by electrons during scattering and is related to the scattering amplitude $F_2[T] = (\pi^2 \nu(0))^{-1} \sin^2 \left(\pi n_f (T = 0)\right)$. Noise suppression sensed near the Anderson impurity is directly proportional to this function which describes the effectiveness of the impurity in scattering electrons near the Fermi surface.}
\label{fig:f2Tplot}
\end{center}
\end{figure}

\begin{align}
F_2[T] &= \frac{\beta \Delta}{2 \pi^3 \nu(0)} \text{Re} \left[ \psi_3 \left( \frac{1}{2} + \frac{\beta \Delta}{2 \pi} + i \frac{\beta \epsilon_f}{2 \pi} \right) \right], \nonumber \\
\label{eq:f2Tcold}
\end{align}
where $\psi_3$ is the Trigamma function (double derivative of the logarithm of the Gamma function). 

In the limit $T \rightarrow 0$, the result of Eq.~(\ref{eq:f2Tcold}) agrees with the Friedel sum-rule~\cite{hewson1997kondo} result: $F_2 [T \rightarrow 0] = \frac{1}{\pi^2 \nu(0)} \sin^2 \left( \pi n_f (T = 0) \right)$. From Eq.~(\ref{eq:noiseprofgeneral}), we see that, in this limit, and for $z_{NV} \ll l_m$, so that $C$ is universal and independent of $l_m$, the modified (due to the impurity) noise amplitude depends solely on the single input parameter $n_f (T = 0)$, the occupation of the impurity at zero temperature. Therefore, noise measurements at $T \ll T_K$, may in principle be used to infer $n_f (T = 0)$ for magnetic impurities with a simple internal structure. Note that in the Kondo regime $n_f (T) = 1/2$ always, however, for more complex magnetic impurities like Fe or Mn, typically several angular momentum channels cooperate to screen the impurity spin.
 The other input parameter, $T_K$, can be more easily determined experimentally by comparing the temperature dependence of the noise near the impurity to the curves of Fig.~\ref{fig:f2Tplot}. Using these parameters the complete form of the Kondo resonance can be accessed experimentally through noise measurements.

For $T \gg T_K$, the resonance model does not provide an accurate description of the system. The full evaluation of the conductivity in this case is beyond the scope of this work but we point the reader to Refs.~\cite{RMPonNRG,CostiHewsonTransportAnderson} where the conductance of quantum dots modeled as Anderson impurities (which is directly proportional to $F_2[T]$) is discussed and evaluated for all temperature ranges.

\subsection{Experimental Protocol and Feasibility}
\label{sec:kondoexpfeas}

We now comment on a possible experimental protocol that could be used to measure the properties of a single Anderson or Kondo resonance. First, we note that the experiment is best carried out at distances $z_{NV} \ll l_m$ since the noise amplitude is larger, but also because the noise does not depend on extrinsic factors such as the mean-free path. Next, from Fig.~\ref{fig:noiseatsmalldist}, we see that the noise contribution of the impurity is maximal at a distance $\rho_{NV} \approx z_{NV}$. Thus,  a possible experiment could involve setting up a range of NV centers at a fixed distance $z_{NV} \ll l_m$ and examining where the noise is most different from the background, as illustrated in Fig.~\ref{fig:noiseprotocol2}. Once this point is found, the temperature can be varied and the temperature variation of the noise can be recorded. 

\begin{figure}
\begin{center}
\includegraphics[width = 3.5in]{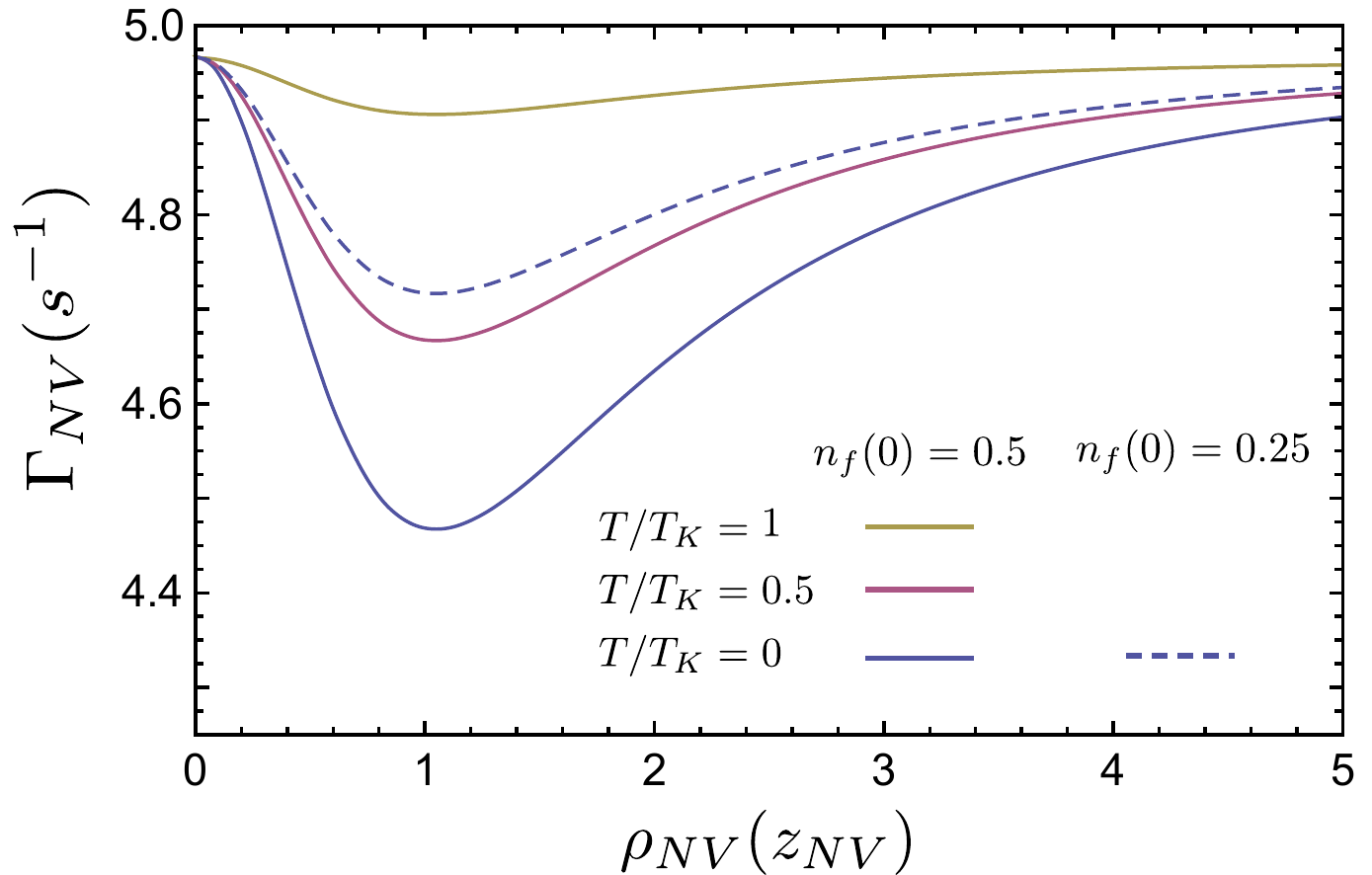}
\caption{The relaxation rate of an NV center measuring noise in the $z$-direction near an Anderson impurity in doped graphene, at a fixed distance $z_{NV} = 10 n$m $\approx 1/k_F$ for different zero-temperature impurity occupation $n_f(0)$ and temperatures $T$ relative to the Kondo temperature $T_K$. We assume an electronic temperature of $100$K for graphene, and a chemical potential bias of $1000$K; for details of conversion of noise into a relaxation rate for the NV-center, see Appendix~\ref{sec:graphenenumerical}.}
\label{fig:graphenekondofig}
\end{center}
\end{figure}

As discussed below Eq.~(\ref{eq:f2Tcold}), the amplitude and the temperature dependence of the measured noise can be 
used to infer the Kondo temperature $T_K$ and the occupation of the $d$ or $f$-level of the magnetic impurity, as described within the slave boson mean-field approximation. Alternatively, one can use the noise measurements to directly infer $F_2[T]$, as per Eq.~(\ref{eq:noiseprofgeneral}) (which the mean-field model only approximates) which describes the scattering properties of the impurity, and is itself a quantity of theoretical interest~\cite{RMPonNRG}. This puts our noise measurement protocol in contrast with tunneling probes which attempt to measure the spectral function of the Kondo resonance (although results are complicated by Fano resonances and the interaction between the impurity and the tunneling electrons~\cite{madhavan1998tunneling}). 

Finally, we estimate the amplitude of the noise due to the scattering off the impurity in comparison to the background noise at distances $z_{NV} \ll l_m$.  We note that, the noise due to the impurity scales as $1/z^2_{NV}$ as opposed to the noise from the material background which scales as $1/z_{NV}$ in this regime. Thus, the modification of the noise due to the impurity is stronger for smaller $z_{NV}$. If we estimate $N_{z,\text{back.}} \approx \mu^2_0 k_B T e^2 v_F \nu(0) / \left( 16 \pi z_{NV} \right)$ [using Eq.~(\ref{eq:nzsimpvalha}), and the conductivity in the ballistic regime, $\sigma_0 / qv_F \tau$], then using Eq.~(\ref{eq:noiseprofgeneral}), and noting the value of maximum value of $C \approx 20$ and $F_2[T] \approx 1/(\pi^2 \nu(0) \hbar)$, we find that the ratio $r = 1-  N_{z}/ N_{z,\text{back.}} $ has the value 

\begin{align}
r = \frac{40}{\pi^4 z_{NV} v_F \nu(0) h} &= \frac{10}{\pi^4} \frac{\hbar}{z_{NV} v_F m_e} \; \text{for a metal} \\  \nonumber
&= \frac{10}{\pi^4} \frac{1}{z_{NV} k_F} \; \text{for doped graphene}
\end{align}

For graphene, the noise suppression due to the impurity can reach (when $z_{NV} = 1/k_F$) $\sim 10.3\%$ of the background noise contribution (below $z_{NV} \sim 1/k_F$ longitudinal current fluctuations become important which we have neglected in our analysis); this contribution can be measured at low temperatures against the relaxation due to noise from the background which is at the Hz level, see Fig.~\ref{fig:graphenekondofig}. For a metal with nearly free electrons and $v_F = 10^6$ m/s, the suppression is about $\sim 1.2\%$ of the background noise at $z_{NV} = 1$ nm.  

\section{Summary and Outlook}
\label{sec:conclude}

To summarize, we studied how the magnetic noise profile above two-dimensional materials can be measured to directly infer the transport properties of the underlying system. We first discussed the theoretical framework for this in the context of spatially homogeneous (upon disorder averaging) materials, and later extended it to non-homogeneous systems. In the first part, we described in detailed how various transport regimes in electronic systems can be observed by measuring the magnetic noise-scaling as a function of the distance from the system; we made the discussion quantitative for the case of graphene. The aim of the latter part was to show that the noise profile near an impurity can be used to directly infer its scattering properties. One of the most striking examples of an impurity with interesting temperature-dependence of its scattering properties is an Anderson impurity. We provided details of how the temperature-dependent spectral properties of the resonance associated with an Anderson impurity can be directly observed in the magnetic noise measurements. We expect that these experiments can be carried out using appropriately placed NV-centers whose relaxation rates can be individually read out to measure the magnetic noise at their position, as illustrated in Figs.~\ref{fig:noiseprotocol1} and~\ref{fig:noiseprotocol2}. 

We further anticipate that NV-center based magnetic noise probes may find many interesting and novel applications in observing unique physical phenomena in two-dimensional materials that have never been established experimentally before. Possible directions include: 1) the observation of localization in two-dimensional electron gases, where one expects the conductivity to scale exponentially to zero at increasing length scales~\cite{localizationscaling}; 2) the observation of Chalker scaling in graphene (with e.g., vector-pseudospin disorder~\cite{chalkergraphene}), or in half-filled Landau Level systems~\cite{chalker1988scaling}, where we expect multi-fractal eigenstates~\cite{multifractalKravtsov} to result in an anomalous power-law scaling of the conductivity with wave-vector $q$; and 3) observation of spin-spin correlations of spinon Fermi surfaces~\cite{spinonFermisurface} in gapless spin-liquid states, to name a few.

\paragraph*{\textbf{Acknowledgments.} -- } We especially thank Javier Sanchez-Yamagishi and Trond Andersen, whose experimental insight into graphene helped closely direct our theoretical efforts on studying hydrodynamic effects in the material. Shimon Kolkowitz, Arthur Safira and Bo Dwyer are thanked for their input on the various experimental challenges of using NV centers to measure magnetic noise. We thank Ivar Martin, Andrew Lucas, and Debanjan Chowdhury for stimulating theoretical discussions. The authors acknowledge support from Harvard-MIT CUA, NSF Grant No. DMR-1308435, AFOSR Quantum Simulation MURI, ARO MURI on Atomtronics, ARO MURI Qusim program, AFOSR MURI Photonic Quantum Matter, NSSEFF, and the Moore Foundation. E.~D. acknowledges support from the Simons foundation, the Humboldt Foundation, Dr.~Max~R\"ossler, the Walter Haefner Foundation, and the ETH Foundation. R.~S. is supported by the NSF through a grant for the Institute for Theoretical Atomic, Molecular, and Optical Physics at Harvard University and the Smithsonian Astrophysical Observatory. V.~O. is grateful to the Harvard-MIT CUA and Harvard University for hospitality and support during his sabbatical, and acknowledges support from the NSF Grant No.~DMR-0955714. B.~H. is supported in part by the STC Center for Integrated Quantum Materials under NSF grant DMR-1231319. 


\appendix
\section{Translationally invariant case}
\label{sec:apptrans}

\begin{figure}
\begin{center}
\includegraphics[width = 3.5in]{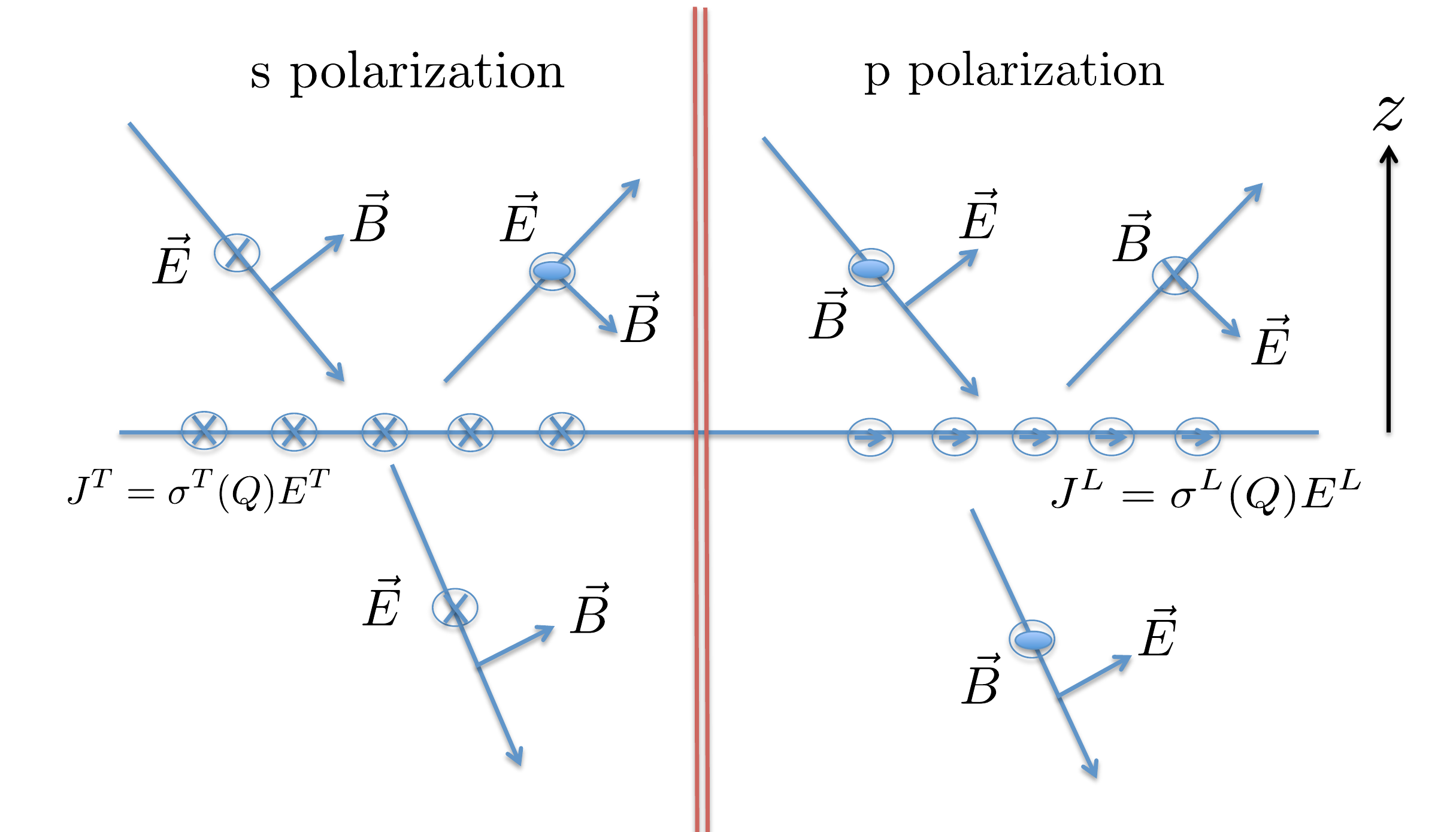}
\caption{Reflection and Transmission of s- and p- polarized waves. The s-polarized waves are seen to generate only transverse currents while p-polarized waves generate only longitudinal currents. The reflection coefficients have the behavior $r_s(Q) \sim \sigma^T(Q)$ while $r_p(Q) \sim \sigma^L(Q)/\epsilon^{RPA}(Q)$.}
\label{fig:rsrpfig}
\end{center}
\end{figure}

In this Appendix, we solve the problem of calculating the noise profile above a homogeneous 2D material. In order to achieve this goal, we need to first discuss the bases of solutions: that of s- and p- polarized waves (see Fig.~\ref{fig:rsrpfig}). These solutions will then be used to solve the problem of calculating the total magnetic field in the presence of a magnetic dipole at the position of the NV center. The response function, whose imaginary part is related to the magnetic noise, is then given by $\partial B_{\text{total}}/ \partial M_{ext}$, where $B_{\text{total}}$ is the total magnetic field in the presence of an external magnetic dipole $M_{ext}$. These parts will be carried out successively in the following subsections.  

\subsection{Bases: magnetic field profile for s- and p- polarized waves}

We assume the following convention: $\vs{B}_{in}$,$\vs{B}_r$,$\vs{B}_t$ correspond to incident (from NV on to the surface), reflected (back to the NV-center from the surface) and transmitted waves (through the surface). Similar conventions hold for the electric fields $\vs{E}_{in}$,$\vs{E}_r$,$\vs{E}_t$. 

\subsubsection{p-polarized waves}

The p-polarized form of these waves is given by

\begin{align}
\vs{B}_{in} &= B_0 \left( \hat{z} \times \hat{Q} \right) e^{i \vs{Q} \cdot \vs{\rho} - i q^\epsilon_z z}, \nonumber \\
\vs{B}_{r} &= r_p (q_z, Q, \omega) B_0 \left(\hat{z} \times \hat{Q} \right) e^{i \vs{Q} \cdot \vs{\rho} + i q^\epsilon_z z}, \nonumber \\
\vs{B}_{t} &= t_p (q_z, Q, \omega) B_0 \left( \hat{z} \times \hat{Q} \right) e^{i \vs{Q} \cdot \vs{\rho} - i q^{\epsilon'}_z z}.\nonumber \\
\end{align}

In the above, $q^\epsilon_z = \sqrt{ \omega^2 \epsilon/c^2 - Q^2}$ and $q^{\epsilon'}_z = \sqrt{ \omega^2 \epsilon'/c^2 - Q^2}$. We will be primarily be interested in the relaxation due to \emph{evanascent} waves for which $q_z$ will be imaginary: the reason for this is that the phase space of these waves is much greater and so they always dominate the noise at the site of the NV. For these waves, we will follow the convention that the imaginary part of $q_z$ is positive. 

One can now calculate the electric field from the magnetic field using the equation $\vs{E} = \frac{\curl{B} c^2}{ - i \omega \epsilon^{(')}}$ where $\epsilon$ or $\epsilon'$ is used according to where the electric field is being calculated. This yields

\begin{align}
\vs{E}_{in} &= \frac{1}{\epsilon} \frac{c^2}{- i \omega} B_0 \left( i q^\epsilon_z \hat{Q} + i Q \hat{z} \right) e^{i \vs{Q} \cdot \vs{\rho} - i q^\epsilon_z z}, \nonumber \\
\vs{E}_{r} &= \frac{r_p}{\epsilon} \frac{c^2}{- i \omega} B_0 \left( -i q^\epsilon_z \hat{Q} + i Q \hat{z} \right) e^{i \vs{Q} \cdot \vs{\rho} + i q^\epsilon_z z}, \nonumber \\
\vs{E}_{t} &= \frac{t_p}{\epsilon'} \frac{c^2}{- i \omega} B_0 \left( i q^{\epsilon'}_z \hat{Q} + i Q \hat{z} \right) e^{i \vs{Q} \cdot \vs{\rho} - i q^{\epsilon'}_z z}. \nonumber \\
\end{align}

Now we solve for the boundary conditions. The electric field's parallel component and magnetic field's perpendicular component must be continuous across the surface. The electric field perpendicular to the plane will depend on the charge accumulated on the 2D system sample; the magnetic field's parallel component will depend on the current in the 2D system sample. These are summarized as

\begin{align}
\vs{E}_\parallel (z = 0^+) &= \vs{E}_\parallel (z = 0^-), \nonumber \\
B_z (z = 0^+) &= B_z (z = 0^-), \nonumber \\
\frac{\rho}{\epsilon_0}  &= \epsilon E_z (z = 0^+) - \epsilon' E_z (z = 0^-), \nonumber \\
\mu_0 \vs{J} &= \hat{z} \times \left( \vs{H}_1 - \vs{H}_2 \right), \nonumber \\
\vs{H} &= \vs{B}/\mu_0, \nonumber \\
\div{J} &= -\pd{\rho}{t}, \nonumber \\
J^\mu &= \sigma^{\mu \nu} E^{\nu} (z = 0),
\label{eq:boundarygraphene}
\end{align}

where $\rho$ and $\vs{J}$ are the charge and current induced on the 2D system. In this case, the current turns out to be entirely in the direction of the in-plane wave-vector of the electric field, $\vs{Q}$. Thus, the current is given by the longitudinal conductivity $\sigma^L (Q, \omega)$ of the 2D system. We get two conditions which determine $r_p$ and $t_p$ uniquely. These are

\begin{align}
\frac{1 - r_p}{\epsilon} &= \frac{t}{\epsilon'} \frac{q^{\epsilon'}_z}{q^\epsilon_z}, \nonumber \\
\frac{1 + r_p - t_p}{t_p} &= \left( \frac{\sigma^L ( Q, \omega) q^{\epsilon'}_z}{\epsilon' \epsilon_0 \omega} \right), 
\end{align}

which can be solved to find

\begin{align}
r_p (q_z, Q, \omega) &=  \frac{ \left( \frac{\sigma^L ( Q, \omega) q^\epsilon_z}{\epsilon \epsilon_0 \omega} \right) + \frac{\epsilon'}{\epsilon} \frac{q^\epsilon_z}{q^{\epsilon'}_z} -1}{ \left( \frac{\sigma^L ( Q, \omega) q^\epsilon_z}{\epsilon \epsilon_0 \omega} \right) + \frac{\epsilon'}{\epsilon} \frac{q^\epsilon_z}{q^{\epsilon'}_z} + 1}, \nonumber \\
r_p  \bigg|_{\epsilon = \epsilon'} &= \frac{1}{1 + \frac{2 \epsilon \epsilon_0 \omega}{\sigma^L q^\epsilon_z}} \nonumber \\
 &\approx 1 - \frac{2 \epsilon \epsilon_0 \omega}{\sigma^L q^\epsilon_z} \; \; (\omega \rightarrow 0 ; q^\epsilon_z \rightarrow \infty).
\end{align}

 \subsubsection{s-polarized waves}

We now solve for the s-polarized case; that is, when the electric field is parallel to the surface. We have:

\begin{align}
\vs{E}_{in} &= E_0 \left( \hat{z} \times \hat{Q} \right) e^{i \vs{Q} \cdot \vs{\rho} - i q^\epsilon_z z}, \nonumber \\
\vs{E}_{r} &= r_s (q_z, Q, \omega) E_0 \left(\hat{z} \times \hat{Q} \right) e^{i \vs{Q} \cdot \vs{\rho} + i q^\epsilon_z z}, \nonumber \\
\vs{E}_{t} &= t_s (q_z, Q, \omega) E_0 \left( \hat{z} \times \hat{Q} \right) e^{i \vs{Q} \cdot \vs{\rho} - i q^{\epsilon'}_z z}. \nonumber \\
\end{align}

We can get the corresponding magnetic fields using Faraday's Law; $\vs{B} = \curl{E}/(i \omega)$. 

\begin{align}
\vs{B}_{in} &= \frac{1}{ i \omega} E_0 \left( i q^\epsilon_z \hat{Q} + i Q \hat{z} \right) e^{i \vs{Q} \cdot \vs{\rho} - i q^\epsilon_z z}, \nonumber \\
\vs{B}_{r} &= r_s \frac{1}{ i \omega} E_0 \left( -i q^\epsilon_z \hat{Q} + i Q \hat{z} \right) e^{i \vs{Q} \cdot \vs{\rho} + i q^\epsilon_z z}, \nonumber \\
\vs{B}_{t} &= t_s \frac{1}{ i \omega} E_0 \left( i q^{\epsilon'}_z \hat{Q} + i Q \hat{z} \right) e^{i \vs{Q} \cdot \vs{\rho} - i q^{\epsilon'}_z z}. \nonumber \\
\end{align}

In this case, $E_z$ is continuous by default, and there is no charge build up on the 2D system. This implies that all the current is in the transverse direction, so that $\vs{J} = \sigma^T \vs{E} (z = 0)$. The continuity of the parallel component of the electric field and perpendicular component of the magnetic field gives the same condition $1 + r_s = t_s$. The discontinuity of the parallel component of the magnetic field depends on the current on the 2D sample [as mentioned in Eq.~(\ref{eq:boundarygraphene})], solving which we get the conditions

\begin{align}
1 + r_s &= t_s, \nonumber \\
\frac{1 -r_s - t_s q^{\epsilon'}_z/q^\epsilon_z}{1 + r_s} &= \frac{\mu_0 \sigma^T \omega}{q^\epsilon_z},
\end{align}

which yield

\begin{align}
r_s (q_z, Q, \omega) &=  \frac{ 1 - \frac{q^{\epsilon'}_z}{q^\epsilon_z} - \frac{\mu_0 \sigma^T (Q, \omega) \omega}{q^\epsilon_z}}{1 + \frac{q^{\epsilon'}_z}{q^\epsilon_z} + \frac{\mu_0 \sigma^T (Q, \omega) \omega}{q^\epsilon_z}}, \nonumber \\
r_s \bigg|_{\epsilon = \epsilon'} &= \frac{-1}{1 + \frac{2 q^\epsilon_z}{\mu_0 \omega \sigma^T}},  \nonumber \\
&\approx - \frac{\omega \sigma^T \mu_0}{2 q^\epsilon_z} \; \; (\omega \rightarrow 0 ; q^\epsilon_z \rightarrow \infty).
\end{align}
 
\subsection{Calculation of the magnetic response function}

As discussed above, our aim is to solve for the problem of the electromagnetic field set up by a local magnetic dipole in the presence of a surface. This will allow us to calculate the magnetic response function via the relation $\chi_\alpha (\vs{r}_{NV}) = B^{\text{total}}_\alpha (\vs{r}_{NV}) / M_\alpha (\vs{r}_{NV})$. Note that here we use the symbols $\chi_\alpha \equiv \chi_{\alpha \alpha}$ since the noise tensor is diagonal if we work in the basis of directions $\hat{\alpha}$ that are perpendicular-to-plane ($\hat{z}$) and parallel-to-plane ($\hat{x}$,$\hat{y}$). We will discuss the cases of the magnetic dipole facing perpendicular and parallel to the 2D system separately. In addition, the symmetry of the problem will make it simpler to first calculate the magnetic field profile due to a sheet of magnetization. The resultant magnetic field profile can be appropriately Fourier-transformed to get the magnetic field profile in the presence of a point dipole. 

\subsubsection{dipole points perpendicular to the surface}

We first start with the case that the magnetic dipole faces in the $z$-direction (see Fig.~\ref{fig:graphenelocal}), but instead of working with a point dipole, we choose $\vs{M} = m_0 \hat{z} \delta(z - z_{NV} ) e^{i \vs{Q} \cdot \vs{\rho}}$; that is, we work with a sheet of magnetization. Once we have the magnetic-field profile due to such a sheet of magnetization, the magnetic-field profile in the presence of a point dipole can be obtained by integrating the result over the measure $\frac{d \vs{Q}}{(2 \pi)^2} e^{-i \vs{Q} \cdot \vs{\rho}_{NV}}$. 

Such magnetization generates only s-polarized waves. The solution of the Maxwell equations are given by

\begin{align}
- \nabla \times {\curl{E}} + \frac{\epsilon \omega^2}{c^2} \vs{E} &= \mu_0 (- i \omega) \curl{M}, \nonumber \\
\vs{E} &= E_0 \left( \hat{z} \times \hat{Q} \right) e^{i \vs{Q} \cdot \vs{\rho} + i q^\epsilon_z |z - z_{NV}|}, \nonumber \\
E_0 &= i \mu_0 \omega m_0 \frac{Q}{2 q^\epsilon_z}, \nonumber \\
q^\epsilon_z &= \sqrt{\epsilon \omega^2/c^2 - Q^2}.
\end{align}

The magnetic field at the site of the NV center (due to the magnetization sheet) can now be found quickly using the solution of the s-polarized case we just considered:

\be
B^{\text{sheet of M}}_{tot,z} (\vs{\rho}, z = z_{NV}) = E_0 \frac{Q}{\omega} e^{i \vs{Q} \cdot \vs{\rho} } \left( 1 + r_s e^{2 i q^\epsilon_z z_{NV}} \right).
\ee

Consequently, the total magnetic field in the z-direction due to a \emph{single} magetic dipole at the NV site pointing in the z-direction will be given by

\be
B_{tot,z} (\vs{r} = \vs{r}_{NV}) = \int \frac{d \vs{Q}}{(2 \pi)^2} e^{-i \vs{Q} \cdot \vs{\rho}_{NV}} B^{\text{sheet of M}}_{tot,z}. 
\ee

Thus, the response function (corresponding to the magnetic field, magnetic field commutator) which is given by $ B_{tot,z} (\vs{r} = \vs{r}_{NV}) / m_0$ is simply

\begin{align}
\chi_{z} (\omega) &= \int^\infty_0 \frac{d Q}{2 \pi} Q^3 \frac{ i \mu_0}{ 2 q^\epsilon_z} \; \text{[vacuum]} \nonumber \\
&+  \int^\infty_0 \frac{d Q}{2 \pi} Q^3 \frac{ i \mu_0}{ 2 q^\epsilon_z} r_s (q_z, Q, \omega) e^{2 i q^\epsilon_z z_{NV} };  \nonumber \\
q^\epsilon_z &= i \sqrt{ Q^2 - \epsilon \omega^2/c^2} \; \; \; \text{for} \; Q > \sqrt{\epsilon} \omega/c, \nonumber \\
q^\epsilon_z &= \sqrt{ \epsilon \omega^2/c^2 - Q^2} \; \; \; \text{for} \; Q < \sqrt{\epsilon} \omega/c,  
\end{align}

from which the magnetic noise spectrum (in the $z$-direction) can be computed via the fluctuation-dissipation relation: $N_{z} (\omega) = \hbar \coth{\left( \beta \hbar \omega / 2 \right)} \text{Im} \left[ \chi_{z} (\omega) \right] \approx \frac{2 k_B T}{ \omega}  \text{Im} \left[ \chi_{z} (\omega) \right]$. 

Note that $\chi_{z} (\omega)$ includes  vacuum fluctuations (that are independent of $r_s$) which also contribute to the noise. However, these contributions come only from undamped waves $Q < \epsilon \omega^2/c^2$. These have a limited phase space, and typically have a much smaller contribution to that total magnetic noise. The non-vacuum contributions are shown in Eqs.~(\ref{eq:xznoisersrp}).

If we assume that the 2D system is sandwiched by the same dielectric material $(\epsilon \approx \epsilon')$, then the noise (neglecting vacuum noise associated with $r_s = 0$) is given by [as also shown in Eq.~(\ref{eq:noisetrans}) of the main text]

\begin{align}
N_{z} (\omega) & \approx \frac{k_B T \mu_0^2}{16 \pi z^2_{NV}} \int_0^\infty \; dx \; x e^{-x} \text{Re} \left[ \sigma^T \left( \frac{x}{2 z_{NV}}, \omega \right) \right]\nonumber \\
&+ \mathcal{O}\left[ \frac{\omega z_{NV}}{c } \right].
\label{eq:chigraphenezz}
\end{align}

Note that, in the above result, we have neglected waves with a real wave-vector $q_z$, which in the limit of low-frequency, have negligible contribution of order $\mathcal{O}\left[ \left( \frac{\omega z_{NV}}{c }\right)^2 \right]$. Thus, the noise is primarily due to evanescent electromagnetic fluctuations above the 2D system. 

If we assume a constant $q$-independent conductivity of the 2D sample, which is valid for a diffusive system at lengths scales greater than the mean free-path, we find the noise to be given by

\be
N_{z} ( \omega) \bigg|_{\epsilon = \epsilon', \frac{\omega z_{NV}}{c} \ll 1} \approx \frac{k_B T \mu^2_0 \sigma^T}{16 \pi z^2_{NV} }.
\label{eq:nzsimpvalha}
\ee 

\subsubsection{dipole points parallel to the surface}

Next we consider a magnetic dipole moment at the NV site facing an in-plane direction, say $x$ (The noise is the same in any in-plane direction). To find the total magnetic field, we solve two simpler problems. We calculate the magnetic field in the presence of a sheet of magnetization with $\vs{M} = m_0 (\hat{z} \times \hat{Q}) \delta(z - z_{NV} ) e^{i \vs{Q} \cdot \vs{\rho} }$ and, separately, in the presence of a sheet of magnetization with $\vs{M} = m_0 \hat{Q} \delta(z - z_{NV} ) e^{i \vs{Q} \cdot \vs{\rho} }$. We note that the total magnetic field due to the point dipole moment can then be found by integrating the magnetic field found (in the $x$-direction, to calculate $N_x \equiv N_{xx}$) in these two separate calculations with the Fourier factor $e^{-i \vs{Q} \cdot \vs{\rho}_{NV}}$ and additionally projection factors $\hat{x}.(\hat{z} \times \hat{Q}) $ and $\hat{x}. \hat{Q} $, respectively. The decomposition into moments parallel to $(\hat{z} \times \hat{Q})$ and $\hat{Q}$ is performed because magnetization sheets in these directions yield exclusively p- and s-polarized waves, respectively.  

First we deal with the part that is in the direction $(\hat{z} \times \hat{Q} )$. 

\paragraph{$\vs{M} \propto  (\hat{z} \times \hat{Q} )$.}

We work with the free field $\vs{H}$ which does not have any implicit dependence on $\vs{M}$. The relevant Maxwell equations and their solutions are given by

\begin{align}
- \nabla^2 \vs{H} - \frac{\epsilon \omega^2}{c^2} \vs{H} &=  \frac{\epsilon \omega^2}{c^2} \vs{M},  \nonumber \\
\vs{H} &= H_0 \left( \hat{z} \times \hat{Q} \right) e^{i \vs{Q} \cdot \vs{\rho} + i q^\epsilon_z |z - z_{NV}|}, \nonumber \\
H_0 &= i m_0 \frac{\epsilon \omega^2/c^2}{2 q^\epsilon_z}, \nonumber \\
q^\epsilon_z &= \sqrt{\epsilon \omega^2/c^2 - Q^2}.
\end{align}

Away from the magnetization strip, $\vs{B} = \mu_0 \vs{H}$. We can easily find the reflected fields because they are in a p-polarized form. Following the methods of the previous section, we find a contribution to the susceptibility which we call $\chi^{\hat{z} \times \hat{Q}}_x$, 

\begin{align}
\chi^{\hat{z} \times \hat{Q}}_{x} &=  \int^\infty_0 \frac{d Q}{2 \pi} Q \frac{\epsilon \omega^2}{c^2} \frac{ i \mu_0}{4 q^\epsilon_z} \left( 1 + r_p (q_z, Q, \omega) e^{2 i q^\epsilon_z z_{NV} } \right). \nonumber \\
\end{align}

\paragraph{$\vs{M} \propto  \hat{Q}$.}

In this case the generated waves are s-polarized. The solution to the electric field in this case can be found solving
\begin{align}
- \nabla \times {\curl{E}} + \frac{\epsilon \omega^2}{c^2} \vs{E} &= \mu_0 (- i \omega) \curl{M}, \nonumber \\
\vs{E} &= E_0 \left( \hat{z} \times \hat{Q} \right) \pd{}{z}e^{i \vs{Q} \cdot \vs{\rho} + i q^\epsilon_z |z - z_{NV}|}, \nonumber \\
E_0 &= \mu_0 \omega m_0 \frac{1}{2 q^\epsilon_z}, \nonumber \\
q^\epsilon_z &= \sqrt{\epsilon \omega^2/c^2 - Q^2}.
\end{align}

Here we find, after calculating the magnetic field (by comparing with the solution we have for the s-polarized case): 
\begin{align}
\chi^{\hat{Q}}_{x} &=  \int^\infty_0 \frac{d Q}{2 \pi} Q q^\epsilon_z \frac{ i \mu_0}{4} \left( 1 - r_s (q_z, Q, \omega) e^{2 i q^\epsilon_z z_{NV} } \right) \nonumber \\
\end{align}
 
In total, we find, for the in-plane susceptibility, and the noise spectrum:

\begin{align}
\chi_{x} &=  \int^{\infty}_0 \frac{d Q}{2 \pi}  \frac{ i \mu_0 Q q^\epsilon_z}{4} \left( 1 +  \frac{\epsilon \omega^2}{(q^\epsilon_z)^2 c^2} \right) \; \text{[vacuum]} \nonumber \\
&+  \int^\infty_0 \frac{d Q}{2 \pi} Q q^\epsilon_z \frac{ i \mu_0}{4} \left( \frac{\epsilon \omega^2}{(q^\epsilon_z)^2 c^2}  r_p  - r_s \right) e^{2 i q^\epsilon_z z_{NV} }, \nonumber \\
&= \chi_{y}; \nonumber \\
q^\epsilon_z &= i \sqrt{ Q^2 - \epsilon \omega^2/c^2} \; \; \; \text{for} \; Q > \sqrt{\epsilon} \omega/c, \nonumber \\
q^\epsilon_z &= \sqrt{ \epsilon \omega^2/c^2 - Q^2} \; \; \; \text{for} \; Q < \sqrt{\epsilon} \omega/c,  \nonumber \\
N_{x} (\omega) &= \hbar \coth{\left( \beta \hbar \omega / 2 \right)} \text{Im} \left[ \chi_{x} (\omega) \right] \approx \frac{2 k_B T}{\omega}  \text{Im} \left[ \chi_{x} (\omega) \right].
\end{align} 

The non-vacuum part is shown in the main text in Eqs.~(\ref{eq:xznoisersrp}). As before, the main contribution to the noise is from evanescent waves ($\text{Im} [q^\epsilon_z ] \neq 0$) owing to their larger phase space; this approximation is correct to order $\mathcal{O} \left[ (\omega z_{NV} / c)^2 \right]$ (Vacuum fluctuations are of a lower order because evanescent vacuum fluctuations do not contribute to the noise). 

The contribution to the noise in the $x$-direction has a component (proportional to $r_s$) that is exactly $N_z / 2$. We define $\tilde{N}_\perp = N_x - N_z/2$ to separately discuss the features of this part of the noise. 

Assuming that $\epsilon \approx \epsilon'$, we find

\begin{align}
\tilde{N}_\perp \omega &= N_x (\omega) - \frac{N_z (\omega)}{2},  \nonumber \\
\tilde{N}_\perp (\omega) &= \frac{k_B T \mu_0^2}{32 \pi z^2_{NV}} \int_0^\infty dx \; x e^{-x} \text{Re} \left[ \frac{\sigma^L \left( \frac{x}{2 z_{NV}}, \omega \right)}{\epsilon^L_{RPA} \left( \frac{x}{2 z_{NV}}, \omega \right)} \right] \nonumber \\
&+ \mathcal{O} \left[ \left( \frac{\omega z_{NV}}{c} \right)^4 \right], \nonumber \\
\label{eq:chigraphenexx}
\end{align}

where we used the result $\epsilon_{RPA} (q, \omega) = 1 + i \frac{q \sigma^L}{\omega \epsilon_0 (\epsilon + \epsilon')}$. 

The above expression is somewhat misleading because for metals (and semi-metals like graphene), screening plays an important role and the imaginary part of $\epsilon_{RPA}$ is large for most $q = x/2 z_{NV}$. Thus, a more appropriate expression is

\begin{align}
\tilde{N}_\perp (\omega) \approx \frac{k_BT}{\epsilon^2 \omega^2}{4 \pi c^2} \int^\infty_0 dx \frac{e^{-x}}{x} \frac{1}{\text{Re}[\sigma^L]}
\end{align}

and consequently, 

\begin{align}
N_{x} &\approx \frac{k_B T \mu^2_0 \sigma^T}{32 \pi z^2_{NV} }   \nonumber\\
&\times\Bigg( 1+\epsilon^2  \frac{z_{NV}^2 \omega^2}{c^2} \frac{\epsilon_0/\mu_0}{\sigma^T (q = \frac{1}{2z_{NV}}) \sigma^L(q = \frac{1}{2z_{NV}})} \Bigg) \nonumber \\
& \approx \frac{N_{zz}}{2}.
\end{align}

The second term, to a very good approximation is always much smaller than $1$ and can be ignored.

\subsection{Calculation of noise from Biot-Savart law}
\label{sec:biotsavart}
In this subsection, we outline the calculation of the magnetic noise as performed using the Biot-Savart law.

The magnetic field $B_\alpha (\vs{r}_{NV}, t)$ due to a current $J_{\beta} (\vs{r}, t)$ in the material can be derived from the Biot-Savart law (assuming speed of light $c \rightarrow \infty$): 

\begin{align}
B_\alpha (\vs{r}_{NV}, t) &= - \frac{\mu_0}{4 \pi} \int  \frac{d^2 \vs{r}J_\beta (\vs{r}, t)}{| \vs{r} - \vs{r}_{NV} |^3} \left( (\vs{r} - \vs{r}_{NV} ) \times \hat{\beta} \right) \cdot \hat{\alpha}. 
\label{eq:biotsavart}
\end{align}

We can calculate the noise $N_{\alpha} (\omega) = \hbar \coth \left(  \beta \hbar \omega / 2 \right) \mathcal{F} \left[ \avg{[B_\alpha(\vs{r}_{NV}, t), B_\alpha(\vs{r}_{NV}, t') ]_+} \right] /2$ from Eq.~(\ref{eq:biotsavart}) and express the result in terms of the imaginary part of the current-current response functions $\chi^J_{\beta \beta'} (\vs{q}, \omega)$. Using the decomposition of these correlations into the transverse and longitudinal parts, that is, $\chi^J_{\beta \beta'} (\vs{q}, \omega) = \chi^J_T (q, \omega) \left( \delta_{\beta \beta'} - \frac{q_\beta q_{\beta'}}{q^2} \right) +  \chi^J_L (q, \omega)  \frac{q_\beta q_{\beta'}}{q^2}$, we can express the noise in the in-plane direction as

\begin{align}
N_x (\omega) &= \frac{k_B T \mu^2_0}{32 \pi z^2_{NV} \omega} \\ 
&\times \int dx \; x e^{-x} \text{Im} \left[ \chi^J_T (\frac{x}{2 z_{NV}}) + \chi^J_L (\frac{x}{2 z_{NV}}) \right]. \nonumber
\label{eq:biosavartnoise}
\end{align}

In order to arrive at the results of Eq.~(\ref{eq:chigraphenezz},\ref{eq:chigraphenexx}), we can substitute $\text{Im} [ \chi^J_T ] = \omega \text{Re} [ \sigma^T]$ and $\text{Im} [ \chi^J_L] = \omega \text{Re} [\sigma^L / \epsilon_{RPA}]$. The reason for the extra factor of $\epsilon_{RPA}$ in the longitudinal case is because the conductivity is defined with respect to the \emph{total} electric field unlike the current-current correlations which measure response due to an external electric field. Note that the main advantage of the method of using reflection-coefficients is the adaptability to cases when the material environment is more complicated. In this case, the reflection coefficients can have a more complicated expression that cannot be captured by a straightforward application of the Biot-Savart law. There are also certain subtle points: the expression for noise Eq.~(\ref{eq:chigraphenexx}) contains $\epsilon_{RPA}(q, \omega)$ only when $Q \gg \omega/c$, since we can then substitute $q_z \approx i Q$.

\subsection{Summary}

The main results of this appendix are given in Eqs.~(\ref{eq:chigraphenezz}) and~(\ref{eq:chigraphenexx}). In principle, these results can be used to measure the entire $q-$dependent transverse and longitudinal conductivities of the system. To extract the transverse part, we can measure the magnetic noise perpendicular to the 2D surface, as a function of the distance from it. To extract the longitudinal part, we observe that we need to measure the noise spectrum in an in-plane direction, and subtract from it half the result obtained from the measurements of noise perpendicular to the surface: this contribution, however, is rather small for most metals as long as the  NV-center is placed at a distance larger than the inverse of the Fermi wave-vector, and we ignore it.

\section{Magnetic noise from spin and current fluctuations}
\label{sec:magfromspins}
In this section, we first show that the magnetic noise from spin fluctuations in a metal becomes significant only in the case of a deeply localized system with $k_F l_m \sim 1$, or when the NV-center is close enough (to the material) to be able to resolve inter-particle distances ($\sim 1/k_F$). Later, we show that noise from spin fluctuations near the Kondo impurity is small compared to the noise from current fluctuations near it. 

\paragraph*{Magnetic fluctuations of free spins in a metal.}We first note that the magnetic noise scales as $|\vs{B}(\vs{r}_{NV})|^2$, where $\vs{B}(\vs{r}_{NV})$ is the magnetic field at the site of the NV center generated by current or spin fluctuations inside the material. The amplitude of the magnetic field $|\vs{B}|$ can be related to currents via the kernels $K_J (z) \sim 1/z^2$ for currents and $K_s (z) \sim 1/z^3$ for spins; this follows from the Biot-Savart law. While both the magnetic field terms (in the correlation function) are evaluated at the same location (at the position of the NV-center), the currents (or spins) producing these fields can themselves originate from different locations. Thus, the magnetic noise scales as $\sim \int \int d \vs{r}_1 d \vs{r}_2 J (\vs{r}_1) J(\vs{r}_2)$. Using the fact that small-momentum current fluctuations have low phase space, and high-momenta current fluctuations cancel each other [as reflected in Eq.~(\ref{eq:noisetrans})], we can argue that the most significant current-current correlations (for noise evaluation) are those at the length scale $\sim z_{NV}$. This reduces one of the spatial integrals in the above calculation, and the remaining spatial integral gives a factor of the volume, $V(z) \sim z^2_{NV}$ from which current fluctuations generate significant magnetic fields at the site of the NV-center. 

Putting the above details together, the magnetic noise from current fluctuations, $N_J (z_{NV}) \sim K^2_J V |J|^2$ while that from spin fluctuations, $N_s(z_{NV}) \sim \mu_B^2. K^2_s V |M|^2$; here, $|J|^2$ and $|M|^2$ are the amplitudes of current and spin fluctuations, respectively. We note that $|J|^2 \sim \sigma_T (q) k_BT $ as discussed in the main text, while, in a single-mode approximation, we can find $|M|^2 \sim \chi^0_M / \Gamma_M$, where $\chi^0_M$ is the static spin susceptibility at wave-vector $q \sim 1/z_{NV}$ and $\Gamma_M$ is a $q-$dependent relaxation rate of magnetic fluctuations. This `relaxation' may be due to diffusion, with $\Gamma_M \sim D q^2$. The result for $|M|^2$ can be arrived at as follows. We assume magnetization dynamics follow from a single pole, that is, $M (q, \omega) \sim \theta(q) / (-i \omega + \Gamma_M)$, where $\theta(q)$ is a function whose amplitude will be determined using a sum-rule. The magnetization fluctuations, $|M(q, \omega)|^2$, are related to the magnetic susceptibility $\chi_M (q, \omega)$ by the fluctuation-dissipation relation: $|M(q, \omega)|^2 = \hbar \text{Im} [ \chi_M (q, \omega)] \coth (\hbar \omega/2 k_BT)$. We can then use the sum-rule $\int^{\infty}_{\infty} \text{Im} [ \chi_M (q, \omega)] / \omega d \omega= \pi \chi^0_M (q)$ to show that $|\theta(q)|^2 = k_BT \chi^0_M (q) \Gamma_M$. Simple manipulations yield $|M|^2 \sim k_BT \chi^0_M / \Gamma_M$ for low frequencies $\omega = \omega_{NV} \ll \Gamma_M$. 

We now note that, away from any magnetic transition, $\chi_M \sim \nu(0)$, or the density of states near the Fermi surface in a conducting system. The conductivity, on the other hand can be estimated as $\sigma \sim e^2 (\nu(0) \epsilon_F) / (\Gamma_J m_e)$, where $\Gamma_J$ is the current relaxation rate, and $\nu(0) \epsilon_F$ is an approximation of the charge density. Putting these details together, we find the ratio $N_s/N_J = 1/(k_F z_{NV})^2 \times (\Gamma_J/ \Gamma_M)$. Here already we note that the factor $1/(k_F z_{NV})^2$ suggests that the magnetic noise from current fluctuations will typically dominate noise from magnetic fluctuations for $z_{NV} \gg 1/k_F$. 

To complete the evaluation of $N_s/N_J$, we now discuss the behavior of $\Gamma_J$ and $\Gamma_M$ as a function of $q$. For $z_{NV} \gg l_m$ or $q \ll 1/l_m$ ($l_m$ being the mean-free path of scattering from impurities), $\Gamma_J (q) \sim \text{const.} = v_F / l_m$, while, magnetization typically diffuses (if the impurities are non-magnetic), so that $\Gamma_M (q) \sim D q^2$, with $D = v_F l_m$. In this limit, $N_s / N_J \approx 1/(k_F l_m)^2$; thus, magnetization fluctuations are only important if the system is strongly localized, that is, $k_F l_m \lesssim 1$. Note that the addition of magnetic impurities or spin-orbit coupling which can relax spin fluctuations only increases $\Gamma_M$, which reduces the ratio $N_s/N_J$ further. Now, for $z \ll l_m$ or $q \gg 1/l_m$, both $\Gamma_J$ and $\Gamma_M$ scale as $v_F q$; this is because the time-scale for both magnetic and current fluctuations sensed by the NV-center is the time taken by electrons to pass through the zone of influence (a region of size $z_{NV}$ in the material closest to the NV-center). In this limit, $\Gamma_J/\Gamma_M = 1$ and $N_s/N_J = 1/(k_F z_{NV})^2$. Thus, the magnetic noise from spin fluctuations becomes comparable to that from current fluctuations only when the NV-center is closer than the inter-particle distance $\sim 1/k_F$ in the system, or when the material is highly localized, $k_F l_m \lesssim 1$.

\paragraph*{Magnetic fluctuations of the Kondo resonance.} Here we investigate the noise from spin fluctuations near the the Kondo impurity and compare it to the noise from modified current fluctuations [second term in Eq.~(\ref{eq:noiseprofgeneral})]. 

We first note that the magnetic noise due to the flipping of the spin of conduction electrons that screen the impurity is much weaker than the noise due to the flipping of the impurity spin. This is because the screening (of the single impurity moment) occurs over a large length scale~\cite{affleck2009kondo} $\xi_K$ and a nearby NV center will pick up only a tiny fraction of this magnetic fluctuation. To estimate the noise due to spin flips at the impurity site, we estimate the amplitude of the magnetic field at the site of the NV center, as produced by dipole fluctuations of the almost perfectly screened magnetic impurity. If we concern ourselves with magnetic noise in the z-direction  and focus on the most relevant direction,   $\rho_{NV} = z_{NV}$, the result for the noise is related to the local spin susceptibility $\chi(\omega)$ as
\be
N_{z, \text{imp.}} = \frac{5 \mu_0^2}{256 \pi^2 z^6_{NV}} \coth \left( \frac{\hbar \omega}{2 k_BT} \right) \text{Im} [ \chi (\omega) ].
\ee 

To estimate the susceptibility of the impurity spin, we use the spin-boson approach. The impurity moment in the $z$ direction is given by $M_z = g \mu_B \sum_\sigma \frac{\sigma}{2} f^\dagger_\sigma f_\sigma$, where $g$ is a Lande factor, and the operators $f_\sigma, f^\dagger_\sigma$ correspond to effective fermion operators whose Green's function $G_f (\omega) = 1/( \omega - \epsilon_f + i \Delta)$ was introduced in the main text. For the purposes of this estimate, we will assume $\epsilon_f = 0$, and as per our definition of $T_K$, this implies $T_K = \Delta (T = 0)$. The time-ordered Green's function $G_z (\tau) = - \avg{T_\tau \{ M_z (\tau) M_z (0) \}}$ is then given by $G_z (\tau) = \frac{(g\mu_B)^2}{2} G_f (\tau) G_f (- \tau)$, and the impurity spin susceptibility can be evaluated from it using $\chi (\omega)= - G_z (\omega + i0^+)$. The calculation yields $\text{Im} [ \chi (\omega) ]  = (g \mu_B)^2 \frac{\hbar \omega}{2 \pi ( k_B T_K)^2}$ and $\text{Re} [ \chi (0)] = (g\mu_B)^2 \frac{1}{2 \pi k_B T_K}$, at zero temperature. Note that a different definition of the Kondo temperature~\cite{hewson1997kondo}, which we introduce here as $\tilde{T}_K$ is through the static, real spin susceptibility $\text{Re} [ \chi (0)] \approx (g\mu_B)^2 \frac{1}{4 \pi k_B \tilde{T}_K}$. Thus, we see $\tilde{T}_K \approx T_K/2 = \Delta(0)/2$. 

We can now compare this to the noise from modified current fluctuations near the Kondo impurity. Using $C \sim 20$ as appropriate for $z_{NV} \ll l_m$ (see Fig.~\ref{fig:noiseatsmalldist}), and $F_2[T] \sim 1/(\pi^2 \nu(0))$, we find that the noise from spin fluctuations at temperatures $T \lesssim T_K$, is smaller than the noise from current fluctuations by a factor $\approx \frac{\pi^3}{64} (\epsilon_z/ k_B \tilde{T}_K)^2$, where $\epsilon_z = \hbar^2 /(z^2_{NV}  2 m_e)$. This defines a crossover scale, $z_c = 0.58\, \hbar/\sqrt{ k_B \tilde{T}_K m_e}$, such that for $z_{NV} < z_c$, the noise due to spin fluctuations dominates over noise due to current fluctuations. Assuming a Kondo temperature in the range of $\tilde{T}_K= 1$ to $10$ K (in metals and graphene~\cite{chen2011tunable} is typically around $10 K$, or greater), we estimate a cross-over scale  between $z_c = 5\, -\, 16$ nm. While such small distances are hard to achieve experimentally, it may be possible to study this regime experimentally in systems with much smaller Kondo temperatures.

\section{Numerical computation of conductivity in graphene}
\label{sec:graphenenumerical}

To obtain the relaxation time due to inter-particle collisions, $\tau_{ee}$, we consider the collision integral in the Born approximation (as in Ref.~\cite{FritzsigmaQ}). The complete evaluation of this integral is beyond the scope of this text. Here we provide a simple order-of magnitude estimate for the value of $\tau_{ee}$. In the Born approximation, the collision rate $\tau_{ee}^{-1}$ of an electron with momentum $\vs{k}$ is $\propto \sum_{\vs{k}',\vs{q}} \delta (|\vs{k'}| + |\vs{k}| - |\vs{k'} + \vs{q}/2| - |\vs{k} - \vs{q}/2|) |V (k_F)|^2 \mathcal{F}$, where $|V(q)|^2 \approx \alpha^2 v_F^2 / (q + \alpha k_F)^2$ is the square of the screened interaction potential between the electrons at wave-vector $q$, $\alpha$ is the effective fine-structure constant in graphene, and $\mathcal{F}$ refers to a factor of Fermi functions which essentially limit the integrals to energies $k_BT$ around the Fermi surface. The integrals can be simplified to yield the scaling $\nu(k_F) T^2 |V(k_F)|^2$ (ignoring factors of $v_F$ and $\hbar$) where $\nu(k_F)$ is the density of states of electrons at the Fermi surface. When $k_BT \gg \mu$, we can set $k_F \sim k_BT/v_F$ to find $\tau_{ee}^{-1} \sim \alpha^2 k_B T/ \hbar$. Alternatively, when $\mu \gg k_BT$, we find $\tau_{ee}^{-1} \sim \alpha^2 (k_BT)^2 / \mu$. The precise numerical factors require a more detailed calculation. For our numerical results, we will use the result from Ref.~\cite{MuellersigmaQ} (which was calculated for $k_B T \gg \mu$), $\tau_{ee}^{-1} \approx (0.27)^{-1} \alpha^2 k_BT / \hbar$ with $\alpha^2$ following the renormalization-group flow as discussed in Ref.~\cite{FritzsigmaQ} and extrapolate the result to general chemical potentials $\mu$ as

\be
\tau_{ee} = 0.27 \frac{\hbar}{k_B T} \frac{1 + (\mu/k_B T)^2}{2}.
\label{eq:tauee}
\ee

Next, we consider the relaxation time $\tau$ due to an external bath. It can be calculated under various approximations depending on whether phonons or charged impurities play a more significant role in the system. Assuming one can engineer extremely clean graphene samples, the mean-free path will be dominated by phonons. Following Ref.~\cite{hwangdassarmaphonons}, one can estimate the instrinsic (due to vibrations in graphene) acoustic phonon contribution at $k_BT = \mu$ to lead to a scattering length $l_m \approx 12 \mu$m $\times \left( \frac{D}{19 \text{eV}}\right)^2 \left( \frac{T}{300 \text{K}} \right)^2$, where $D = 19 \text{eV}$ is a typical deformation potential constant. (Note that the phonons that cause momentum relaxation have momentum $\sim 2 k_F$, and due to the low sound velocity, correspond to an energy much smaller than the chemical potential or the temperature.) There is another contribution to momentum relaxation due to scattering by surface phonons~\cite{chen2008intrinsic} residing in the substrate. For a $\text{SiO}_2$ substrate, this channel is, in fact, dominant above $T \approx 200 K$, which corresponds to the frequency of these surface phonons. Since this contribution is Arrhenius activated, it is strongly temperature dependent. The surface phonons play less of a role in hBN substrates which have higher optical phonon frequencies ($\sim 0.1$eV); Ref.~\cite{hBNgraphenepolarphonon} estimates the relaxation time due to surface polar phonons in $\text{hBN}$ substrates to both be around $\sim 10$ ps., at $T = 300$K and $\mu \approx 1000 K$ (or smaller), which translates to a scattering length $l_m \sim 10 \mu$m. Thus, on hBN substrates, the phonon scattering should be limited by acoustic phonons; thus, $l_m \approx 12 \mu$m $\times \left( \frac{D}{19 \text{eV}}\right)^2 \left( \frac{T}{300 \text{K}} \right)^2$.  

In order to measure the viscosity in the NV-center experiments, we would like $\sigma^T$ to be dominated by the term inversely proportional to the viscosity. Setting $\mu = k_B T$, we find $\frac{\sigma^T}{\sigma_0} \approx 1 + 11.05 / (q l_{ee})^2$ [from Eq.~(\ref{eq:hydroqtranscond})]. Thus, for $q > l_{ee}$ we expect the conductivity to be dominated by the viscous flow of the net charge $\rho_0$. 

Finally, we estimate the relaxation rate of the NV-center due to these hydrodynamic currents. We first note that, in the hydrodynamic regime, the NV-center should detect magnetic noise that is largely independent of its distance from the graphene surface. This is because the $1/q^2$ dependent conductivity in this regime precisely compensates for the distance dependence of the strength of electromagnetic fluctuations from a material surface. This behavior is distinct from both ballistic and diffusive behavior, as illustrated in Fig.~\ref{fig:noiseprotocol1} of the main text. The relaxation rate in this regime can be estimated as $\Gamma_{NV} \approx \sqrt{S(S+1)} g^2 \mu_B^2 N_z / 2 \hbar^2$, where $N_z = \mu^2_0 k_B T \sigma^T (q = 1/2z_{NV}) / (16 \pi z^2_{NV})$ and $g = 2$ is the Lande-factor of the NV center and $S = 1$ is the spin-size. We find $T_1 \approx 0.5 \text{s}  \left(\frac{\text{300 K}}{T}\right)^3$ which can be easily be detected since NV centers can be operated with a life-time of many seconds.

\section{Translational-symmetry breaking case}
\label{sec:appnontrans}

In this section, we examine the situation when a single impurity, associated for instance with a Kondo resonance, can significantly enhance the magnetic noise when the NV center is brought close to it. These single impurities can have interesting physics by themselves (temperature dependent resonance energy and line width, scattering properties etc.) that can be probed by the NV center as a novel spectroscopic probe. 

To deal with such a situation, we must modify our existing formalism to allow for conductivity and dielectric response functions that are, in principle, objects depending on two momenta. Our solution is to deal with these translationally variant response functions perturbatively. We assume that the conductivity comprises of a background single-momentum part, $\sigma_0 (\vs{Q})$ besides a smaller, two-momentum part $\sigma (\vs{Q}, \vs{Q}')$. 

In the reflection/transmission problem discussed in App.~\ref{sec:apptrans}, the magnetic dipole generates an electric field (at all wave-vectors, but we first consider a single wave-vector $\vs{Q}$), say $\vs{E}_0 (\vs{Q}, q_z)$, which impinges on the 2D material, and will now generate weaker ``source" currents with a different in-plane wave-vector $\vs{J}_s (\vs{Q'}) =  \sum_{\alpha \beta} \hat{\alpha} \sigma_{\alpha \beta} ( \vs{Q}', \vs{Q}) E_{0,\beta} (\vs{Q})$. These source currents will then generate additional outgoing electromagnetic waves $\vs{E}_1 (\vs{Q}', q'_z)$ that will modify the electromagnetic noise coming from the material. The amplitude of this additional field $\vs{E}_1$ must be determined consistently to first order in perturbation theory, in particular, in the presence of the additional ``induced" current $\vs{J}_1 (\vs{Q}') = \sigma_0 (\vs{Q}') \vs{E}_1 (\vs{Q}')$. Note that, in our reflection/transmission problem, both the ``induced" current $\vs{J}_1 (\vs{Q}')$ and ``source" current $\vs{J}_s (\vs{Q}')$ are of the same, higher order in the corrections to the conductivity, and must be treated together. At the same time, it is important to note that these higher order corrections can be treated as an independent electrodynamics problem on top of the solution that was found in previous sections when the conductivity was assumed to be given by $\sigma_0$. Thus, to find the corrections to the noise due to the two-momentum correction to the conductivity, we examine the electromagnetic problem of radiation in the presence of a ``source" current $\vs{J}_s( \vs{Q})$ (note that we have now dropped the prime superscript for ease of notation) in the material (and without any external incoming radiation). This current may be transverse or longitudinal. The two cases will be seen to decouple from one other with the transverse (longitudinal) source current producing only $s$-($p$-)polarized radiation. We consider them next. 

\subsection{Transverse source current / s-wave polarized outgoing waves}

The `outgoing' solutions comprise only of `reflected' and transmitted waves (using earlier nomenclature). For s-polarized waves, we have

\begin{align}
\vs E_r &= r^{(o)}_s \left( \hat{z} \times \hat{Q} \right) e^{i \vs{Q} \cdot \vs{\rho} - i q^\epsilon_z z }, \nonumber \\
\vs E_t &= t^{(o)}_s \left( \hat{z} \times \hat{Q} \right) e^{i \vs{Q} \cdot \vs{\rho} - i q^{\epsilon'}_z z}, \nonumber \\
\vs B_r &= \frac{r^{(o)}_s}{i \omega} \left( -i q^\epsilon_z \hat{Q} + i Q \hat{z} \right) e^{i \vs{Q} \cdot \vs{\rho} + i q^\epsilon_z z}, \nonumber \\
\vs B_t &= \frac{t^{(o)}_s}{i \omega} \left( i q^{\epsilon'}_z \hat{Q} + i Q \hat{z} \right) e^{i \vs{Q} \cdot \vs{\rho} - i q^{\epsilon'}_z z}, 
\label{eq:transout}
\end{align}

where the $o$ superscript is used to indicate that these are all outgoing waves. There is also a source current which we can quickly surmise to be transverse. The electric field $\vs{E}_r (z = 0)$ [or, by continuity, $\vs{E}_t (z = 0$)] generates an induced current within the 2D material which is transverse: $\vs{J}_1 (\vs{Q}) =  \sigma^T_0 (\vs{Q}) \vs{E} (z = 0) \propto \left(\hat{z} \times \hat{Q} \right)$. Moreover, since the perpendicular electric field is continuous across the surface (by construction), no charge is induced which, owing to the transverse nature of the induced current, implies that the source current must also be transverse. Thus, we assume $\vs{J}_s (\vs{Q}) = J^T_{s} \left(\hat{z} \times \hat{Q} \right) e^{i \vs{Q}. \rho}$.

Applying the usual boundary conditions, one can find:

\begin{align}
r^{(o)}_s (\vs{Q}) = \frac{- J^T_s}{\sigma^T_0 (\vs{Q}) + \frac{q^{\epsilon} + q^{\epsilon'}}{\mu_0 \omega}} \approx - \frac{\mu_0 \omega}{q^{\epsilon} + q^{\epsilon'}} J^T_s.
\end{align} 

\subsection{Longitudinal source current / p-wave polarized outgoing waves}

Here we will consider a longitudinal source current. (A transverse current will generate s-polarized radiation as described above.) The solutions will be of a p-polarized form:

\begin{align}
\vs B_r &= r^{(o)}_p \left( \hat{z} \times \hat{Q} \right) e^{i \vs{Q} \cdot \vs{\rho} - i q^\epsilon_z z }, \nonumber \\
\vs B_t &= t^{(o)}_p \left( \hat{z} \times \hat{Q} \right) e^{i \vs{Q} \cdot \vs{\rho} - i q^{\epsilon'}_z z}, \nonumber \\
\vs E_r &= \frac{r^{(o)}_p}{\epsilon} \frac{c^2}{-i \omega} \left( -i q^\epsilon_z \hat{Q} + i Q \hat{z} \right) e^{i \vs{Q} \cdot \vs{\rho} + i q^\epsilon_z z}, \nonumber \\
\vs E_t &= \frac{t^{(o)}_p}{\epsilon'} \frac{c^2}{-i \omega} \left( i q^{\epsilon'}_z \hat{Q} + i Q \hat{z} \right) e^{i \vs{Q} \cdot \vs{\rho} - i q^{\epsilon'}_z z}.
\label{eq:longout}
\end{align}

As before, the electric fields generate an induced current $\vs{J}_1 = \sigma^L_0 (\vs{Q}) \frac{r_p}{\epsilon} \frac{c^2}{-i \omega} ( - i q^\epsilon_z ) \hat{Q} e^{i \vs{Q} \cdot \vs{\rho}}$.  The currents generate a surface charge density $\rho  = \vs{Q}. (\vs{J}_1 + \vs{J}_s)  / \omega$ by the continuity relation. If we assume $\vs{J}_{s} = J^L_s \hat{Q} e^{i \vs{Q} \cdot \vs{\rho}}$, solving the electromagnetic boundary conditions yields the result

\begin{align}
r^{(o)}_p (\vs{Q}) = \frac{- \mu_0 J^L_s}{1 + \frac{\epsilon'}{\epsilon} \frac{q^{\epsilon}_z}{q^{\epsilon'}_z} + \frac{\sigma^L_0 (Q) }{ \epsilon \epsilon_0} \frac{q^\epsilon_z}{\omega} }.
\end{align} 

\subsection{two-momentum Conductivity}

The two-momentum conductivity $\sigma_{\alpha \beta} (\vs{Q}_1, \vs{Q}_2)$ is better represented in the basis where the left (right) sub-index corresponds to a direction either perpendicular ($T$) or parallel ($L$) to  $\vs{Q}_1$ ($\vs{Q}_2$): 

\begin{align}
\sigma_{\begin{Bmatrix}T \\ L \end{Bmatrix}, \begin{Bmatrix} T \\ L \end{Bmatrix}} & (\vs{Q}_1, \vs{Q}_2 ) \nonumber \\ &= \sum_{\alpha \beta} \begin{Bmatrix} \hat{z} \times \hat{Q}_1 \\ \hat{Q}_1 \end{Bmatrix}_{\alpha}  \sigma_{\alpha \beta} (\vs{Q}_1, \vs{Q}_2) \begin{Bmatrix} \hat{z} \times \hat{Q}_2 \\ \hat{Q}_2 \end{Bmatrix}_{\beta}, \nonumber \\
\end{align}

from which follows

\begin{align}
J_\alpha (\vs{Q}_1) &= \sum_\beta \sigma_{\alpha \beta} (\vs{Q}_1, \vs{Q}_2) E_\beta (\vs{Q}_2), \nonumber \\
\text{Re} [ \sigma_{\alpha, \beta} (\vs{Q}_1, \vs{Q}_2 )] & = \frac{\text{Im} [\Pi_{\alpha \beta} (\vs{Q}_1, - \vs{Q}_2, \omega)]}{\omega}, \nonumber \\
J_{\begin{Bmatrix}T \\ L \end{Bmatrix}} (\vs{Q}_1) &= \sum_\alpha \begin{Bmatrix} \hat{z} \times \hat{Q}_1 \\ \hat{Q}_1 \end{Bmatrix}_{\alpha} J_\alpha, \nonumber \\
J_{\begin{Bmatrix}T \\ L \end{Bmatrix}} (\vs{Q}_1) &= \sigma_{\begin{Bmatrix}T \\ L \end{Bmatrix}, \begin{Bmatrix} T \\ L \end{Bmatrix}}  (\vs{Q}_1, \vs{Q}_2 ) E_{\begin{Bmatrix}T \\ L \end{Bmatrix}} (\vs{Q}_2), \nonumber \\
\sigma_{T/L,T/L} (\vs{Q}_1, \vs{Q}_2 ) &= \frac{\text{Im} [\Pi_{T/L T/L} (\vs{Q}_1, - \vs{Q}_2, \omega)]}{\omega},
\end{align} 

where $\Pi_{\alpha \beta} (\vs{Q}_1, -\vs{Q}_2)$ is the retarded response function defined as a commutator of the currents $J_\alpha (\vs{Q}_1, \omega)$ and $J_\beta (- \vs{Q}_2, -\omega)$. The correlator $\Pi_{T/L T/L}$ corresponds to taking the transverse and longitudinal parts of these currents, and is defined with an additional negative sign which follows from our definition of $\sigma_{T/L,T/L}$. 

Transverse and longitudinal currents are fundamentally different: transverse currents are not mitigated due to strong Coulomb forces which greatly modify the dielectric constant and suppress longitudinal fluctuations. The conductivity is written in this basis to make this distinction obvious. As we will see in the results below, longitudinal conductivity corrections are always accompanied by a suppression due to a large dielectric constant. Thus, for the purposes of calculating noise corrections, the fully transverse part of the conductivity $\sigma_{T,T}$ will be most relevant. 

\subsection{Corrections to magnetic noise in the $z$-direction}

The electromagnetic field generated by a sheet of magnetic moment $\vs{M} = m_0 \hat{z} \delta(z - z_{NV}) e^{i \vs{Q}_0 \cdot\vs{\rho}}$ pointing in the $z$-direction is $s$-polarized. Neglecting two-momentum corrections to the conductivity, the form of the electric and magnetic fields for $z \le z_{NV}$ is

\begin{align}
\vs{E} &= E_0 e^{i q^\epsilon_{z,0} z_{NV}} (\hat{z} \times \hat{Q}_0 ) e^{i \vs{Q}_0 \cdot\vs{\rho}} \left( e^{-i q^\epsilon_{z,0} z} + r_s (Q_0) e^{i q^\epsilon_{z,0} z}  \right), \nonumber \\
\vs{B} &= E_0 e^{i q^\epsilon_{z,0} z_{NV}} \frac{q^\epsilon_{z,0} \hat{Q}_0 }{\omega} e^{i \vs{Q}_0 \cdot\vs{\rho}}  \left( e^{-i q^\epsilon_{z,0} z} - r_s (Q_0) e^{i q^\epsilon_{z,0} z}  \right), \nonumber \\
E_0 &= i \mu_0 \omega m_0 \frac{Q_0}{2 q^\epsilon_{z,0}},
\end{align}

where $q^\epsilon_{z,0} = \sqrt{\epsilon \omega^2/c^2 - Q^2_0}$. Now we consider the corrections to the electromagnetic field due to the corrections to the conductivity. First, the electric field at the surface (which is transverse, and at wave-vector $\vs{Q}_0$) will generate two `source' transverse and longitudinal currents at wave-vector $\vs{Q}_1$ with amplitude

\begin{align}
J^{T/L}_s (\vs{Q}_1, \vs{Q}_0) &= \sigma_{T/L,T} (\vs{Q}_1, \vs{Q}_0) E_0 e^{i q^\epsilon_z z_{NV}} \left(1 + r_s (Q_0) \right). \nonumber \\
\end{align}

Noting [from Eqs.~(\ref{eq:longout})] that longitudinal currents do not generate magnetic fields in the $z$-direction (and hence, do not affect magnetic noise in the $z$-direction), we ignore such fluctuations. The transverse source-current amplitude directly yields the corrections to the magnetic field using Eqs.~(\ref{eq:transout}); the magnetic field corrections in the case of the magnetization sheet are given by 

\begin{align}
B^{sheet}_z (\vs{\rho},z_{NV}) &= - \int \frac{d^2 \vs{Q}_1}{(2 \pi)^2} e^{i \vs{Q}_1 \cdot\vs{\rho} } \frac{\vs{Q}_1}{\omega} E_0 (Q_0) \left( 1 + r_s ( Q_0) \right) \nonumber \\
& e^{i (q^{\epsilon}_{z,1} + i q^{\epsilon}_{z,0}) z_{NV} } \frac{\sigma_{T,T} ( \vs{Q}_1, \vs{Q}_0 )}{ \sigma^T_0 (Q_1) + \frac{q^{\epsilon}_{z,1} + q^{\epsilon}_{z,0}}{ \mu_0 \omega} }.
\end{align}

where $q^\epsilon_{z,1} = \sqrt{\epsilon \omega^2/c^2 - Q^2_1}$ and we have neglected the contribution from traveling waves with momentum $Q_1,Q_0 < \sqrt{\epsilon} \omega / c$, so that both $q^\epsilon_{z,1}, q^\epsilon_{z,0}$ are imaginary (with positive imaginary parts). The magnetic field in the case of a single magnetic moment can now be found by integration: $B_{z} (\vs{\rho}_{NV}, z_{NV}) = \int \frac{d^2 \vs{Q}_0}{(2 \pi)^2} B^{sheet}_z ((\vs{\rho}_{NV} ,z_{NV}) e^{- i \vs{Q}_0 \cdot\vs{\rho}_{NV} })$. Finally, the response function $\chi_{zz} (\omega) = B_z (\vs{\rho}_{NV}, z_{NV}) / m_0$ from which we can calculate the noise using the fluctuation-dissipation relation. 

The result above can be simplified greatly by realizing that $\sigma^T_0 (Q_1) \ll i (q_1 + q_2) / (\mu_0 \omega) $ for momenta ($\sim 1\mu m^{-1}$) and frequencies ($\sim 1$ GHz) of interest. This yields

\begin{align}
&N_{z} (\vs{r}_{NV}) = \frac{\mu^2_0 k_B T}{2} \int_0^\infty \frac{q_1 d q_1 d \theta_1}{(2\pi)^2}  \int_0^\infty \frac{q_2 d q_2 d \theta_2}{(2\pi)^2}  \nonumber \\
&\quad e^{i \rho_{NV} (q_1 \cos \theta_1 - q_2 \cos \theta_2 ) - (q_1 + q_2 ) z_{NV} } \text{Re} \left[ \sigma_{T,T} (\vs{q}_1,\vs{q}_2) \right].
\label{eq:nzzkondo}
\end{align}

Let us note that this result (which holds under previously justified approximations) is a straightforward generalization of the result in Eq.~(\ref{eq:chigraphenezz}). In particular, setting $\text{Re} \left[ \sigma_{T,T} (\vs{q}_1,\vs{q}_2) \right] = \text{Re} \left[ \sigma_{T,T} (\vs{q}_1) \right]  \delta(\vs{q}_1 - \vs{q}_2)$ immediately reproduces the translationally-invariant result. Thus, one may think of the Eqs.~(\ref{eq:nontransnoise}) as the more general result for noise due to a material with translationally non-invariant conductivity. 

\subsection{Corrections to in-plane magnetic noise}

As before, we consider this case by first finding the form of the electromagnetic fields in the presence of a sheet of magnetization which fluctuates at a frequency $\omega$ and fixed in-plane wave-vector $\vs{Q}_0$. In line with our previous calculations, it helps to consider the two cases where the magnetization direction is either $\hat{Q}_0$ or $\hat{z} \times \hat{Q}_0$. When $\vs{M} \parallel \hat{z} \times \hat{Q}_0$, we know that the amplitude of the magnetic field is suppressed by a factor $\omega^2 /c^2 q^2$ which largely suppresses the contribution to the noise from such radiation. Therefore, we do not consider it further here. 

We now analyze the case $\vs{M} = m_0 \hat{Q}_0 \delta(z - z_{NV}) e^{i \vs{Q}_0 \cdot\vs{\rho}}$. The in-plane components of the electromagnetic fields for $z < z_{NV}$, before accounting for two-momentum corrections to the conductivity, are

\begin{align}
\vs{E} &= E_0 e^{i q^\epsilon_{z,0} z_{NV}} (\hat{z} \times \hat{Q}_0 ) e^{i \vs{Q}_0 \cdot\vs{\rho}} \left( e^{-i q^\epsilon_{z,0} z} + r_s (Q_0) e^{i q^\epsilon_{z,0} z}  \right), \nonumber \\
\vs{B} &= E_0 e^{i q^\epsilon_{z,0} z_{NV}} \frac{q^\epsilon_{z,0} \hat{Q}_0}{\omega} e^{i \vs{Q}_0 \cdot\vs{\rho}}  \left( e^{-i q^\epsilon_{z,0} z} + r_s (Q_0) e^{i q^\epsilon_{z,0} z}  \right), \nonumber \\
E_0 &= \frac{-i \mu_0 \omega m_0}{2}.
\end{align}

This produces both transverse and longitudinal currents in the 2d system:
\begin{align}
J^{T/L}_s (\vs{Q}_1, \vs{Q}_0) &= \sigma_{T/L,T} (\vs{Q}_1, \vs{Q}_0) E_0 e^{i q^\epsilon_z z_{NV}} \left(1 + r_s (Q_0) \right) \nonumber \\
\end{align}

The transverse and longitudinal source-current amplitudes directly yield the corrections to the in-plane magnetic field using Eqs.~(\ref{eq:transout}) and~(\ref{eq:longout}) given by

\begin{align}
&\vs{B}^{sheet}_{(T)} (\vs{\rho},z_{NV}) = \int \frac{d^2 \vs{Q}_1}{(2 \pi)^2} e^{i \vs{Q}_1 \cdot\vs{\rho} } \frac{\hat{Q}_1 q^\epsilon_{z,1}}{\omega} E_0 (Q_0)  \nonumber \\
&\quad\quad\quad e^{i (q^{\epsilon}_{z,1} + i q^{\epsilon}_{z,0})z_{NV} } \frac{ \left( 1 + r_s ( Q_0) \right)\sigma_{T,T} ( \vs{Q}_1, \vs{Q}_0 )}{ \sigma^T_0 (Q_1) + \frac{q^{\epsilon}_{z,1} + q^{\epsilon}_{z,0}}{ \mu_0 \omega} }, \nonumber \\
&\vs{B}^{sheet}_{(L)} (\vs{\rho},z_{NV}) = - \mu_0 \int \frac{d^2 \vs{Q}_1}{(2 \pi)^2} e^{i \vs{Q}_1 \cdot\vs{\rho} } (\hat{z}\times\hat{Q}_1) E_0 (Q_0) \nonumber \\
 &\quad\quad\quad  e^{i (q^{\epsilon}_{z,1} + i q^{\epsilon}_{z,0})z_{NV} } \frac{\sigma_{L,T} ( \vs{Q}_1, \vs{Q}_0 )\left( 1 + r_s ( Q_0) \right) }{1 + \frac{\epsilon'}{\epsilon} \frac{q^{\epsilon_{z,1}}}{q^{\epsilon'}_{z,1}} + \frac{\sigma^L_0 (Q_1)}{\epsilon \epsilon_0} \frac{q^\epsilon_{z,1}}{\omega} }.
\end{align}

Note that the out-of-plane magnetic field from $J^{T/L}_s (\vs{Q}_1, \vs{Q}_0)$ comes with an extra factor of $i$; this implies that the noise correlations $N_{xz}$ and $N_{yz}$ are negligible as they are proportional to the imaginary part of the conductivity which comes with an additional factor of $\omega \tau$.  Also note that, for a clean conducting sample, the conductivity is typically large enough that the contribution from the longitudinal current $B^{sheet}_{(L)}$ can be neglected. Thus, the total magnetic field correction $\vs{B}^{sheet} \sim \vs{B}^{sheet}_{(T)}$. If we again make the approximation of neglecting $\sigma^T_0 (Q_1)$ in the denominator, we arrive at the result for the in-plane magnetic noise $N_{\hat{n}_1 \hat{n}_2}$ provided in the main text in Eq.~(\ref{eq:nontransnoise}).  

Note that we have assumed that the single impurity that generates two-momentum corrections to the conductivity resides at the origin. This ensures that the conductivity $\sigma (\vs{q}_1, \vs{q}_2)$ does not pick up phase factors $e^{i (\vs{q}_1 - \vs{q}_2). \vs{r}_f}$ where $\vs{r}_f$ is the impurity position. This allows the conductivity to be symmetric in the two momenta, and simplifies the final form of the result. We have also assumed, without loss of generality that $\vs{\rho}_{NV} \parallel \hat{x}$; this implies that $N_{x} \neq N_{y}$ in contrast to the previous, translationally invariant case.

Finally, we note that above result reduces to the translationally invariant case [as in Eq.~(\ref{eq:chigraphenexx})] by setting $\text{Re} \left[ \sigma_{T,T} (\vs{q}_1,\vs{q}_2) \right] = \text{Re} \left[ \sigma_{T,T} (\vs{q}_1) \right]  \delta(\vs{q}_1 - \vs{q}_2)$. 

\section{two-momentum conductivity in the presence of a Kondo impurity}
\label{sec:appkondo}

We would like to evaluate the corrections to the current-current correlator) given by 

\begin{align} 
\Pi_{\alpha \beta} (\vs{q}_1 , -\vs{q}_2; \tau - \tau') &= - \frac{1}{\mathcal{V} \beta }\avg{ T_\tau [J_\alpha (\vs{q}_1, \tau) J_\beta (- \vs{q}_2, \tau')]} \nonumber \\
J_\alpha (\vs{q}_1, \tau) &= -\frac{e}{m \mathcal{V}} \sum_{\vs{q}_1} \left( \vs{q} + \frac{\vs{q}_1}{2} \right)_\alpha \times\nonumber\\
&\quad\quad\quad\quad c^\dagger_{\vs{q}} (\tau + 0^+) c_{\vs{q} + \vs{q}_1} (\tau). \nonumber \\
\end{align}

Note that for graphene operated away from the charge neutrality point, we can make the replacement $\vs{q}/m \rightarrow v_F \hat{\vs{q}}$. The results discussed below will apply to graphene operated in this regime as well.

In particular, the corrections to the current-current correlator comes from the diagrams in Fig.~\ref{fig:currentcorrdiags}. These diagrams can be expanded as:

\begin{align}
\Pi_{\alpha \beta}^{(1)} (\vs{q}_1 , -\vs{q}_2, i k_n ) &= \frac{e^2}{\beta m^2 \mathcal{V}} \sum_{\vs{q} \; i q_n} \left( \vs{q} + \frac{\vs{q}_1}{2} \right)_\alpha \left( \vs{q} + \frac{\vs{q}_2}{2} \right)_\beta \times \nonumber \\
& G_0(\vs{q},\vs{q}, iq_n) G_1 (\vs{q} + \vs{q}_1, \vs{q} + \vs{q}_2, iq_n + ik_n ), \nonumber \\
\Pi_{\alpha \beta}^{(2)} (\vs{q}_1 , -\vs{q}_2, i k_n ) &= \frac{e^2}{\beta m^2 \mathcal{V}} \sum_{\vs{q} \; i q_n} \left( \vs{q} + \frac{\vs{q}_1}{2} \right)_\alpha \left( \vs{q} + \frac{\vs{q}_2}{2} \right)_\beta \times \nonumber \\
& G_0(\vs{q},\vs{q}, iq_n + ik_n) G_1 (\vs{q} + \vs{q}_1, \vs{q} + \vs{q}_2, iq_n ), \nonumber \\
\Pi_{\alpha \beta}^{(3)} (\vs{q}_1 , -\vs{q}_2, i k_n ) &= \frac{e^2}{\beta m^2 \mathcal{V}^2} \sum_{\vs{q}, \vs{q}' \; i q_n} \left( \vs{q} + \frac{\vs{q}_1}{2} \right)_\alpha \left( \vs{q}' + \frac{\vs{q}_2}{2} \right)_\beta \times \nonumber \\
& G_1(\vs{q}',\vs{q}, iq_n + ik_n) G_1 (\vs{q} + \vs{q}_1, \vs{q}' + \vs{q}_2, iq_n ), \nonumber \\
\Pi_{\alpha \beta}^{(2)} (\vs{q}_1 , -\vs{q}_2, i k_n ) &= \Pi_{\alpha \beta}^{(1)} (\vs{q}_1 , -\vs{q}_2, - i k_n ),
\end{align}

where $G_0$ and $G_1$ are the bare electron Green's function and the correction to it due to the scattering off of an impurity, respectively, and as described in Eq.~(\ref{eq:gcgreen}). The sum over indices such as spin for a metal, or spin and valley for graphene are implied and will be absorbed in the density of states $\nu(0)$. Note that $\Pi^{(3)}$ evaluates to zero for the transverse current-current correlations we are interested in. An argument for this will be provided below. 

Before we proceed with the calculation, we make a small note on the presence of exponential factors inside $\text{Re} [...]$ and $\text{Im} [...]$ operations that may concern the attentive reader. We note that the noise is always proportional to the imaginary part of the response functions given in real-space. That is, to evaluate the noise, we have to compute $\text{Im} \left[ \frac{1}{\mathcal{V}^2} \sum_{\vs{q}_1 \; \vs{q}_2} \Pi^{(1/2)} (\vs{q}_1, - \vs{q}_2, \omega + i 0^+ )  e^{i \vs{q}_1 \cdot \vs{r}_1 - i \vs{q}_2 \cdot \vs{r}_2 } \right] $. It is typically possible to bring the exponential factors outside of the operation $\text{Im} [...]$ because of the symmetry $ \vs{q}_1 \leftrightarrow \vs{q}_2$. However, if the position of the impurity $\vs{r}_f \neq 0$, then the Green's function $G_1 (\vs{q}_1,\vs{q}_2)$ will contain an explicit factor of $e^{i (\vs{q}_1 - \vs{q}_2). \vs{r}_f}$ that will destroy this symmetry. To deal with this, we can define $\Pi^{(1/2)'}_{\alpha \beta} (\vs{q}_1, -\vs{q}_2 , ik_n) = e^{i \vs{q}_1 \cdot \vs{r}_f - i \vs{q}_2 \cdot \vs{r}_f} \Pi^{(1/2)}_{\alpha \beta} (\vs{q}_1, -\vs{q}_2 , ik_n)$ which does have the aforementioned symmetry. In this case, 

\begin{align}
\frac{1}{\mathcal{V}^2} \sum_{\vs{q}_1 \; \vs{q}_2} \text{Im} \bigg[ \Pi^{(1/2)}_{\alpha \beta} & (\vs{q}_1, - \vs{q}_2, \omega + i 0^+ ) e^{i \vs{q}_1 \cdot \vs{r}_1 - i \vs{q}_2 \cdot \vs{r}_2 } \bigg] = \nonumber \\
\frac{1}{\mathcal{V}^2} \sum_{\vs{q}_1 \; \vs{q}_2}  e^{i \vs{q}_1 \cdot (\vs{r}_1 - \vs{r}_f)} & e^{- i \vs{q}_2 \cdot (\vs{r}_2 - \vs{r}_f) } \nonumber\\
&\times\text{Im} \bigg[ \Pi^{(1/2) '}_{\alpha \beta} (\vs{q}_1, - \vs{q}_2, \omega + i 0^+ ) \bigg].
\end{align} 

One can think of $\Pi^{(1/2) '}_{\alpha \beta} (\vs{q}_1, -\vs{q}_2 , ik_n) $ as simply the result for current-current fluctuations when $\vs{r}_f = 0$. Note that these extra exponential factors are precisely the same as those in Eq.~(\ref{eq:nontransnoise}) (due to $\vs{r}_{NV} - \vs{r}_f$) but there we set $\vs{r}_f = 0$ for convenience; we will assume $\vs{r}_f = 0$ in what follows. 

\begin{center}
\begin{figure}
\includegraphics[width=1.7in]{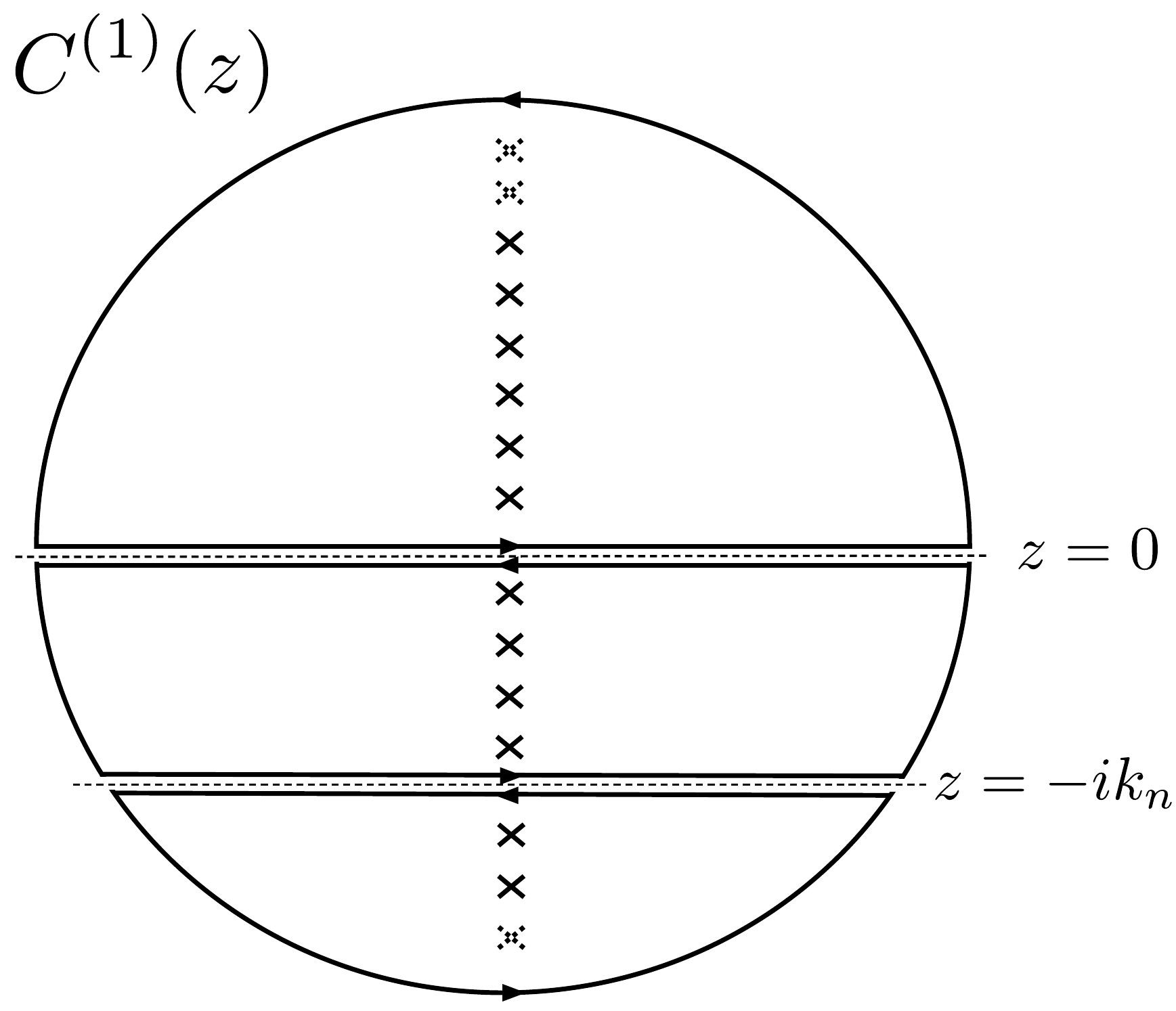}
\caption{The contour $C^{(1)} (z)$ for evaluating the Matsubara sum in $\Pi^{(1)}$.}
\label{fig:contour}
\end{figure}
\end{center}

We first evaluate the Matsubara sum by an integration over the contour shown in Fig.~\ref{fig:contour}. For the bare bubble (without the impurity line), integration quite straightforwardly yields 
\begin{align}
&\text{Im} \big[ \Pi^{(0)}_{\alpha \beta} (\vs{q}_1, - \vs{q}_2, \omega + i0^+) \big] = \pi e^2 \sum_{\vs{q}} \int^\infty_{-\infty} d \omega' \frac{ (\vs{q}+\vs{q}_1/2)_\alpha}{m^2} \nonumber \\
& \times (\vs{q}+\vs{q}_2/2)_\beta [ n_F (\omega') - n_F (\omega + \omega') ] A_0 ( \vs{q}, \omega') A_0 ( \vs{q}, \omega + \omega'), \nonumber \\
&A_0 (\vs{q}, \omega) = \frac{1}{\pi} \frac{\frac{1}{2 \tau}}{(\omega - \epsilon_{\vs{q}})^2 + \left(\frac{1}{2\tau}\right)^2}, \nonumber \\
&\text{Im} \big[ \Pi^{(0)}_{\alpha \beta} (\vs{q}_1 =0, \vs{q}_2 = 0, \omega = 0) \big] \bigg|_{T \rightarrow 0} =  \delta_{\alpha \beta} \; \mathcal{V} \delta_{\vs{q}_1 \vs{q}_2} \; \omega \sigma_0, \nonumber \\ 
&\sigma_0 = \frac{\nu(0) e^2 v^2_F \tau}{2},
\end{align}
where we checked that current-current noise due to just the metal agrees with the Drude result.  Next, we perform the calculation for the impurity-related diagrams in Fig.~\ref{fig:currentcorrdiags}. This yields 

\begin{align}
& \text{Im} \big[ \Pi^{(1)}_{\alpha \beta} (\vs{q}_1, -  \vs{q}_2, \omega + i0^+) \big] = \frac{\pi e^2}{\mathcal{V}} \sum_{\vs{q}} \int^\infty_{-\infty} d\omega' \frac{ (\vs{q}+\vs{q}_1/2)_\alpha }{m^2} \nonumber \\
& \times (\vs{q}+\vs{q}_2/2)_\beta [ n_F (\omega') - n_F (\omega + \omega') ] A_0 ( \vs{q}, \omega') \nonumber \\
& \times A_1 (\vs{q} + \vs{q}_1, \vs{q} + \vs{q}_2, \omega + \omega'), \nonumber \\
& \text{Im} \big[ \Pi^{(2)}_{\alpha \beta} (\vs{q}_1, - \vs{q}_2, \omega + i0^+) \big] = \frac{\pi e^2}{\mathcal{V}} \sum_{\vs{q}} \int^\infty_{-\infty} d\omega' \frac{ (\vs{q}+\vs{q}_1/2)_\alpha }{m^2} \nonumber \\
& \times  (\vs{q}+\vs{q}_2/2)_\beta [ n_F (\omega') - n_F (\omega + \omega') ] A_0 ( \vs{q}, \omega + \omega') \nonumber \\
& \times A_1 (\vs{q} + \vs{q}_1, \vs{q} + \vs{q}_2, \omega'), \nonumber \\
& A_1 ( \vs{q} + \vs{q}_1, \vs{q} + \vs{q}_2, \omega) = \frac{-1}{\pi} \text{Im} \bigg [ G_1 (\vs{q} + \vs{q}_1 , \vs{q} + \vs{q}_2, \omega + i0^+) \bigg]. \nonumber \\
\end{align}

Note that, in the limit of interest, $\omega \rightarrow 0$ (that is, $\omega = \omega_{NV}$ is the smallest scale in our problem), $\Pi^{(1)}$ and $\Pi^{(2)}$ are identical. Another simplification is that we only require the transverse parts $\Pi^{(1/2)}_{T,T} (\vs{q}_1, -\vs{q}_2, \omega)$. Thus, the vertex terms $(\vs{q} + \vs{q}_1/2)_\alpha = q_\alpha$, $(\vs{q} + \vs{q}_2/2)_\beta = q_\beta$. We can also safely replace $n_F (\omega') - n_F (\omega + \omega') \approx - n'_F(\omega') \omega$. Finally, as long as $q_1, q_2 \ll k_F$, we can replace the integral $\frac{1}{\mathcal{V}} \sum_{\vs{q}} \approx \int^{\infty}_{-\infty} d \xi \; \nu(0) \int^{2\pi}_0 \frac{d\theta}{2\pi}$, and set $\epsilon_{\vs{q} + \vs{q}_1} \approx \xi + v_F q_1 \cos \theta_1$, $\epsilon_{\vs{q} + \vs{q}_2} \approx \xi + v_F q_2 \cos \theta_2$, and $q_\alpha q_\beta \approx k^2_F \sin(\theta_1 - \theta) \sin (\theta_2 - \theta)$. Here $\theta$, $\theta_1$, and $\theta_2$ are the azimuthal angles of wave-vectors $\vs{q}$, $\vs{q}_1$ and $\vs{q}_2$ respectively. Finally, with these assumptions, and setting $\omega \rightarrow 0$ in the Green's functions, we find

\begin{widetext}
\begin{align}
\text{Im} \left[\Pi_T (\vs{q}_1,-\vs{q}_2,\omega^+) \right] &= \frac{\omega \nu(0) e^2 v^2_F}{2} \int d\omega' \int d \xi \int d \theta \sin(\theta-\theta_1) \sin (\theta- \theta_2) \left(-n'_F (\omega') \right) \frac{1}{\pi} \frac{\frac{1}{2 \tau}}{(\omega' - \xi)^2 + \left( \frac{1}{2 \tau} \right)^2} \frac{-1}{\pi} \nonumber \\
&\times \text{Im} \left[ \frac{1}{(\omega' - \xi  - v_F q_1 \cos (\theta - \theta_1) + \frac{i}{2 \tau})} \frac{1}{(\omega' - \xi  - v_F q_2 \cos (\theta - \theta_2) + \frac{i}{2 \tau})} T_f (\omega') \right], \nonumber 
\end{align}
where performing $\xi \rightarrow \xi + \omega' $ and integrating out $ \xi $ yields
\begin{align}
\text{Im} \left[\Pi_T (\vs{q}_1,-\vs{q}_2,\omega^+) \right] &= \frac{\omega \nu(0) e^2 v^2_F}{2} \int d\omega' \int d \theta \sin(\theta-\theta_1) \sin (\theta- \theta_2) \left(-n'_F (\omega') \right) \nonumber \\
& \frac{-1}{\pi} \times \text{Im} \left[ \frac{1}{\left(v_F q_1 \cos (\theta - \theta_1) + \frac{i}{\tau} \right)\left(v_F q_2 \cos (\theta - \theta_1) + \frac{i}{\tau} \right)} T_f (\omega') \right]. \nonumber
\end{align}

One may now integrate over $\theta$ analytically to find

\begin{align}
\text{Im} \left[\Pi_T (\vs{q}_1,-\vs{q}_2,\omega^+) \right] &= \omega \nu(0) e^2 v^2_F \tau^2 F_1[x_1 = v_F q_1\tau ,x_2 = v_F q_2 \tau ,\theta_1-\theta_2] F_2 [T],  \nonumber \\
F_1\left[x_1 ,x_2, \theta \right] &=  2 \pi \frac{x^2_2 ( 1 - \sqrt{1 + x^2_1} ) + x^2_1 (1 - \sqrt{1 + x^2_2}) + x^2_1 x^2_2 \sin^2 \theta + x_1 x_2 \cos \theta (\sqrt{1 + x^2_1} + \sqrt{1 + x^2_2} - 2) }{4x^2_1 x^2_2 \cos \theta - 2 x_1 x_2 (x^2_1 + x^2_2) - x^3_1 x^3_2 (1 - \cos 2 \theta )}, \nonumber \\
F_2 [T] &= \frac{-1}{\pi} \int d \omega (-n'_F (\omega)) \text{Im} [ T_f (\omega) ], \nonumber \\
&\text{Re} \left[ \sigma_{T} (\vs{q}_1, \vs{q}_2, \omega) \right] = 2 \omega \nu(0) e^2 v^2_F \tau^2 F_1 [x_1,x_2, \theta_1 - \theta_2] F_2 [T] = 4 \sigma_0 \tau F_1 [x_1,x_2, \theta_1-\theta_2] F_2 [T].
\label{eq:kondocond}
\end{align}

For the purposes of completeness, we also mention the result for the longitudinal current fluctuations (which are calculated analogously):

\begin{align}
\text{Im} \left[\Pi_L (\vs{q}_1,-\vs{q}_2,\omega^+) \right] &= \frac{\omega \nu(0) e^2 v^2_F}{2} \int d\omega' \int d \xi \int d \theta \cos(\theta-\theta_1) \cos (\theta- \theta_2) \left(-n'_F (\omega') \right) \frac{1}{\pi} \frac{\frac{1}{2 \tau}}{(\omega' - \xi)^2 + \left( \frac{1}{2\tau} \right)^2} \frac{-1}{\pi} \nonumber \\
&\times \text{Im} \left[ \frac{1}{\omega' - \xi  - v_F q_1 \cos (\theta - \theta_1) + \frac{i}{\tau}} \frac{1}{\omega' - \xi  - v_F q_2 \cos (\theta - \theta_2) + \frac{i}{\tau}} T_f (\omega') \right], \nonumber \\
\text{Re} \left[ \sigma_{L} (\vs{q}_1, \vs{q}_2, \omega) \right] &= 2 \nu(0) e^2 v^2_F \tau^2 F_{1,L} [x_1,x_2, \theta_1 - \theta_2] F_2 [T] = 4 \sigma_0 \tau F_{1,L} [x_1,x_2, \theta] F_2 [T], \nonumber \\
F_{1,L}\left[q_1 (x_1) ,q_2 (x_2), \theta \right] &=  \frac{ 2 \pi}{-4x^2_1 x^2_2 \cos \theta + 2 x_1 x_2 (x^2_1 + x^2_2) + x^3_1 x^3_2 (1 - \cos 2 \theta )}\nonumber \\
&\times\Bigg[x^2_1 + x^2_2 - \frac{x^2_1}{\sqrt{1+x^2_2}} - \frac{x^2_2}{\sqrt{1+x^2_1}} + x_1 x_2 \cos \theta \left(\frac{1}{\sqrt{1+x^2_2}} +  \frac{1}{\sqrt{1+x^2_1}} -2 \right) \nonumber\\
&\quad\quad+x^2_1 x^2_2 \sin^2 \theta \left(1 -  \frac{1}{\sqrt{1+x^2_2}} -\frac{1}{\sqrt{1+x^2_1}} \Bigg)\right].
\label{eq:kondocondlong}
\end{align}
\end{widetext}

Let us briefly note why $\Pi^{(3)} \propto G_1 (\vs{q}, \vs{q}') G_1 (\vs{q} + \vs{q}_1, \vs{q}' + \vs{q}_2)$, represented by the diagram corresponding to scattering of both the particle and the hole, vanishes. The introduction of an additional momentum $\vs{q}'$ due to another scattering event makes the whole term factorizable in terms of parts $G_0 (\vs{q}) G_0 (\vs{q} + \vs{q}_1)$ and $G_0 (\vs{q}) G_0 (\vs{q} + \vs{q}_2)$. For the transverse current-current correlations, integration over the angle $\theta$ of $\vs{q}$ now proceeds as $\int d \theta \sin (\theta - \theta_1) / (\omega' - v_F k_1 \cos (\theta - \theta_1) + \frac{i}{2 \tau})$ which vanishes identically. Thus, this diagram does not contribute to the \emph{transverse} current-current correlations. We also note that this diagram does contribute to longitudinal correlations (the $\sin(\theta- \theta_1)$ above is replaced by $\cos(\theta - \theta_1)$ and the integral is finite) but we do not evaluate it since the noise does not depend on longitudinal correlations. 

\subsection{Noise from a Kondo impurity in the far-field}

The limiting forms of the function $F_1[x,x',\theta]$ are

\begin{widetext}
\begin{align}
F_1[x_1 \ll 1, x_2 \ll 1, \theta]  &\approx -\pi/2 \cos \theta, \nonumber \\
F_1[x_1 \gg 1, x_2 \gg 1, \theta] &\approx \begin{Bmatrix} - \frac{2\pi}{x_1 x_2} \; ; \; \theta \notin \left[\pi - \frac{2}{\sqrt{x_1 x_2}} , \pi + \frac{2}{\sqrt{x_1 x_2}} \right] \\ + \frac{2\pi}{4} \left( \frac{1}{x_1} + \frac{1}{x_2} \right) \; ; \; \theta \in \left[\pi - \frac{2}{\sqrt{x_1 x_2}} , \pi + \frac{2}{\sqrt{x_1 x_2}} \right]  \end{Bmatrix}. \nonumber \\
\label{eq:f1forms}
\end{align}
\end{widetext}

The noise spectrum can now, in principle, be calculated using the results in Eqs.~(\ref{eq:nontransnoise}) and using the two-momentum conductivity in Eq.~(\ref{eq:kondocondsimp}). In general, the integrals must be evaluated numerically, but they are analytically tractable in the far-field regime (the impurity distance is much greater than the mean-free path of the electrons in the sample: $z_{NV} \gg l_m$). Here we present these analytical results for this case. For momenta $q \sim 1/z_{NV} \ll 1/l_m$, the two momenta conductivity reduces to (by taking the limit $x_1 \rightarrow 0$, $x_2 \rightarrow 0$ in $F_1[x_1,x_2,\theta]$) $\sigma(\vs{q}_1, \vs{q}_2, \omega_{NV}) \approx -2 \pi \cos(\theta) \sigma_0 F_2 [T] \tau$, where $\theta$ now is the angle between $\vs{q}_1$ and $\vs{q}_2$. If we assume that the $x$-axis corresponds to the line joining the impurity (at $\vs{r}_f = \vs{0}$) to the NV center (projected on to the plane), then $N_{xy} = 0$ in this limit. The diagonal components of the noise tensor are
\begin{widetext}
\begin{align}
N_{z} (z_{NV} \gg l_m) &= \frac{\mu^2_0 k_B T \sigma_0}{16 \pi z^2_{NV}} - \frac{\mu^2_0 k_B T \sigma_0 \tau}{4 \pi} F_2 [T] \frac{\rho_{NV}^2}{(\rho^2_{NV} + z^2_{NV} )^3}, \nonumber \\
N_{x} (z_{NV} \gg l_m) &= \frac{\mu^2_0 k_B T \sigma_0}{32 \pi z^2_{NV}} - \frac{\mu^2_0 k_B T \sigma_0 \tau}{4 \pi} F_2 [T] \frac{\left( 2 \rho_{NV}^2 z_{NV} + z_{NV}^3 - (\rho^2_{NV} + z^2_{NV} )^{3/2} \right)^2}{\rho^4_{NV} (\rho^2_{NV} + z^2_{NV} )^3}, \nonumber \\
N_{y} (z_{NV} \gg l_m) &= \frac{\mu^2_0 k_B T \sigma_0}{32 \pi z^2_{NV}} - \frac{\mu^2_0 k_B T \sigma_0 \tau}{4 \pi} F_2 [T] \frac{(\sqrt{\rho^2_{NV} + z^2_{NV}} - z_{NV})^2}{\rho^4_{NV} (\rho^2_{NV} + z^2_{NV} )}. \nonumber \\
\label{eq:Kondofar}
\end{align} 
\end{widetext}

%

\end{document}